\documentclass[]{aa}  

\usepackage{graphicx}
\usepackage{txfonts}
\usepackage{siunitx}
\usepackage{ulem}
\usepackage[hyphenbreaks]{breakurl} 

\begin{document}

   \title{COSMOS2020: manifold learning to estimate physical parameters in large galaxy surveys}

   \author{I.~Davidzon
          \inst{1,2}\fnmsep\thanks{Corresponding authors, \email{iary.davidzon@nbi.ku.dk, keerthana.jegatheesan@mail.udp.cl}}
          \and
          K.~Jegatheesan\inst{3,4}\fnmsep$^{\star}$
          \and
          O.~Ilbert\inst{4,5,6}
          \and 
          S.~de la Torre\inst{4}
          \and
          S.~K.~Leslie\inst{7}
          \and 
          C.~Laigle\inst{8}
          \and
             S.~Hemmati\inst{9}
          \and
            D.~C.~Masters\inst{9}
          \and
          D.~Blanquez-Sese\inst{1,10}
          \and
          O.~B.~Kauffmann\inst{4}
          \and
          G.~E.~Magdis\inst{1,10,2}
          \and 
          K.~Ma{\l}ek\inst{11,4}
          \and
          H.~J.~McCracken\inst{8}
          \and 
          B.~Mobasher\inst{12}
          \and
          A.~Moneti\inst{8}
          \and
          D.~B.~Sanders\inst{13,1}
          \and
          M.~Shuntov\inst{8}
          \and
          S.~Toft\inst{1,2}
          \and
          J.~R.~Weaver\inst{1,2}
          }

   \institute{
        Cosmic Dawn Center (DAWN), Denmark   
         \and
        Niels Bohr Institute, University of Copenhagen, Jagtvej 128, 2200, Copenhagen N, Denmark 
        \and 
        N\'ucleo de Astronom\'ia de la Facultad de Ingenier\'ia y Ciencias, Universidad Diego Portales, Av. Ej\'ercito Libertador 441, Santiago, Chile  
        \and 
        Aix Marseille Univ, CNRS, CNES, LAM, Marseille, France   
        \and
        California Institute of Technology, Pasadena, CA 91125, USA 
        \and
        Jet Propulsion Laboratory, California Institute of Technology, Pasadena, CA 91109, USA 
        \and
        Leiden Observatory, Leiden University, PO Box 9513, NL-2300 RA Leiden, the Netherlands 
        \and
        Sorbonne Universit\'e, CNRS, UMR 7095, Institut d'Astrophysique de Paris, 98 bis bd Arago, 75014, Paris, France  
        \and
        Infrared Processing and Analysis Center, California Institute of Technology, Pasadena, CA 91125, USA   
        \and
        DTU-Space, Technical University of Denmark, Elektrovej 327, 2800, Kgs. Lyngby, Denmark 
        \and
        National Centre for Nuclear Research, ul.\  Pasteura 7, 02-093 Warszawa, Poland  
        \and
        Physics and Astronomy Department, University of California, 900 University Avenue, Riverside, CA 92521, USA  
        \and 
        Institute for Astronomy, University of Hawaii, 2680 Woodlawn Drive, Honolulu, HI 96822, USA  
             }

   \date{Received ---; accepted ---}

  \abstract
 {
 We present a novel method  to estimate galaxy physical properties from spectral energy distributions (SEDs), alternate to  template fitting techniques  and based on self-organizing maps (SOM) to learn the high-dimensional manifold of a photometric galaxy catalog. The method 
  has been previously tested with hydrodynamical simulations in \citet{davidzon19} while here is applied to real data for the first time. It is crucial for its implementation to build the SOM with a high quality, panchromatic data set, which we elect to be the ``COSMOS2020'' galaxy catalog. After the training and calibration steps with COSMOS2020, other galaxies can be  processed through SOM to obtain an estimate of their stellar mass and star formation rate (SFR). Both quantities  result to be in good agreement with independent measurements derived from more extended photometric baseline, and also their combination (i.e.,  the SFR vs.\ stellar mass diagram) shows a main sequence of star forming galaxies consistent with previous studies. 
  We discuss the advantages of this method  compared to traditional SED fitting, highlighting the impact of having, instead of the usual synthetic templates, a collection of empirical SEDs built by the SOM in a ``data-driven'' way. Such an approach also allows, even for extremely large data sets, an efficient visual inspection  to identify photometric errors or peculiar galaxy types. Considering in addition the computational speed of this new estimator, we argue that it will play a valuable role in the analysis  of oncoming large-area surveys like \textit{Euclid} or the Legacy Survey of Space and Time at the \textit{Vera Cooper Rubin} Telescope. }


   \maketitle
%

\section{Introduction}
\label{sec:intro}

Redshift ($z$), stellar mass ($M$), and star formation rate (SFR) are fundamental parameters for studying galaxy evolution. Measuring these quantities has been instrumental to the discovery of the main sequence of star forming galaxies \citep{brinchmann04, Noeske:2007p5802,Elbaz:2007p11978,Daddi:2007p2924}, a tight correlation between $M$ and SFR pointing to a  ``steady'' mechanism of gas-to-stars conversion at work in systems that went through different evolutionary paths. Other examples are the SFR and stellar mass functions, two demographics that can be computed for galaxies in different redshift bins and are commonly used to either calibrate or validate cosmological simulations \citep[e.g.,][]{henriques13,furlong15,katsianis17,dave17}. 

In absence of spectroscopic data, which demand more telescope time than broad-band photometry, the three quantities can be derived by fitting galaxy templates to the observed spectral energy distribution (SED). Those  templates are built via stellar population synthesis models  \citep[e.g.,][]{bruzual&charlot03,maraston05,Conroy2009a}  assuming a grid of ages and metallicity values, different star formation histories (SFHs) and one or more options for dust attenuation \citep[for a review, see][]{conroy13}. With respect to redshift and stellar mass, SED fitting is a standard practice that produces robust results even when the input photometry is limited to a few bands in the optical and near-infrared (NIR) regime \citep[for a review, see ][]{salvato19}. The method is sub-optimal for constraining SFR, unless far-infrared (FIR) data are included \citep{pannella09,buat10,riccio21}. However, sky coverage and depth of FIR ancillary data should match those of surveys at shorter wavelengths, a condition which is satisfied  only in a few extragalactic fields. 

Another way to estimate SFR is to rely on techniques based on machine learning (ML). For example, a galaxy catalog can be processed through a neural network previously  trained with a sample of objects whose physical properties are already known \citep{davidzon19,surana20,gilda21,simet21}. This means that the targets are compared to other observed galaxies instead of synthetic templates, with the advantage of adhering more coherently to the observational parameter space. In many cases, the training sample is also more accurate than standard templates because its features directly come from high-quality data  \citep[e.g., a  spectroscopic sample like the Sloan Digital Sky Survey,][]{acquaviva16} instead of approximated models. On the other hand, such a data-driven approach provides limited physical insight and may introduce an observational bias if the ML algorithm is not trained on a fully representative sample. A possible solution to this drawback is to use a training sample built from cosmological simulations \citep[like mock photometry or spectra,][]{lovell19,simet21} even though providing galaxy models with ``observational-like'' properties -- and realistic error bars -- is a complex task \citep[see discussion in][]{laigle19}.

The present study shows the advantages of a ML-based method  built on previous work published in \citet{masters15}, \citet{hemmati19}, and \citet{davidzon19} to predict galaxy properties. The method involves dimensionality reduction through  self-organizing maps \citep[SOM,][]{kohonen81} and has been previously tested in \citet{davidzon19} with a mock galaxy catalog  \citep[derived from hydrodynamical simulations, see][]{laigle19}. Knowing the intrinsic properties of those simulated galaxies,  \citeauthor{davidzon19} proved the SOM to be an effective ``manifold learning'' tool, able to explore the complex, non-linear galaxy parameter space and reproduce an approximated but accurate version in a lower-dimensional space. 
We are now ready to apply the same method to real data and judge its performance in a practical ``user case'' situation.

In a nutshell, we use the SOM to analyze a multi-dimensional space made by the combination of galaxy colors in the observer's frame between 0.3 and 5\,$\mu$m. This data set is a combination of two photometric catalogs in distinct survey fields, already described in previous work \citep{mehta18,weaver22} and summarized here in Sect.~\ref{sec:data}.   The SOM  returns an \textit{unsupervised} classification of the input galaxy sample, i.e., the various classes (which will be called ``cells'') are formed without comparing to already classified examples. The second step does require supervision, since at least a fraction of the galaxies need a label (from ancillary data, especially \textit{Spitzer}/MIPS) to  determine the physical properties of each SOM cell. This set-up allows to assign $z$, $M$, SFR to any new object projected onto the SOM. More  details about the method are provided in Sect.~\ref{sec:methods}. 

In Sect.~\ref{sec:results} we verify that SOM-based estimates are more precise than what would result from fitting a standard template library to the same optical-NIR photometry. The same section also shows the resulting main sequence of star-forming galaxies up to $z=2$, and the comparison with previous studies where SFRs have been derived in different ways. Further extension and other potential applications of our SOM-based method are discussed in Sect.~\ref{sec:discussion} along with a few caveats. We summarize the work and draw our conclusions in Sect.~\ref{sec:conclusions}. 

Throughout this work we use a flat $\Lambda$CDM cosmology with 
$H_{0}=70$\,km\,s$^{-1}$\,Mpc$^{-1}$,  
$\Omega_{m}=0.3$,  $\Omega_{\Lambda}=0.7$. 
All magnitudes are in the AB \citep{Oke:1974p12716} system, and the initial mass function (IMF) assumed here is the one proposed in \citet{chabrier03}. In the quoted scaling relations between SFR and rest-frame luminosities, the latter ones are in units of $L_\odot\equiv3.826\times10^{33}$\,erg\,s$^{-1}$ for the former to be in $M_\odot\,\mathrm{yr}^{-1}$.

\section{Data}
\label{sec:data}

\begin{figure}
    \includegraphics[width=0.9\columnwidth]{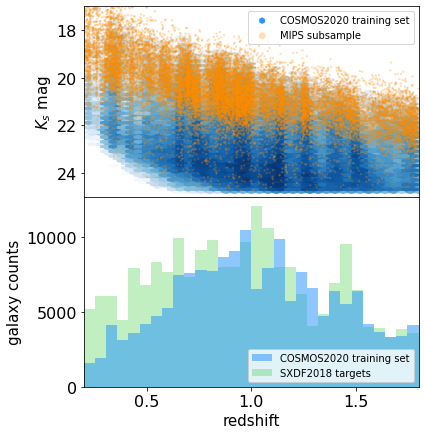}
    \caption{\textit{Upper panel:} The magnitude-redshift distribution of the COSMOS2020 galaxy sample used to train the SOM (blue color map with hexagonal pixels) after pre-selection at $K_\mathrm{s}<24.8$ and $\log(M/M_\odot)>8.5$. Individual sources with a 5$\sigma$ detection in MIPS are overplotted as orange dots.  \textit{Lower panel:} photometric redshift distribution of the COSMOS2020 training set (blue histogram) and the  SXDF2018 target sample (green). These data sets are described in Sect.~\ref{subsec:data_photom_SXDF} and \ref{subsec:data_photom_COSMOS}. }
    \label{fig:dataset}
\end{figure}

The present study relies on two catalogs built from deep photometric surveys in the Subaru XMM/Newton  and Cosmic Evolution Survey fields  \citep[][SXDF and COSMOS, see respectively]{Furusawa08,Scoville:2007p12720}. The two extragalactic fields, each of them of the order of a square degree, are particularly suited to our ML application: they can be easily used in tandem because they have in common a subset of galaxy colors spanning from UV to mid-infrared (MIR). In the present analysis, we are especially interested in broad-band photometry since the galaxy colors derived from it are the features used as input by the SOM. In the last two decades COSMOS has been targeted by numerous telescopes, producing a wealth of ancillary data beyond MIR. The observations in the COSMOS field are also extremely deep, and likely to increase depth even further in the near future owing to planned surveys with both existing and oncoming facilities. For these reasons, a catalog of COSMOS galaxies is ideal for \textit{training and calibrating} the SOM (Sect.~\ref{subsec:meth_basics} and \ref{subsec:meth_calib}). On the other hand, SXDF galaxies shall serve as a \textit{test sample} to verify the reliability of our method  (Sect.~\ref{subsec:meth_sxdf}). The advantages of such a strategy, relying on multiple layers of independent data with increasing depth, have been illustrated e.g.\ in \citet{myles21}.

\subsection{The SXDF2018 photometric catalog}
\label{subsec:data_photom_SXDF}

Published in  \citet{mehta18}, the ``SXDF2018'' photometric catalog  includes optical images from the first data release (DR1) of the Hyper-SuprimeCam Subaru Strategic Program \citep[HSC-SSP,][]{aihara18_ssp} and MUSUBI, the MegaCam Ultra-deep Survey with U-Band Imaging\footnote{W.-H.\ Wang et al., in preparation.} carried out at the Canada-France-Hawaii Telescope (CFHT). In the NIR regime, images come from the ultra-deep area of the UKIRT Infrared Deep Sky Survey \citep[UKIDSS-UDS,][]{Lawrence:2007p13267} and from the VISTA Deep Extragalactic Observations \citep[VIDEO,][]{jarvis13}. The entire area is also covered by the \textit{Spitzer} space telescope, with  several IRAC programs documented in \citet{moneti21}.  The broad bands used here, along with their 5$\sigma$ sensitivity limits, are listed in Table~\ref{tab:filters}. We refer to \citet{mehta18} for further details.

\subsection{The COSMOS2020 photometric catalog}
\label{subsec:data_photom_COSMOS}

COSMOS2020 is the latest photometric catalog released by the COSMOS collaboration   \citep{weaver22}. Two versions of the catalog have been produced through different source extraction methods. Here we use the \textsc{Classic} version based on \texttt{SourceExtractor} \citep{Bertin:1996p13615} and \texttt{IRACLEAN} \citep{hsieh12} since that pipeline is similar to the one applied to SXDF2018.\footnote{We also obtained the SOM from COSMOS2020-\textsc{Farmer}, and  have verified that the ML training in our analysis is stable against different photometric versions.} The COSMOS field is covered by the same instruments and bands mentioned in Sect.~\ref{subsec:data_photom_SXDF}, although here they are generally deeper than in SXDF (see Table~\ref{tab:filters}). In particular, HSC-SSP images come from an updated release \citep[DR2,][]{aihara19_ssp2} while the NIR is covered by UltraVISTA \citep{mccracken12} with an exposure time per pixel $>$150h (compared to the 10$\!-\!$20h of VIDEO). The UltraVISTA depth is not homogeneous, i.e., there are ``Deep'' and ``Ultra-Deep'' areas  but \citet{shuntov22} show that at low redshift they can be safely combined in a joint analysis, if this is limited to magnitudes brighter than the Deep 3$\sigma$  limit (see Table \ref{tab:filters}). 
Medium- and narrow-band images from Subaru and VISTA telescopes are also available in COSMOS, as well as GALEX data from \citet{zamojski07}. These additional bands are taken into account for template fitting (see below) whereas they are not part of the SOM analysis.  VISTA observations in COSMOS have been carried out with a dual-layer strategy  therefore the table provides two values  for $Y$, $J$, $H$, and $K_\mathrm{S}$ depths. 
More details about COSMOS2020 can be found in \citet[][]{weaver22}.

\subsection{Physical properties of COSMOS2020 galaxies} 
\label{subsec:data_phys}

Being the reference sample for SOM training and calibration, COSMOS2020 galaxies must be provided not only with broad-band colors but also with measurements of key physical properties. Most of this additional information is extracted by fitting the observed SED with galaxy models (also called templates). Namely, the fitting code \texttt{LePhare} \citep{arnouts99,ilbert06} is used with three different configurations to derive 
\begin{enumerate}[1)]
    \item  photometric redshifts ($z_\mathrm{LePh}$); \label{est1}
    \item stellar masses ($M_\mathrm{LePh}$)  and absolute magnitudes; \label{est2}
    \item star formation rates only for sources observed in FIR. \label{est3}
\end{enumerate}
Each step is summarized below; estimates \ref{est1}) and \ref{est2}) are described more extensively  in \citet{weaver22} while we implement the third step  following \citet{ilbert15}. Since information in the FIR regime -- namely a 24\,$\mu$m detection with \textit{Spitzer}/MIPS -- is required to constrain star formation, \ref{est3}) can be applied only to a subsample of galaxies (see Fig.~\ref{fig:dataset}).

   \paragraph{Redshift.} 
   In this first run of \texttt{LePhare} the observed SED is interpolated by a library of 33 galaxy templates \citep[originally proposed in][]{ilbert13}. The intrinsic spectrum of these templates is modified by a dust screen ranging from $E(B-V)=0$ to 0.5, with different options to model the extinction curve (\citealp{prevot84},  \citealp{calzetti2000}, and two variations of Calzetti's law including the 2175\,{\AA} bump). A complementary library of stellar templates is used, together with other criteria\footnote{Namely, a $g-z$ vs.\ $z-K_\mathrm{s}$ selection and the stellar locus in half-light radius vs.\ magnitude in HST/ACS and Subaru/HSC bands \citep[see][]{weaver22}.}, to remove stars from the catalog. After evaluating the $\chi^2$ of each fit, the resulting $z_\mathrm{LePh}$ estimates are defined as the median of the redshift likelihood function, i.e.\ the combined probabilities from the various models (each one $\propto e^{-\chi^2}$) . 

   \paragraph{Stellar mass and other physical properties.} 
   After fixing the redshift of each source to $z_\mathrm{LePh}$, the \texttt{LePhare} code is used to measure galaxy stellar mass.   
   Templates optimized for such a task are derived from \citet[][]{bruzual&charlot03} models (BC03 hereafter) as done in \citet{laigle16}, i.e.\ with a grid of 44 time steps for eight star formation histories, two non-evolving metallicity values ($Z_\odot$ and 0.2$Z_\odot$), and a range of dust extinction parameters similar to the one adopted in the first configuration.  Absolute magnitudes in different broad-band filters, required to compute rest-frame colors, are an additional output. They are derived from the apparent magnitude that most closely samples the desired rest-frame filter. For example, the observed $i$ band is used to calculate the rest-frame $u$ magnitude for a galaxy at $z_\mathrm{LePh}\simeq1$. When needed, a k-correction is calculated from the best-fit template, which also provides the apparent magnitude of the object when   \texttt{SourceExtractor} cannot measure it in any band. 
   
   The same run of \texttt{LePhare} would also provide SFR$_\mathrm{LePh}$, but we deem the FIR-based estimates more robust for the calibration of our method as will be discussed in Sect.~\ref{subsec:meth_calib}.  Nonetheless, the SFR$_\mathrm{LePh}$ values are stored in output to compare to the SOM-based estimates derived from the same photometric baseline. 
   
\begin{table}
\caption{Optical and NIR data building the observer-frame color space explored by the SOM. For SXDF2018 galaxies, this is also the baseline for \texttt{LePhare} template fitting, while for COSMOS2020  additional 14 bands are used \citep[see][]{weaver22}. VISTA observations in COSMOS have been carried out with a dual-layer strategy (``Deep'' and ``Ultra-Deep'' regions) therefore the table provides two values  for $Y$, $J$, $H$, and $K_\mathrm{S}$ depth.  
}
\footnotesize
\setlength{\tabcolsep}{2pt}
\begin{tabular}{lccccc}
 \hline \hline
Instrument  & Band  & Central$^\mathrm{a}$  & COSMOS depth$^\mathrm{b}$ & SXDF depth$^\mathrm{b}$ \\
            &       & $\lambda$ [\AA{}] &  2\arcsec aper. [mag]&  2\arcsec aper. [mag]  \\
 \hline 
MegaCam/CFHT    & $u$   & 3858 & 27.7   & 27.93  \\
 \hline
Subaru/HSC      & $g$       & 4847 & 28.1   & 27.39  \\
                & $r$       & 6219 & 27.8   & 26.91  \\
                & $i$       & 7699 & 27.6   & 26.66  \\
                & $z$       & 8894 & 27.2   & 26.07  \\
                & $y$       & 9761 & 26.5   & 25.34  \\
 \hline
VISTA/VIRCAM    & $Y$  & 10216 & 25.3/26.6  & 25.41  \\
                & $J$  & 12525 & 25.2/26.4  & 24.89  \\
                & $H$  & 16466 & 24.9/26.1  & 24.56  \\
                & $K_s$& 21557 & 25.3/25.7  & 24.23  \\
 \hline
\textit{Spitzer}/IRAC & ch1 & 35686 &  26.4 & 25.94  \\
                      & ch2 & 45067 &  26.3 & 25.68  \\
 \hline
\end{tabular}
\\
$^\mathrm{a}$ Median of the transmission curve.\\
$^\mathrm{b}$ Depth at $3\sigma$ computed on PSF-homogenized images (except for IRAC images) in empty apertures with 2\arcsec diameter. \\ 
\label{tab:filters}
\end{table}

   \paragraph{Star formation from infrared continuum.} 
   
   A major role in our analysis is played by the star formation tracer that leverages the correlation between SFR and rest-frame emission in UV and infrared  (hereafter SFR$_\mathrm{UVIR}$).
   Interstellar dust absorbs stellar light from UV to optical and re-emits  those photons mostly between 8 and 1\,000\,$\mu$m, i.e.\ in the infrared (IR) and sub-mm regime. The IR integrated luminosity ($L_\mathrm{IR}$)  is therefore closely -- albeit indirectly --  linked with star formation activity. 
   We derive $L_\mathrm{IR}$ for COSMOS2020 galaxies as in \citet{ilbert15} using FIR  data as observational constraint. In brief, we assign a 24\,$\mu$m flux to 28\,347 objects over the whole COSMOS area via cross-correlation between COSMOS2020 sources and the \textit{Spitzer}/MIPS catalog from \citet{lefloch09}. We cut the parent sample at $z_\mathrm{LePh}<1.8$ and   signal-to-noise ratio $S/N>5$ in the 24\,$\mu$m band.  The redshift upper limit ensures that the polycyclic aromatic hydrocarbons (PAH) complex at rest-frame 6.2--7.7\,$\mu$m  does not contaminate the 24\,$\mu$m measurements. We also removed sources with an X-ray counterpart to minimize contamination from their active galactic nucleus (AGN). The remaining 22\,571  MIPS galaxies constitute the calibration sample to enable SFR estimation with the SOM.

  The  $L_\mathrm{IR}$ estimates are obtained by means of \texttt{LePhare} by fitting \citet{dale&helou02} templates to the MIPS data\footnote{ Since \citet{dale&helou02}  models start from 3\,$\mu$m (rest frame), the SED fitting may include IRAC data point as well. However, we prefer to restrict the photometry to the FIR range for a more homogeneous constraint.}, then integrating the best-fit model  in the rest frame between 8 and 1\,000\,$\mu$m. 
  For 7\% of them there is also a 5$\sigma$ detection in \textit{Herschel}/PACS   \citep[either 100 or 160\,$\mu$m from][]{lutz11} while 9\% have a counterpart in \textit{Herschel}/SPIRE \citep[250, 350, or 500\,$\mu$m from][]{oliver12}.  
  During  this procedure, the redshift is fixed to the $z_\mathrm{LePh}$ value from the  optical-NIR fitting. No systematic effect is introduced when the 24\,$\mu$m flux is complemented with \textit{Herschel} measurements up to 500\,$\mu$m \citep[figure 1 of][]{ilbert15}. 
   Unobscured star formation is traced by the luminosity in the near-UV band ($L_\mathrm{NUV}$), which is obtained from the  BC03 template fitting described above. Accounting for both contributions, we eventually compute the total SFR as in \citet{arnouts13}:
   \begin{equation}
    \mathit{SFR_\mathrm{UVIR}} = \mathcal{K} (L_\mathrm{IR}+2.3L_\mathrm{NUV}) \;,
   \label{eq:SFR_UVIR}
   \end{equation}
   where $\mathcal{K}=8.6\times 10^{-11}$. This  equation traces star formation on a $\sim$100\,Myr time scale. The capability  of MIPS 24\,$\mu$m to provide reliable SFR estimates is also thoroughly discussed in \citet{rieke09}.

Alternate SFR$_\mathrm{UVIR}$ estimates for both COSMOS and SXDF galaxies are available from \citet{barro19}, although only in the sub-region that was also observed by the Cosmic Assembly Near-infrared Deep Extragalactic Legacy Survey  \citep[CANDELS,][]{Grogin:2011p13008,Koekemoer:2011p12718}. In addition to ground-based facilities, CANDELS includes data from  the \textit{Hubble} Space Telescope (HST) in both optical and near-IR bands, which may generate some difference in template fitting results (especially photometric redshifts). The ancillary \textit{Spitzer} and \textit{Herschel} data in \citeauthor{barro19} are similar to the one used here.  At $z<1.8$ there are 509 (608) matches between \citet{barro19} and our COSMOS (SXDF) catalog if we require their photometric redshifts to be within $\pm$30\% from $z_\mathrm{LePh}$. In COSMOS, after converting the two sets of SFR$_\mathrm{UVIR}$ estimates to a common framework\footnote{Same IMF, luminosity-SFR calibration, etc. (see Appendix~\ref{appendix2}).} we find a scatter of 0.17\,dex and a small systematic offset ($-0.03$\,dex) due to a handful of objects overestimated by \citeauthor{barro19}  (Fig.~\ref{fig:sfr_barro_vs_ilbert}).   Although the SOM will be calibrated only with \texttt{LePhare} outcomes to ensure consistency, these alternate estimates will play an important role for testing (Sect.~\ref{subsec:results_sfr}). 

\subsection{Selection functions}

The COSMOS2020 training sample is limited to galaxies outside masked areas of the survey (i.e., avoiding the surroundings of bright stars\footnote{Masking is done by  selecting \texttt{FLAG\_COMBINED==0} in the catalog.}) and with  $z_\mathrm{LePh}<1.8$ to be consistent with the MIPS-detected sample. In principle, it would be possible to extend the SOM to higher redshift, but in that case the calibration would require a different SFR proxy (see Sect.~\ref{subsec:discuss_othersoms}). We also cut the training sample at $K_\mathrm{s}<24.8$ to remove objects with a low signal-to-noise ratio. Eventually, the COSMOS training sample includes 174\,522 galaxies, with 13\% of them being MIPS-detected. 
 
We apply a 3$\sigma$ cut also to the SXDF2018 catalog, corresponding to a $K_\mathrm{s}<24.2$ limit (see Table~\ref{tab:filters}). Since the goal is to replace \texttt{LePhare} physical properties with SOM estimates, and not the photometric redshifts obtained beforehand, we have the latter ones at our disposal to limit the SXDF2018 sample to $z_\mathrm{LePh}<1.8$ and exclude stellar-like objects. As a result, 208\,404 galaxies in the SXDF field will be used in the following.

\section{Methods}
\label{sec:methods}

\subsection{Training the SOM with COSMOS2020}
\label{subsec:meth_basics}

\begin{figure*}
    \includegraphics[width=0.3\textwidth]{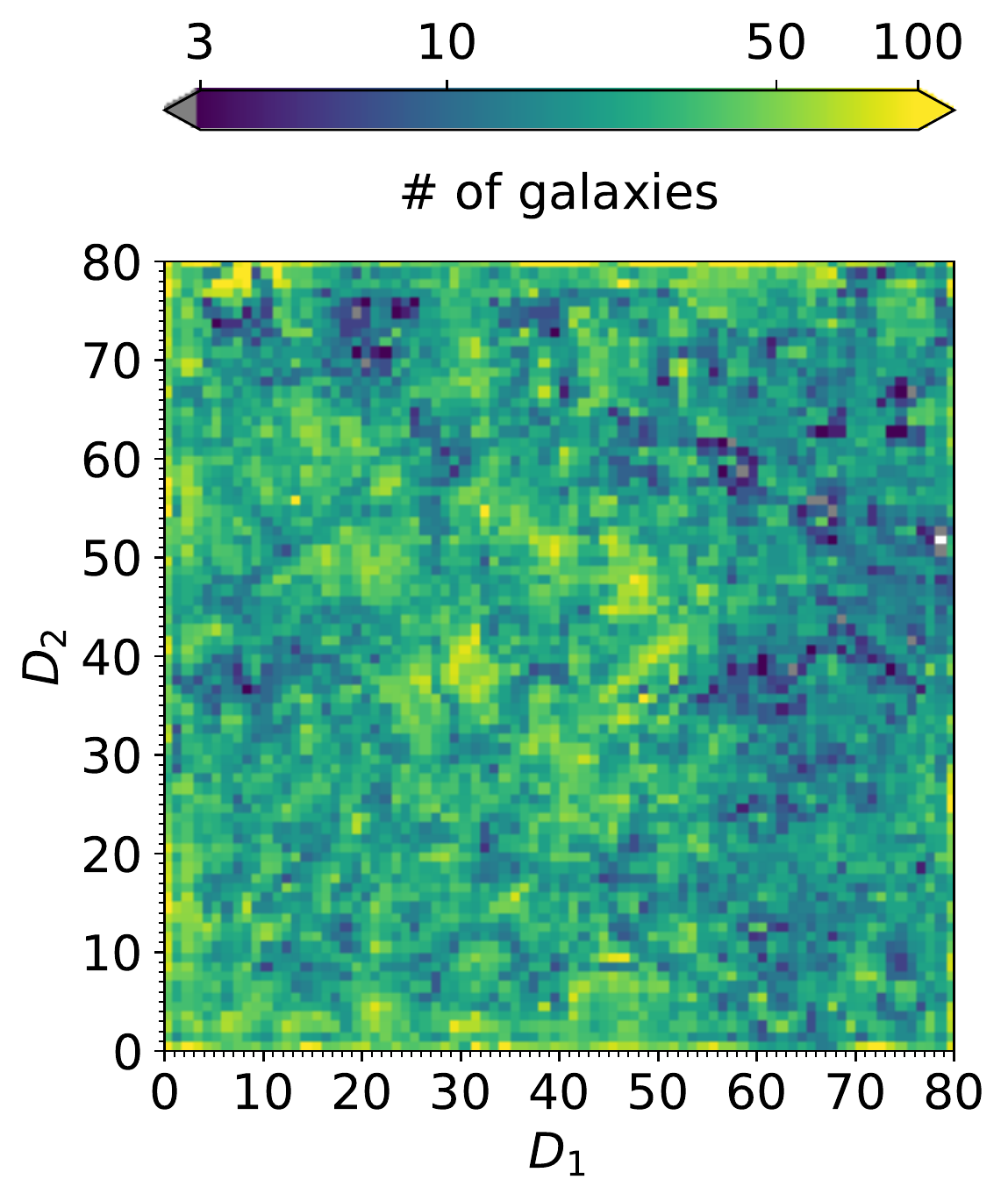}\hspace{5mm}
    \includegraphics[width=0.3\textwidth]{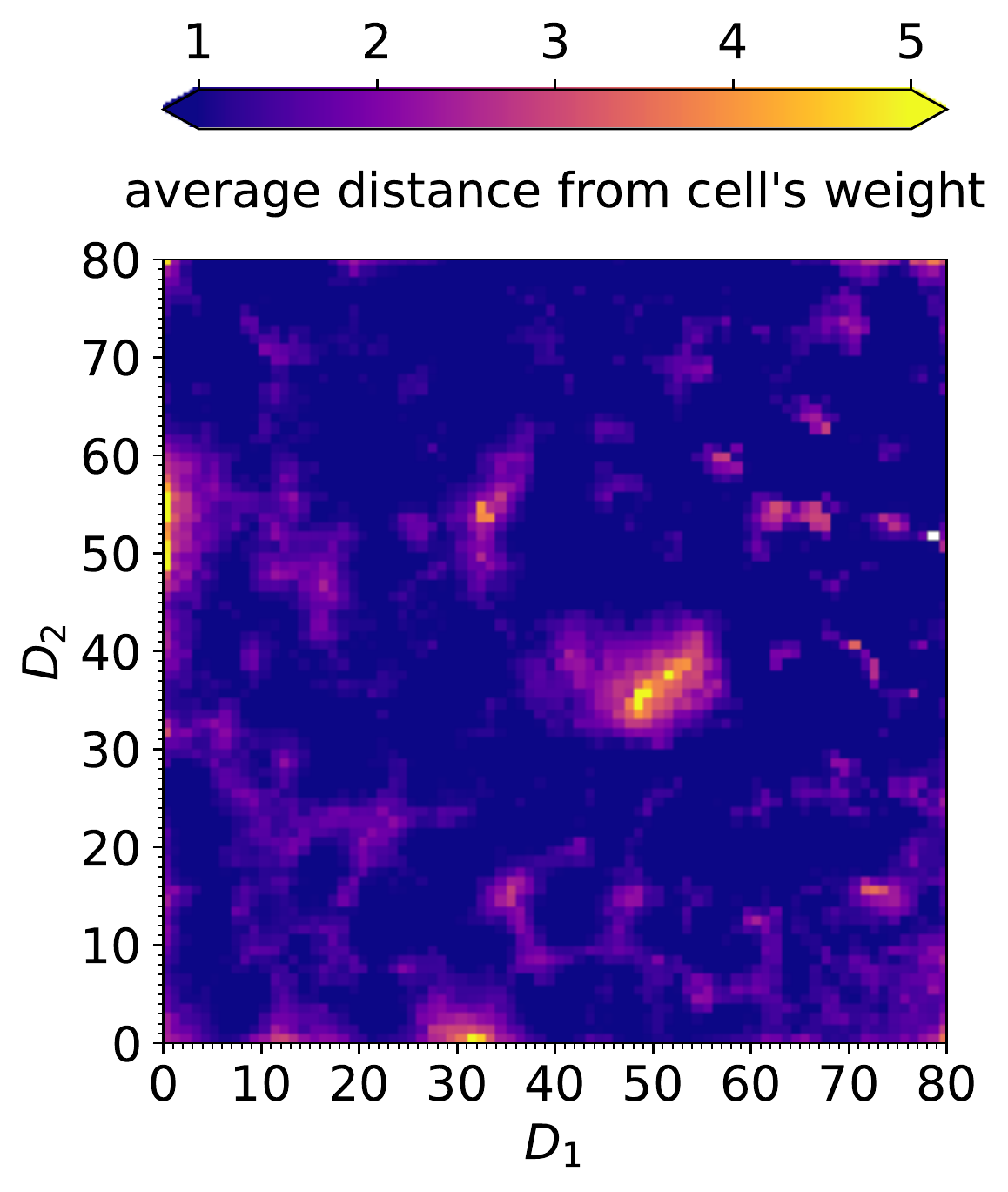}\hspace{5mm}
    \includegraphics[width=0.3\textwidth]{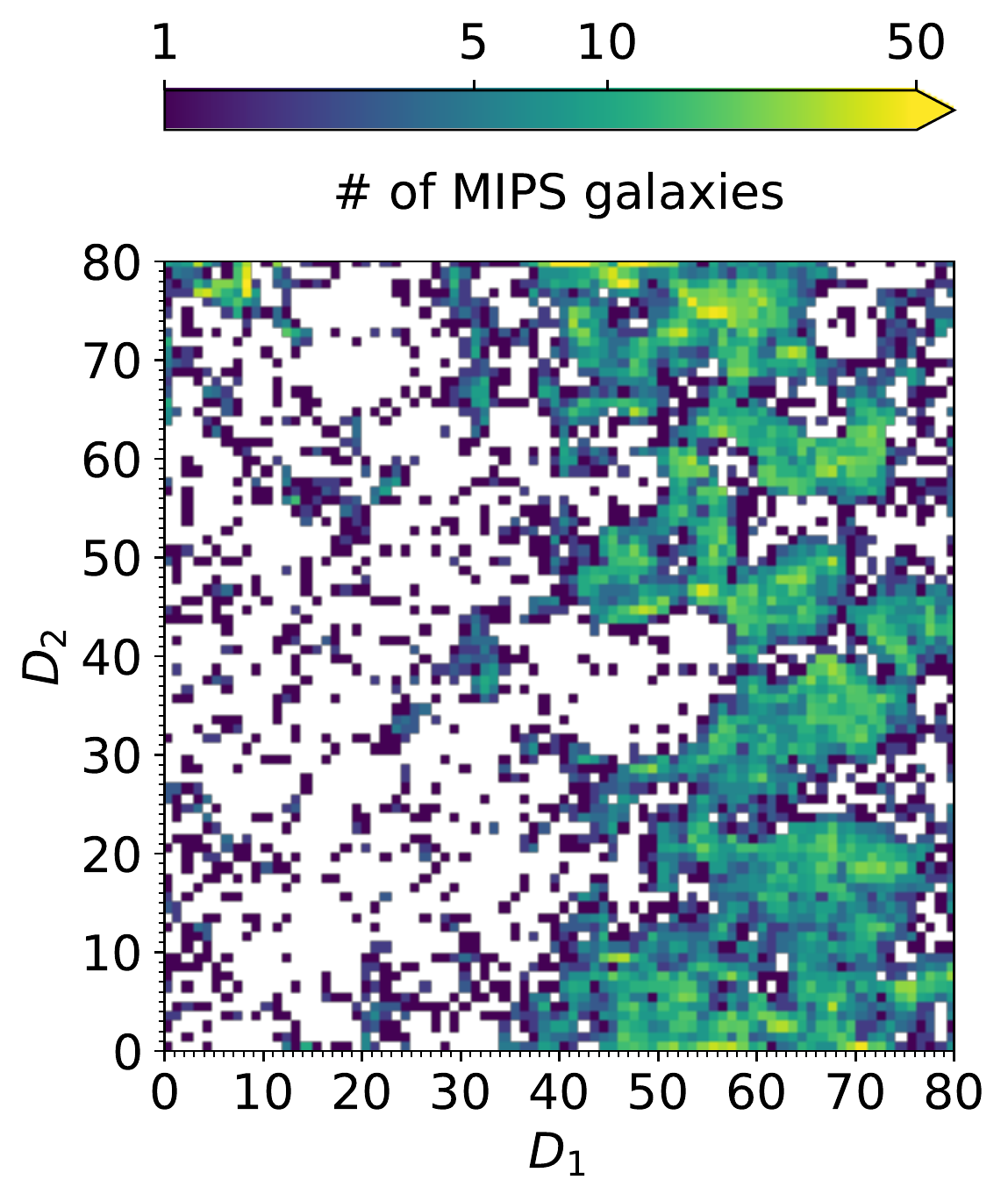}  
    \caption{The SOM of COSMOS2020 galaxies at $z<1.8$, selected as described in Sect.~\ref{subsec:meth_basics}. The same grid is color-coded in three different ways, to show the number of galaxies per cell (\textit{left panel}), the similarity between galaxies in a given cell and the  correspondent SOM weight (\textit{middle panel}), and the MIPS-detected objects with $M_\mathrm{LePh}>10^{10}\,M_\odot$ (\textit{right panel}). In each grid, empty cells are white. In the \textit{left panel}, the handful of  cells with only one or two objects are colored in gray. In the middle panel, similarity is quantified by means of Eq.~\ref{eq:d_som}. Throughout this work, the dimensions of the 2-d grid are arbitrarily labeled $D_1$ and $D_2$ since they have no physical meaning.  }
    \label{fig:som_train}
\end{figure*}

\begin{figure}
    \centering
    \includegraphics[width=0.99\columnwidth]{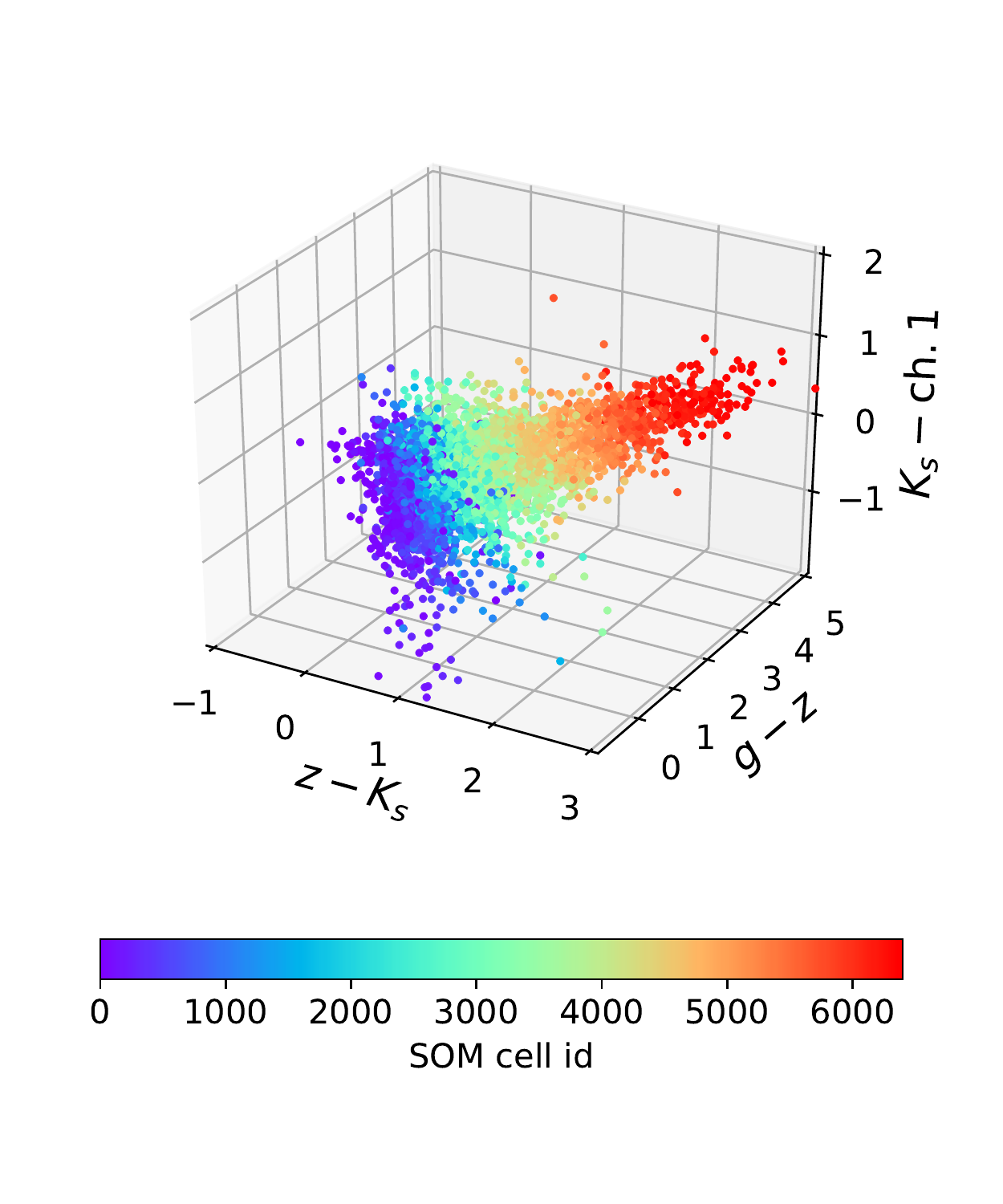}
    \caption{An illustration of the relationship between input features  and the SOM classification (made by the so-called ``cells''). This simplified example shows only three broad-band colors in the observer's frame, for a random subsample of 10\,000 COSMOS2020 galaxies.}
    \label{fig:som_didactic}
\end{figure}

The SOM can identify data patterns in a multi-dimensional feature space by sampling it with an adaptive distribution of points called  ``weights''.\footnote{Although it would be formally correct to define them as a network of artificial neurons, we decided to use the term ``weight'' not to mislead the majority of readers, who commonly associate ``neural networks'' to a  different class of machine learning architectures.  } During a training phase, the coordinates of these weights in the multi-dimensional space are adjusted to move them close to the input data, according to a convergence criterion detailed in \citet{davidzon19}. After that, the SOM reproduces not only the `shape' of the input manifold but also its `density', i.e.\ a larger number of test particles is located where data are more clustered. The algorithm is self-organizing because the training is unsupervised. The SOM maps the input data set into a lower-dimensional space: first, its network is arranged in a 2D geometry then data will be re-arranged as well, each entry being  associated to a given test particle. Despite the discretization due to the finite number of test particles,  the dimensionality reduction preserves most of the topological structure of the original manifold. The 2D geometry we choose\footnote{Other  renderings (3D or even higher dimensions) are also allowed but do not bring  the same advantages, e.g.\ for human eye inspection. More exotic geometries can also be used as long as they are equipped with appropriate metrics. } is a square grid (see  Fig.~\ref{fig:som_train}); therefore throughout this study we refer to the SOM test particles as ``cells''.  With the manifold's topology mostly preserved, elements that are close to each other in the higher-dimensional space are also neighbors in the grid, i.e.\ belong to the same cell or two contiguous  cells   (see a pedagogical illustration in Fig.~\ref{fig:som_didactic}). We refer to the seminal work of \citealp{kohonen81} for a more extensive explanation. 

The feature space consists of eleven colors in the observer's frame, derived from the broad-band filters listed in Table~\ref{tab:filters} and paired in sequential order: $u-g$, $g-r$, ..., $K_\mathrm{s}-\mathrm{ch1}$,  $\mathrm{ch1}-\mathrm{ch2}$. Galaxies are associated to their nearest weight according to their (11-dimensional) Euclidean  distance. The ones linked to the same weight are expected to have similar SEDs by construction  \citep[see figure 1  in][]{masters15} modulo their brightness (i.e., a normalization factor) and the scatter due to photometric errors. The COSMOS2020 SOM is presented in Fig.~\ref{fig:som_train}.  

The weight's coordinates in the multi-dimensional space are actually a set of eleven colors, which can be regarded as the ``prototype'' SED of the cell. This terminology is the same used in \citet[][see their table 1]{davidzon19}.   Similarly to that study, shape and size of the grid are chosen in order to minimize the dispersion of data within a given cell, but at the same time avoid a large fraction of them to be empty. The result is a configuration with $80\times 80$ cells, nearly 94\% of them containing at least ten objects at the end of the training   (Fig.~\ref{fig:som_train}, left panel). Number counts vary across the grid by maximum an order of magnitude, and only 14 cells have less than three galaxies (one of them being empty).  

To quantify the accuracy of the $80\times 80$ SOM  grid in describing the COSMOS2020 manifold, we calculate the Euclidean distance between galaxies and the weight of the cell associated to them. That is,
\begin{equation}
    \Delta = \sqrt{ \sum_{N_\mathrm{dim}} (f_i - w_i)^2 },
\label{eq:d_som} 
\end{equation}
where $f_i$ is a galaxy's feature in the $i$-th dimension (out of $N_\mathrm{dim}=11$) and $w_i$ is one component of the weight representing the cell where the galaxy lies. Across the grid the average $\Delta$ per cell is $<1$  (Fig.~\ref{fig:som_train}, middle panel) as it should be if the scatter of observed SEDs  around their weight is dominated by photometric uncertainties, which for the chosen broad-band colors are typically 0.05$-$0.1\,mag. A few cells, however, present larger $\Delta$ values, an indication they contain either a wider variety of galaxy types or SEDs with similar shapes but low $S/N$ ratio. As an example, a couple of those cells will be inspected in Sect.~\ref{subsec:discuss_soundness}. 
We also verify that galaxies that are close to each other in the feature space turn out to be in the same (or nearby) cells, as partially shown in Fig.~\ref{fig:som_didactic}. 
Another metrics to quantify goodness of fit is the reduced $\chi^2$ distance, used e.g.\ in \citet{masters15}. This is further discussed in Appendix~\ref{appendix2} (see in particular Fig.~\ref{fig:som_chi2}).  

To conclude the presentation of the training sample, the third (right-hand) panel of Fig.~\ref{fig:som_train} shows the MIPS-detected  galaxies in COSMOS. This subsample is spread over 3\,561 cells (55.6\% of the total) which are mainly located on the right half of the grid (see Fig.~\ref{fig:som_train}, right panel). One third of the cells probed by MIPS galaxies contain only one of them. Most of these undersampled cells define the boundaries of star-forming regions in the SOM, while others are scattered across the left side of the grid, an area that is scarcely occupied because of the MIPS selection function. In fact, those are cells containing galaxies with $i>23$\,mag and  $M_\mathrm{LePh}<10^{10}\,M_\odot$ (see discussion in Appendix \ref{appendix2} and Fig.~\ref{fig:som_mstar_map}). Most of them are expected to be fainter than the survey sensitivity limit at 24\,$\mu$m, as can be deduced e.g.\ from \citet{lefloch09} and  \citet{kokorev21}.

\subsection{SOM calibration with physical properties}
\label{subsec:meth_calib}

\begin{figure*}[h!]
\includegraphics[width=0.45\textwidth]{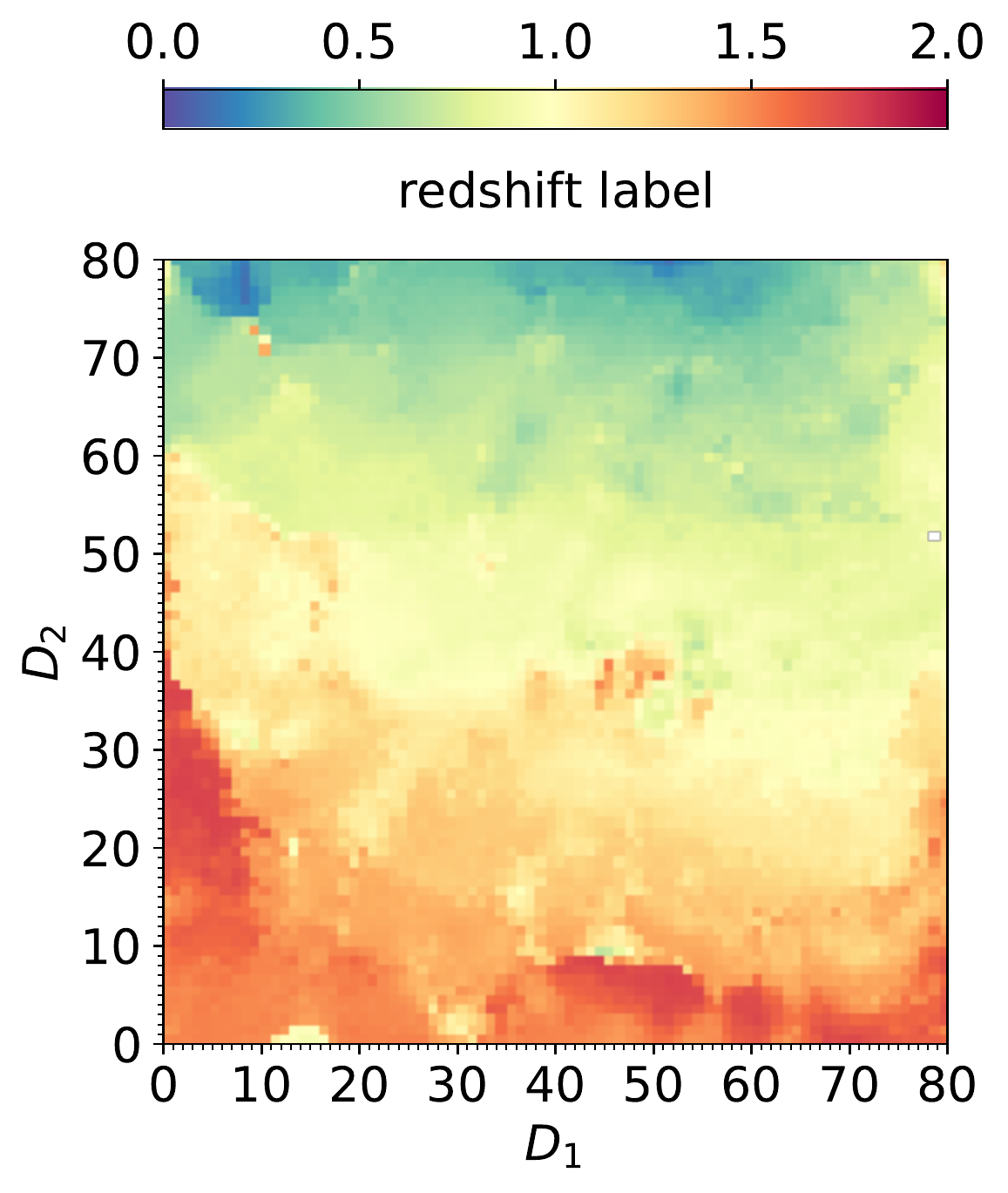}
\includegraphics[width=0.45\textwidth]{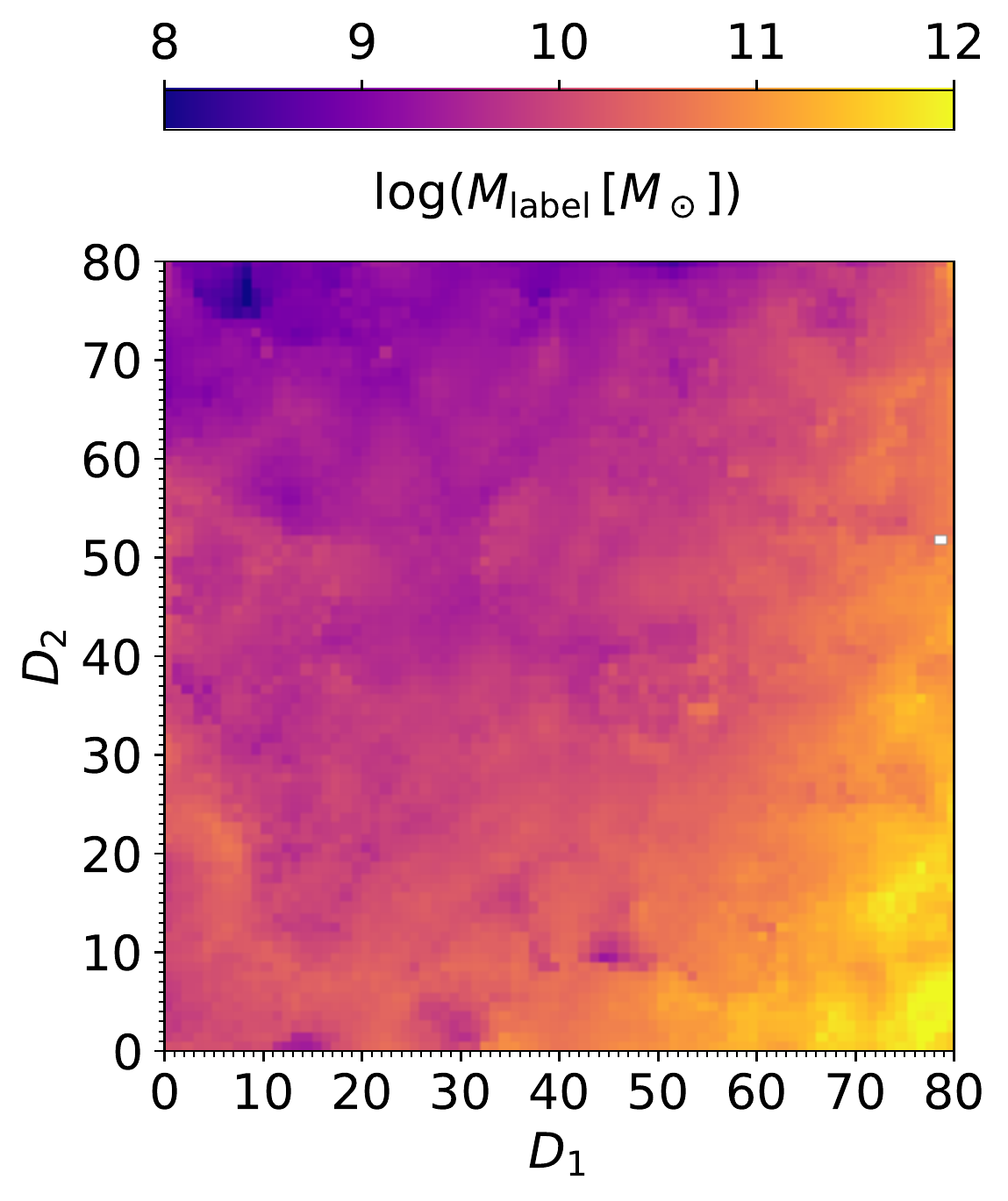}\\
\includegraphics[width=0.45\textwidth]{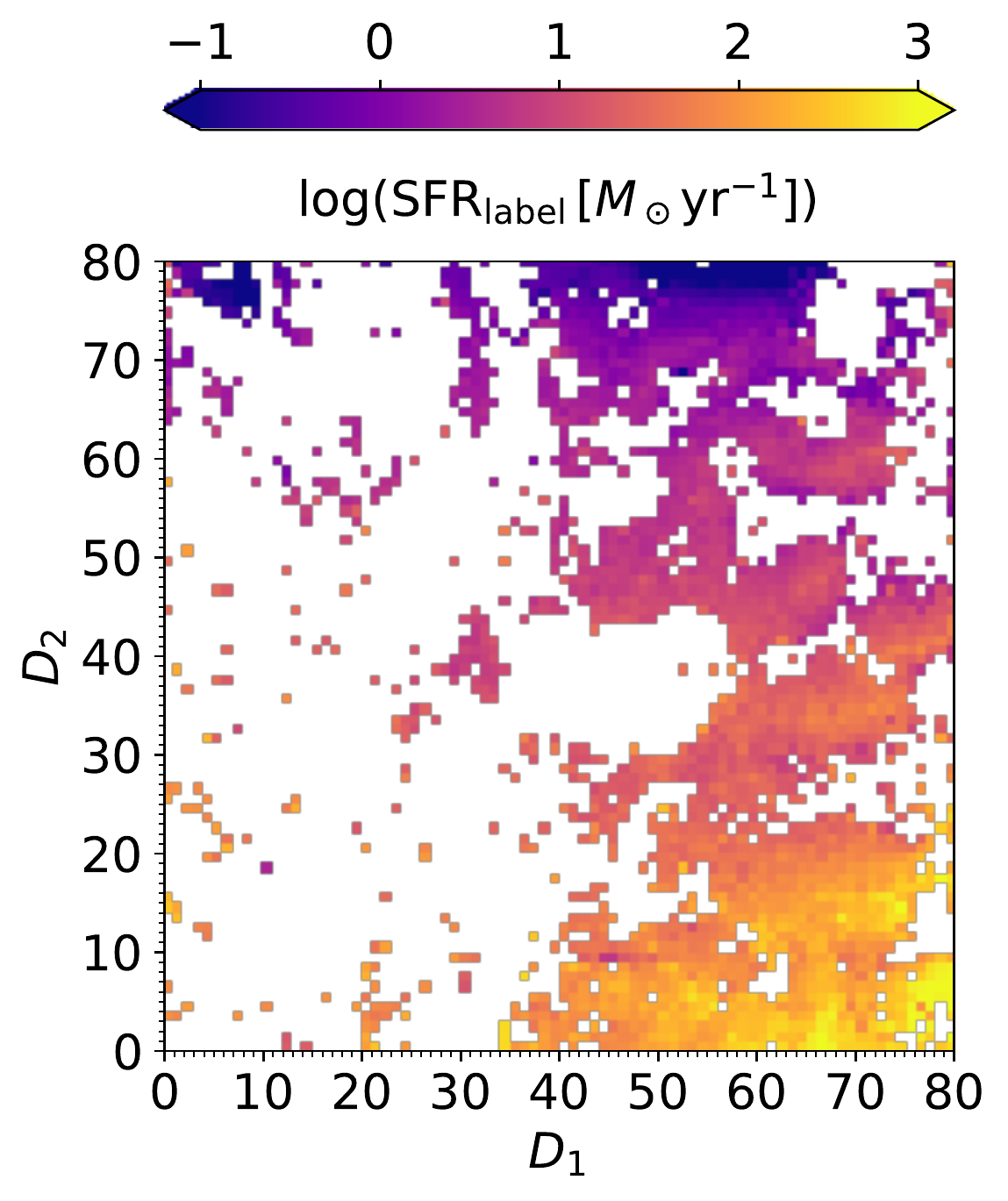}
\includegraphics[width=0.45\textwidth]{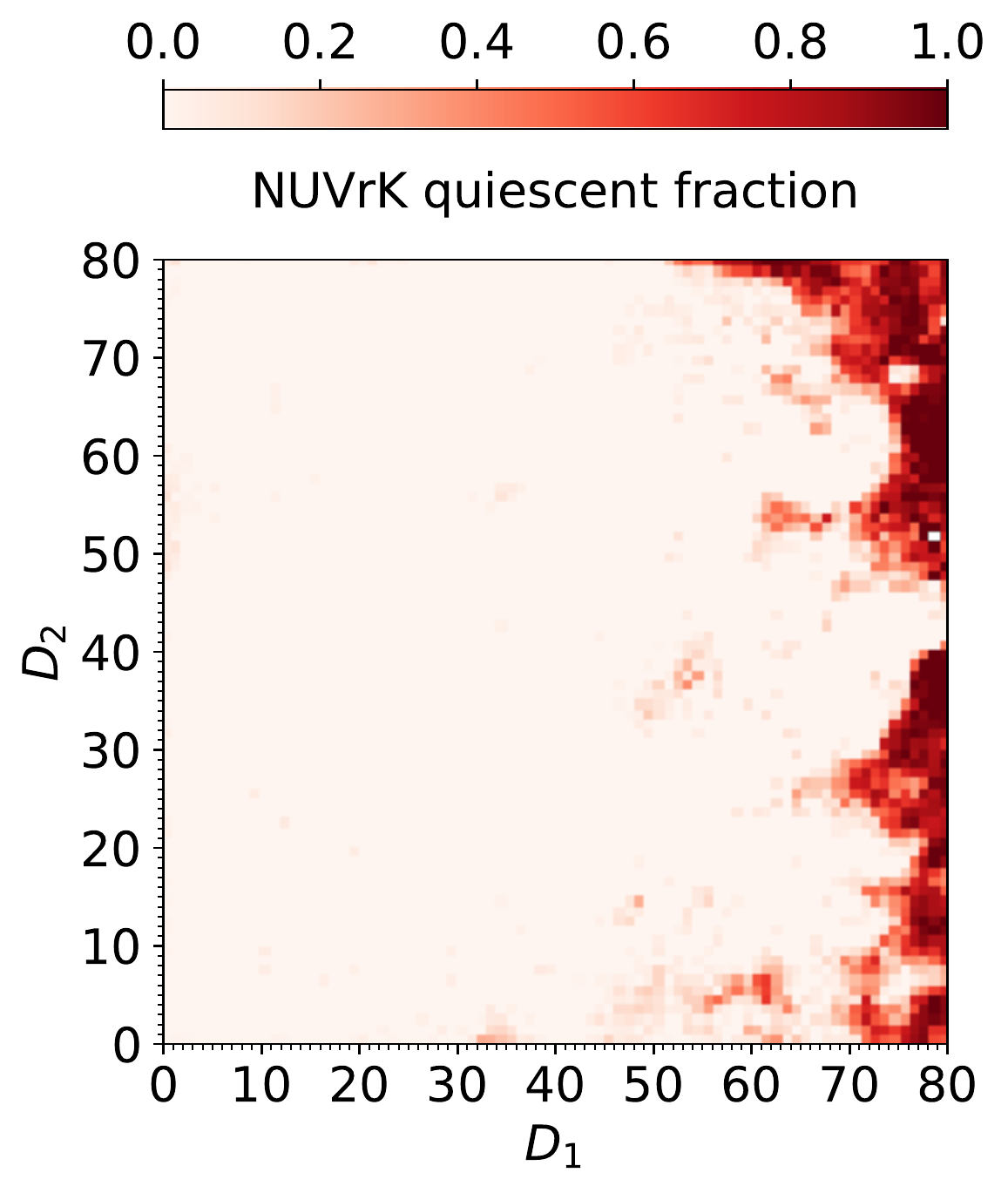}
\caption{Different sets of labels added to the SOM, after training it with COSMOS2020 data. \textit{Upper left:} redshift labels assigned from the median $z_\mathrm{LePh}$ of galaxies in the given cell. 
\textit{Upper right:}  median $\log(M_\mathrm{label}/M_\odot)$ per cell, where the  $M_\mathrm{label}$ values are galaxy stellar masses obtained  after rescaling every SED to $i=22.5$  (Eq.~\ref{eq:mass_resc}). \textit{Lower left:} labels from the median SFR$_\mathrm{label}$ per cell applying the same 22.5\,mag  rescaling as for stellar mass (Eq.~\ref{eq:sfr_resc}); with respect to Fig.~\ref{fig:som_train} a larger portion of the SOM grid is empty because we did not assign SFR$_\mathrm{label}$ to cells containing only one MIPS galaxy. 
\textit{Lower right:} fraction of quiescent galaxies per cell, according to NUVrK classification (Eq.~\ref{eq:nuvrk}). }
\label{fig:som_calib}
\end{figure*}

The cells in a trained SOM can be calibrated \textit{a posteriori} with redshift values   \citep[see e.g.][]{geach2012,masters15} or other physical parameters  \citep{hemmati19,davidzon19}. We choose the labels to be redshift, stellar mass, SFR, and quiescent vs.\ star-forming galaxy type; these four ``layers'' of classification are shown with different color codes in Fig.~\ref{fig:som_calib}.  

For a given cell, the label for each of these quantities is calculated using galaxies inside it. For redshift and stellar mass, this is defined  as the median of  the  $z_\mathrm{LePh}$ and the normalized $M_\mathrm{LePh}$ distribution respectively  (Fig.~\ref{fig:som_calib}, upper panels). The reason for normalizing the mass is discussed in \citet{davidzon19}. Their simulations show that when the SOM is trained with colors, objects in the same cell have similar mass-to-light ratios while still spanning a range of stellar masses (see Appendix \ref{appendix2}).  We proceed as in \citeauthor{davidzon19} by rescaling  $M_\mathrm{LePh}$ estimates to 22.5\,mag in the $i$ band, i.e.
\begin{equation}
    M_\mathrm{i=22.5} =  10^{+0.4(m_i-22.5)} M_\mathrm{LePh} \,,
\label{eq:mass_resc}
\end{equation}
where $m_i$ is the original $i$-band magnitude of the given galaxy.  The median of the $M_\mathrm{i=22.5}$ distribution in a given cell is used as its label ($M_\mathrm{label}$). In the upper-right panel of Fig.~\ref{fig:som_calib} the SOM is color-coded according to that label set, showing a smooth transition from $10^8\,M_\odot$ in the upper-left corner of the grid, corresponding to $z\sim0$, up to $10^{12}\,M_\odot$ in the bottom right. This trend follows, as expected, the evolution in the $z$ label: the higher the redshift, the larger the stellar mass at the fixed 22.5\,mag reference.  

Regarding galaxy star formation, the calibration is done by using only objects with an available SFR$_\mathrm{UVIR}$ estimate. A rescaling similar to Eq.~\ref{eq:mass_resc} is applied for the same reason: galaxies in any given cell have similar colors by SOM construction, but this does not prevent them to freely span a certain magnitude range.  Therefore, we define
\begin{equation}
    \mathit{SFR}_\mathrm{i=22.5} =  10^{+0.4(m_i-22.5)} \mathit{SFR}_\mathrm{UVIR} 
\label{eq:sfr_resc}
\end{equation}
and attach a SFR$_\mathrm{label}$ value to the cell based on the median of the SFR$_\mathrm{i=22.5}$ distribution. We disregard cells containing only one MIPS-detected galaxy. The resulting labels are shown in Fig.~\ref{fig:som_calib} (lower-left panel). 

Each COSMOS2020 galaxy can also be classified as quiescent or active depending on its location in the rest-frame plane $\mathit{NUV}-r$ vs.\ $r-K_\mathrm{s}$  \citep[NUVrK,][]{arnouts13}. This diagnostic is one of the most common in the literature, together with  $U-V$ vs.\ $V-J$  \citep[UVJ,][]{williams09}. Either diagram can single out quiescent galaxies from other types as the former ones occupy a delimited locus in the color-color spaces.   Advantages of $NUV-r$ with respect to $U-V$ are discussed e.g.\ in \citet{siudek18} and  \citet{leja19c}. We use the rest-frame NUVrK classification  as in  \citet[][]{ilbert15}, where quiescent galaxies have
\begin{equation}
\begin{split}
    & (NUV-r)+C>2.6 \\ 
    & \mathrm{and} \\
    & (NUV-r)+C>2(r-K_\mathrm{s})+1.7,   
\end{split}
  \label{eq:nuvrk}
\end{equation}
 with $C\equiv0.17(t-t_\mathrm{z=2})$ being a cosmic time-dependent correction to maintain   consistent criteria across the whole redshift range. 
From Eq.~\ref{eq:nuvrk} we derive the fraction of quiescent galaxies per cell ($f_\mathrm{Q}$)  and use it as the fourth label of the  SOM  (Fig.~\ref{fig:som_calib}, lower-right panel). In most of the cases the NUVrK classification agrees well with the direct assessment of star formation: 86\% of the cells with a large quiescent fraction ($f_\mathrm{Q}>0.5$) have either a specific SFR (sSFR, defined as SFR$/M$) lower than  $10^{-11}\,\mathrm{yr}^{-1}$ or do not contain MIPS-detected objects.
Some discrepancy between the two indicators (i.e., MIPS-detected objects in cells with low sSFR) can be due to an active galactic nucleus (AGN) which has not been removed\footnote{COSMOS2020 includes a flag to clean the catalog from AGN.} and  is boosting the 24\,$\mu$m signal \citep[see e.g.][]{delvecchio17}. Another possibility is a break of the  $L_\mathrm{IR}$-SFR correlation for post-starburst galaxies  \citep{hayward15}.   The 68 cells (out of 496) that have $f_\mathrm{Q}>0.5$ but show significant star formation activity   ($\mathrm{sSFR}>10^{-10}\,\mathrm{yr}^{-1}$) are poorly sampled by the MIPS galaxies: most of them contain only one MIPS galaxy, which may be an interloper in that cell e.g.\ owing to the similarity between the SEDs of dusty, star-forming galaxies and the  ``red and dead'' ones  \citep[e.g.,][]{arnouts13}. 
In the following we will opt for a conservative approach and disregard SFR measurements from cells with $f_\mathrm{Q}>0.9$.

\subsection{Predictions for a new data set}
\label{subsec:meth_sxdf}

The key advantage of the SOM is that after the training, other data sets can be ``projected'' onto the grid in a fast and efficient way. Training is the most time-consuming part (0.3 CPU hour in the COSMOS2020  case) while  several thousands galaxies can be subsequently projected in a few seconds. The additional data set used here is the SXDF2018 galaxy catalog described in Sect.~\ref{sec:data}.  For sake of consistency, the properties of the SXDF galaxies are derived  as done in COSMOS2020  \citep[][see also  Sect.~\ref{subsec:data_phys}]{weaver22}. However, the only quantity the SOM method needs from the SXDF2018 catalog, besides the observed colors, is redshift for the pre-selection mentioned above. Other properties derived by \texttt{LePhare}, namely stellar mass and SFR, will be only used a posteriori to compare the two methods. 

The procedure described in the following is not strictly designed for SXDF. In principle, it can be applied to any survey that  includes the same features used to build the SOM. However, the mapping can be affected by non-negligible systematics if the new survey has a very different set-up with respect to COSMOS2020, especially regarding its noise properties. This is not an issue for SXDF2018, which has been built using the same instruments, filters, and pipeline of the training data set.  

We map the SXDF2018 sample using a $k$-dimensional tree\footnote{Implemented in \texttt{scikit-learn}  \citep{scikit-learn}.} with $k=5$ to find the five cells with the closest weight to each galaxy (including the cell the galaxy is associated to). As in \citet{davidzon19} we prefer to work with $k$ nearest neighbors instead of considering only the best matching cell, to take into account observational uncertainties. Physical properties can be predicted for any individual galaxy using the labels of its five nearest cells. For example, the stellar mass estimate $M_\mathrm{SOM}$ is derived from 
 the weighted mean of the $M_\mathrm{label}$ labels and readjusting it from the reference $i=22.5$\,mag  to the magnitude of the target galaxy: 
\begin{equation}
    M_\mathrm{SOM} = \frac{ \sum_{j=1}^5  \frac{1}{\Delta_j} M_{\mathrm{label},j}}{\sum_{j=1}^5\frac{1}{\Delta_j}} \times  10^{-0.4(m_i-22.5)} \,,
    \label{eq:m_som}
\end{equation}
where $\Delta_j$ is defined as in Eq.~\ref{eq:d_som} and corresponds to the $j$-th neighbor among the five nearest cells (including the one to which the galaxy is attributed). 
The same applies to SFR$_\mathrm{SOM}$:
\begin{equation}
    SFR_\mathrm{SOM} = \frac{ \sum_{j=1}^5  \frac{1}{\Delta_j} SFR_{\mathrm{label},j}}{\sum_{j=1}^5\frac{1}{\Delta_j}} \times  10^{-0.4(m_i-22.5)} \,,
    \label{eq:sfr_som}
\end{equation}
with the caveat that one or more cells may not contain MIPS objects and therefore have no SFR$_{\mathrm{label},j}$ defined  (Fig.~\ref{fig:som_calib}, lower-left panel). When all the five neighbors are not labeled, a SOM estimate cannot be assigned. The choice of the number of neighbors is arbitrary, but increasing to more than five cells does not change the results as the additional  neighbors  are further away from the target galaxy’s cell and  weighted in   Eq.~(\ref{eq:m_som})-(\ref{eq:sfr_som}) with a much smaller $1/\Delta_j$ factor.

We also derive $z_\mathrm{SOM}$ even though in principle we could use $z_\mathrm{LePh}$, as the latter is required in input. This is intended to avoid inconsistency between the various properties of any given galaxy. Moreover, the choice allows us to generalize the method, which could be used also when high-quality photometric redshifts are not available:  in fact, the $z<1.8$ cut could have been implemented with inferior redshift measurements or even a color selection \citep[][]{Guzzo:2014eb}.

\begin{figure}
    \centering
    \includegraphics[width=0.99\columnwidth]{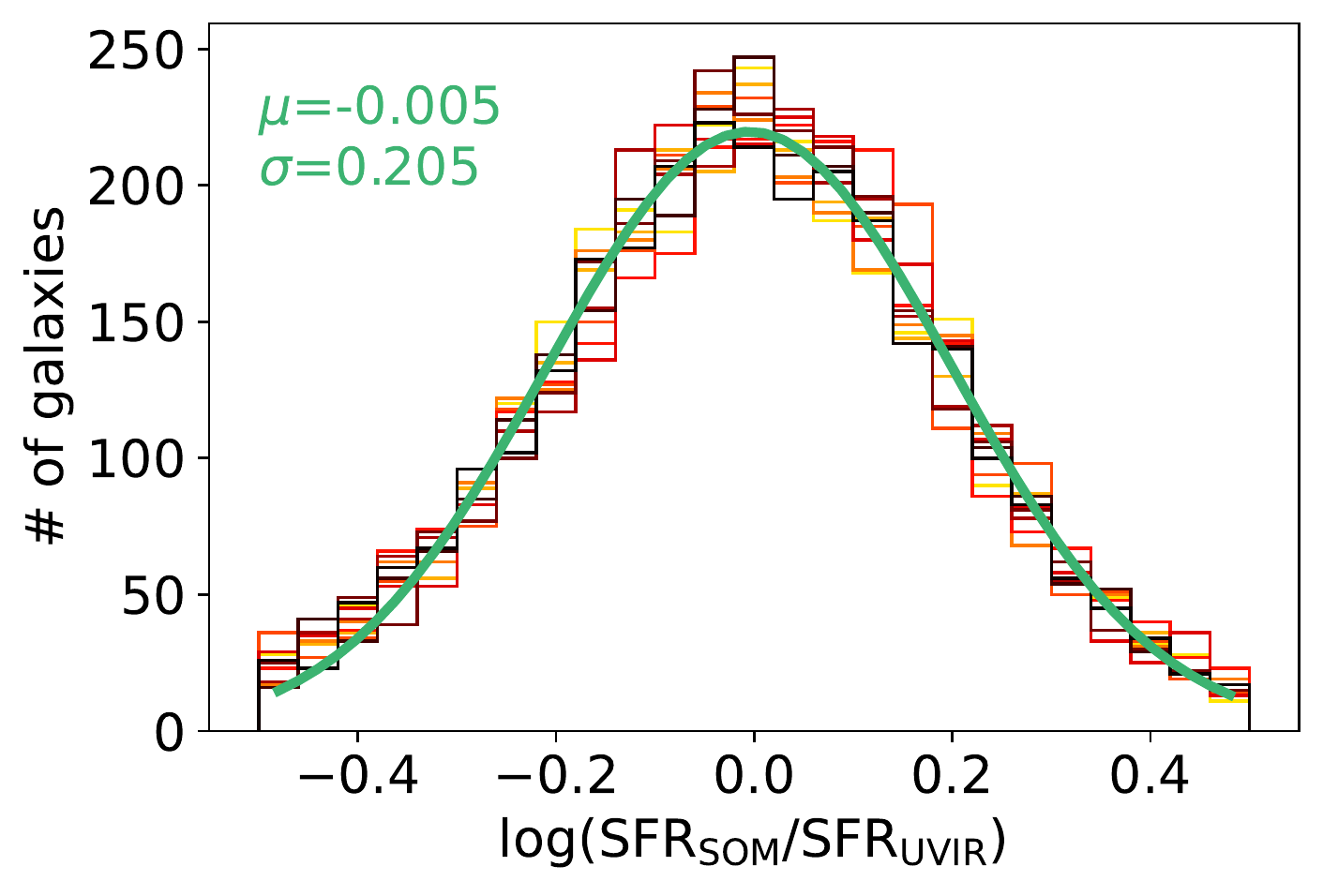}
    \caption{Validation of the SOM estimates with ``out-of-bag'' objects. The COSMOS2020 SOM described in Sect.~\ref{sec:methods} is built again ten times, excluding every time  $\sim$3,000 galaxies randomly extracted from those with a SFR$_\mathrm{UVIR}$. The figure compares SFR$_\mathrm{SOM}$ and SFR$_\mathrm{UVIR}$ for each SOM realization (histograms of different colors). The Gaussian fit to the ensemble of distributions (green line) has mean $\mu=-0.005$ and standard deviation $\sigma=0.205$\,dex. }
    \label{fig:validation_oob}
\end{figure}

For validation purposes, we also build ten SOMs  trained with COSMOS2020 after removing 3\,000 MIPS-detected galaxies randomly extracted each time.\footnote{
In general, every training of the SOM results in a different outcome when the weights are initialized at random positions. For this task however, we force the iterative adaptation to start from the same configuration of the reference SOM. Moreover,  the number of galaxies removed is so small compared to the whole training sample that the ten new SOMs are almost indistinguishable from the original one.  } These ``out-of-bag'' objects belong to the MIPS parent sample, meaning that some of them have $S/N<5$ and would not contribute to the calibration phase anyway. By construction, all the out-of-bag objects have a SFR$_\mathrm{UVIR}$ estimate that can be compared to the SFR$_\mathrm{SOM}$ obtained by projecting them on the respective SOM, similarly to what is done for the SXDF2018 catalog. The $\log(\mathrm{SFR_\mathrm{SOM}/SFR_{UVIR}})$
distribution is the same in the ten realizations, modulo stochastic fluctuations. By fitting them with a Gaussian function we find a standard deviation of 0.205\,dex and a negligible offset ($-0.005$) with respect to the reference MIPS estimates (see Fig.~\ref{fig:validation_oob}). We repeat the Monte Carlo experiment only in the left side of the SOM that is undersampled by MIPS: for the galaxies that do receive a SOM prediction -- i.e., at least one of their nearest neighbors is labeled -- we find that offset and scatter ($\mu=-0.005$, $\sigma=0.205$ respectively) are almost the same as the global experiment. The analog procedure is performed to establish the $M_\mathrm{SOM}$ accuracy, dividing the out-of-bag sample in bins of stellar mass and comparing predictions with $M_\mathrm{LePh}$. In this case the uncertainties are $\mu=-0.018$ and $\sigma=0.097$\,dex.  Since they are calculated for targets coming from the same parent sample of the training galaxies, these values can be interpreted as lower limits for the $M_\mathrm{SOM}$ and SFR$_\mathrm{SOM}$ uncertainties  that would result when projecting other data sets. 

Error bars on an object-by-object basis can be provided in different ways. \citet{hemmati19} propose  Monte Carlo re-mapping of the target galaxy, each time perturbing its colors within the corresponding 1$sigma$ uncertainty; the object  is scattered in nearby cells and the minimum and maximum (mass or SFR) values it reaches are used as lower and upper limits of the error bar. \citet{speagle19} adopt a different approach, which assigns to each cell a probability of hosting the target galaxy proportional to its $\chi^2$ (see Eq.~\ref{eq:chi2_som} in the Appendix). We tried both methods and found that they often provide significantly different error bars. A possible explanation is the fact that i) in the Monte Carlo we perturbed colors independently (i.e., without taking into account covariance) and/or ii) the color error bars do not correspond precisely to a 68\% uncertainty, resulting in a biased $chi^2$. We preferred to invest more time exploring this issue,  postponing the ascription of error bars to $M_\mathrm{SOM}$ and SFR$_\mathrm{SOM}$ to a future study.

\section{Results}
\label{sec:results}
In  this section, our SOM-based analysis in SXDF is compared to other methods that independently derived stellar masses and SFR for the same objects. The alternate measurements for stellar masses are produced by means of  \texttt{LePhare}, which fits BC03 templates to SEDs that includes not only the data used in the SOM but also UV photometry from GALEX, medium- and narrow-bands in the optical, and IRAC channel 3 and 4. These estimates are originally published in \citet{mehta18}, but we re-run \texttt{LePhare} to update some of its parameters to the actual configuration used for COSMOS2020 (the redshift being fixed at the same value used in \citeauthor{mehta18}).  For the SFR comparison, we first compare to  \citet{barro19}, a study  already introduced above which is restricted to the CANDELS $\sim$200 square arcmin area within SXDF. Like in  \citet[][]{mehta18}, \citeauthor{barro19} also exploit a set of observations larger than the 12 broad-bands colors used to feed the SOM.  Their study assumes the same cosmological parameters and IMF used here, but other adjustments related to  Eq.~\ref{eq:SFR_UVIR} are required to convert their SFR estimates to a common ground. This  homogenization procedure is described in Appendix~\ref{appendix1}. Another set of independent  estimates comes from spectroscopic data, where nebular emission lines can be used as a proxy for SFR. 

\subsection{Stellar mass estimates}
\label{subsec:results_mass}

\begin{figure}
    \centering
    \includegraphics[width=0.99\columnwidth]{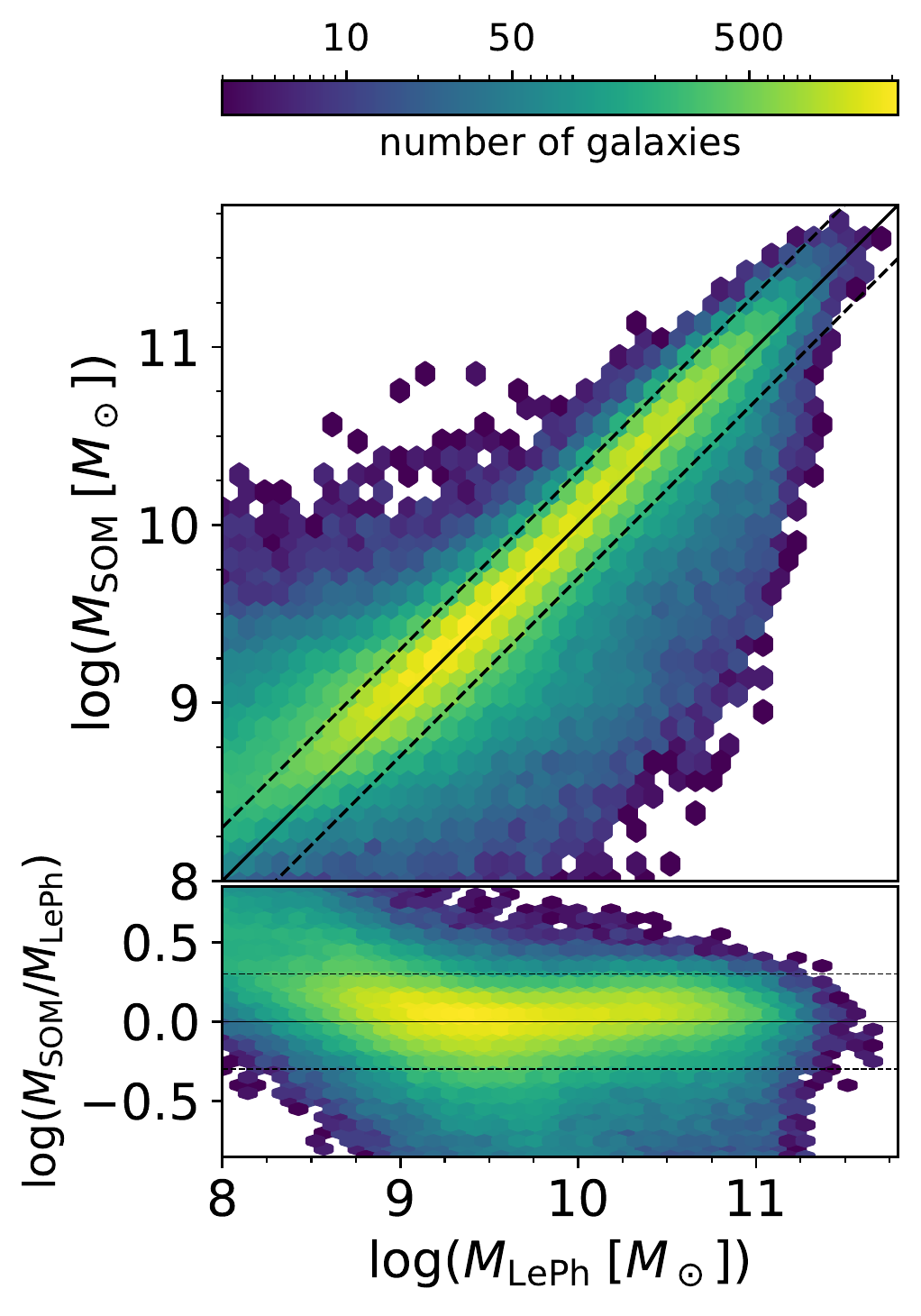}
    \caption{Comparison  between stellar masses obtained through standard template fitting ($M_\mathrm{LePh}$) and the new method presented in this work ($M_\mathrm{SOM}$). The Figure shows the density map of 208\,404 galaxies in the SXDF field with the magnitude cut at $K_\mathrm{s}<24.2$.   In the \textit{upper panel}, a solid line marks the 1:1  relationship. In the \textit{bottom panel}, showing the logarithmic ratio  $\log(M_\mathrm{SOM}/M_\mathrm{LePh})$ as a function of $\log(M_\mathrm{LePh}/M_\odot)$, the solid line marks the zero offset while the two dotted  lines are set at  $\pm0.3$\,dex. }
    \label{fig:results_msom-msed}
\end{figure}

\begin{figure*}
    \centering
    \includegraphics[width=0.80\columnwidth]{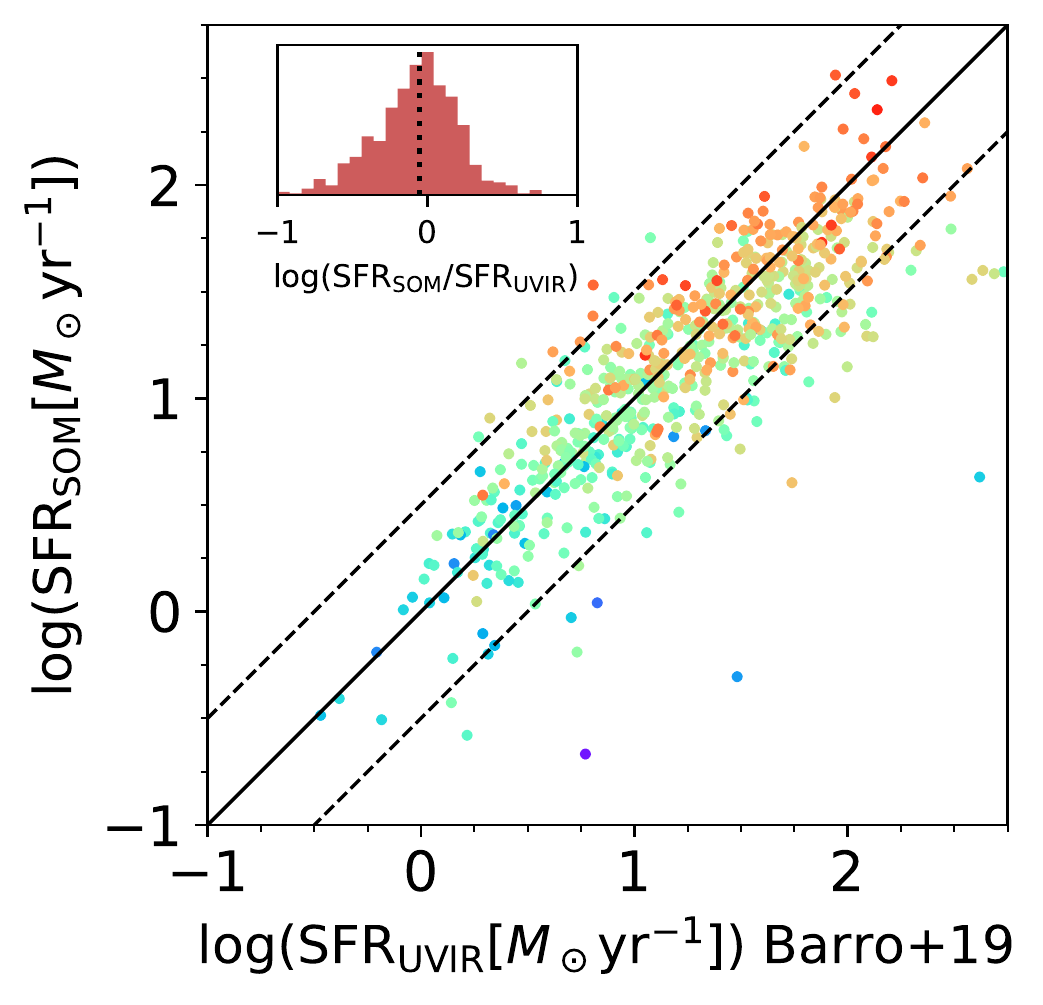} \hspace{5mm} \includegraphics[width=0.99\columnwidth]{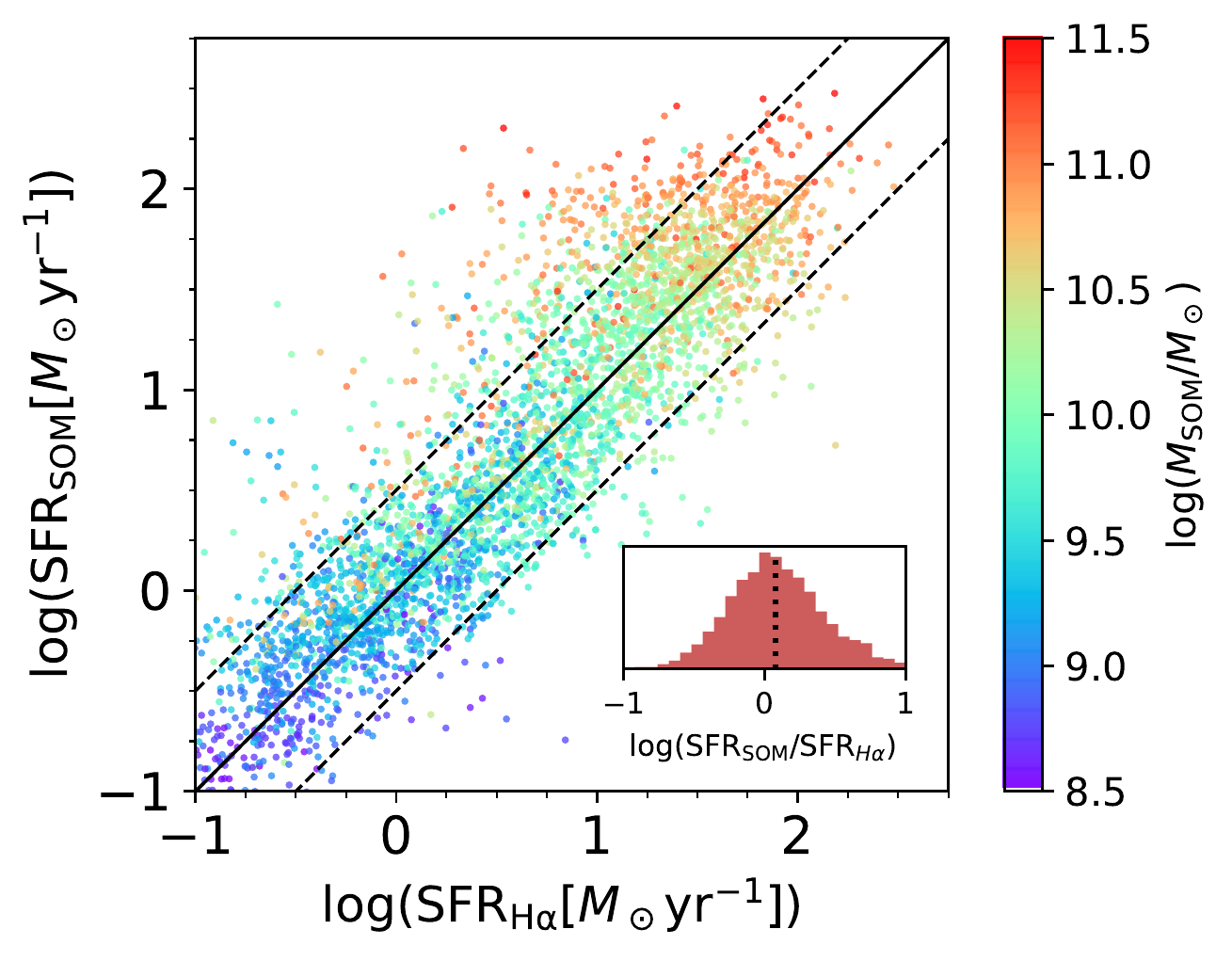}
    \caption{
    \textit{Left panel:} 
    Comparison between SFR$_\mathrm{SOM}$ and the template-based estimates from \citet[][SFR$_\mathrm{UVIR}$]{barro19}  for 608 galaxies at $z<1.8$ in the SXDF-CANDELS area (colored dots). For sake of consistency, \citeauthor{barro19} estimates are converted to the same reference system of \citet{ilbert15}, since the latter has also been used to calibrate the SOM method (see Appendix~\ref{appendix1}).   
    \textit{Right panel:}
    Comparison to another star formation estimator, i.e., H$\alpha$ nebular emission line. In this case the comparison is performed in COSMOS, for 3\,718 galaxies  (colored dots). 
    In both panels, the color palette from violet to red indicates the stellar mass of each galaxy according to the SOM method; the solid line is the 1:1 relation from which two  dashed lines are set at a distance of $\pm0.3$\,dex. Each panel also includes an inset that quantifies the dispersion between SFR$_\mathrm{SOM}$ and the alternate estimate (red histogram); the vertical dotted line marks the median offset, which is $-0.05$\,dex for $\log(\mathrm{SFR_{SOM}/SFR_{UVIR}})$ and 0.07\,dex for $\log(\mathrm{SFR_{SOM}/SFR_{UVIR}})$.
    }
    \label{fig:sfr_barro_vs_som}
\end{figure*}

The stellar mass comparison (Fig.~\ref{fig:results_msom-msed}) shows a good agreement  between  $M_\mathrm{LePh}$ and $M_\mathrm{SOM}$. The systematic offset between the two, measured as $\log(M_\mathrm{SOM}/M_\mathrm{LePh})$, is $<$0.02\,dex. The relative scatter of 0.25\,dex is nearly symmetric, with the exception of a $\sim$3\% for which $M_\mathrm{SOM}$ is severely underestimated (i.e., more than a factor 3). The majority of that subsample has $1.2<z_\mathrm{LePh}<1.8$ and is located either along the borders of the grid (i.e., they suffer from ``boundary effects'') or in the central area characterized by large $\Delta$ values (see Sect.~\ref{subsec:discuss_soundness}). 

At a first glance, such an agreement may seem obvious because the set of $M_\mathrm{label}$  used to derive individual masses in the SOM is built from \texttt{LePhare} estimates (Eq.~\ref{eq:mass_resc}). However, there are several deviations from the SED fitting performed by \texttt{LePhare} that make the SOM method substantially different.
First, the smaller number of photometric data points used in the latter, as emphasized above. In addition, the available ``solutions''  in the SOM fit are much smaller, given the limited number of cells used to describe the data manifold (6\,400 in total). On the contrary,  the  synthetic templates used by \texttt{LePhare} are between five and eight thousands at every fixed $z$ step\footnote{The precise number of templates that \texttt{LePhare} can use at any given $z$ step depends on how many of them represent galaxies older than the age of the universe at that redshift; in fact, those models are excluded by default.}  and over two millions in total.

We also notice that in the SOM method  only the  weights in the  surroundings of the galaxy target  are relevant for the   $M_\mathrm{SOM}$ calculation (see  Eq.~\ref{eq:m_som}) whereas there  are no mass priors in \texttt{LePhare}. In the  library of the latter, the   BC03 galaxy models are all defined at  $1\,M_\odot$, but each  of them can be rescaled to fit the observed SED. This means that each template can span the entire stellar  mass range of the target pool, i.e., is eligible to fit any observed galaxies irrespective of its mass (or other characteristics).  Moreover, the choice of either using or disregarding a SOM weight for a certain galaxy is modulated by the adaptive resolution of neighboring cells, i.e.\ the fact that in dense areas of the parameter space the SOM concentrates more weights. 
Another distinctive feature of the  SOM, which may be regarded as an advantage or a shortcoming depending on the scientific goal, is that its collection of  \textit{empirical} weights is limited by the survey's characteristics (e.g., its selection function).  On the other hand a synthetic template library, at least theoretically, may include models of known galaxy types that are not observed in that survey (e.g., low-mass objects fainter than the detection limit) but also miss real objects that have not been correctly simulated as templates. Given these differences, the small fraction of catastrophic errors and the overall tight 1:1 correlation are a remarkable confirmation of the effectiveness of the SOM method.

\subsection{Star formation rate estimates}
\label{subsec:results_sfr}

First, we compare the SOM estimates to the SFR$_\mathrm{UVIR}$ ones from \citet{barro19}. The authors only analyze CANDELS-UDS, which is about one twentieth of the whole SXDF area. As an additional restriction we select only the CANDELS galaxies with UV-to-FIR data available. Given also the SOM cut at $z<1.8$ and $K_\mathrm{s}<24.23$, there are only 608 galaxies in common with \citet{barro19}. For AGN contamination, we consider the classification from \citet{mehta18} which is similar to the one applied to the COSMOS2020 training sample. We also discard sources with discrepant redshift estimates, i.e.\ the absolute value of the  difference between $z_\mathit{LePh}$ and the photometric redshift from \citet{barro19} is larger than $0.3(1+z_\mathrm{LePh})$. The 608 sources are plotted in the left panel of  Fig.~\ref{fig:sfr_barro_vs_som}, showing a fairly good agreement: the $-0.05$\,dex offset in $\log(\mathrm{SFR_{SOM}/SFR_{UVIR}})$ is comparable with the $-0.03$\,dex initially found in the CANDELS-COSMOS comparison.\footnote{This is the residual offset after removing other sources of discrepancy such as IMF and IR calibration 
 (see Fig.~\ref{fig:sfr_barro_vs_ilbert}).} The SFR scatter (i.e., the standard deviation of the distribution in the left panel of Fig.~\ref{fig:sfr_barro_vs_som}) is 0.28\,dex, comparable to that of the stellar mass distribution. The $\log(\mathrm{SFR_{SOM}/SFR_{UVIR}})$ histogram (inset in the left panel of Fig.~\ref{fig:sfr_barro_vs_som}) features a tail of a few objects that the SOM severely underestimates with respect to \citet{barro19}, many of them being very dusty ($A_V>2$\,mag) according to the latter. Dust correction is indeed one of the major parameters responsible for this kind of discrepancy, as discussed e.g.\ in \citet{speagle14}.

 Another star formation proxy that can be used for comparison is the luminosity of H$\alpha$ nebular emission line ($L_\mathrm{H\alpha}$). We assume SFR$_\mathrm{H\alpha} = 2.1 \times 10^{-8} \times L_\mathrm{H\alpha}$ \citep{kennicutt98} after correcting for dust attenuation in the host galaxy. The formula for such a correction is the same as in \citet{kashino13}, based on the stellar color excess $E(B-V)$ parametrized by \texttt{LePhare} while measuring the photometric redshift. 
 
 It is more convenient to perform this test in the COSMOS field instead of SXDF, because we have access to an unparalleled amount of spectroscopic data in the former. The most relevant surveys to our purposes are $z$COSMOS \citep{Lilly:1996p12057}, 3DHST \citep{brammer12,momcheva16}, KROSS \citep{harrison17}, and FMOS \citep{kashino19}, targeting different galaxy types from $z\sim0.1$ up to 1.7. They have been selected and merged into a single spectroscopic catalog in \citet{saito20}. Because of the overlap between these line emitters and the SOM calibration sample, we prefer to compare SFR$_\mathrm{H\alpha}$ vs.\ SFR$_\mathrm{SOM}$ for the out-of-bag galaxies of the SOM bootstrap test (see Sect.~\ref{subsec:meth_sxdf} and Fig.~\ref{fig:validation_oob}). Assuming SFR$_\mathrm{H\alpha}$ to be the ``ground truth'', the comparison is a further validation of the SFR$_\mathrm{SOM}$ estimates, presented in the right panel of Fig.~\ref{fig:sfr_barro_vs_som}. The resulting scatter (0.4\,dex) and  systematics ($+0.07$\,dex)  are relatively contained considering the differences between the two methods, especially the shorter time scale of star formation (on the order of 10\,Myr) probed by $L_{H\alpha}$. We also note that the 0.07\,dex offset shown in the figure is mainly due to the most massive galaxies, which may have a biased SFR$_\mathrm{H\alpha}$: in fact, their dust content is usually high and susceptible to be underestimated by \texttt{LePhare}, whose templates are limited to $E(B-V)\leq 0.7$ and may not assume the correct extinction law \citep{ilbert10,walcher11}.  Moreover, in (local) massive galaxies,  nebular extinction often adds a greater contribution to stellar dust  extinction  \citep{vandergiessen22} and BC03 models  do not take this into account.

\subsection{Main sequence of star-forming galaxies}
\label{subsec:results_ms}

\begin{figure}
    \centering
    \includegraphics[width=0.99\columnwidth]{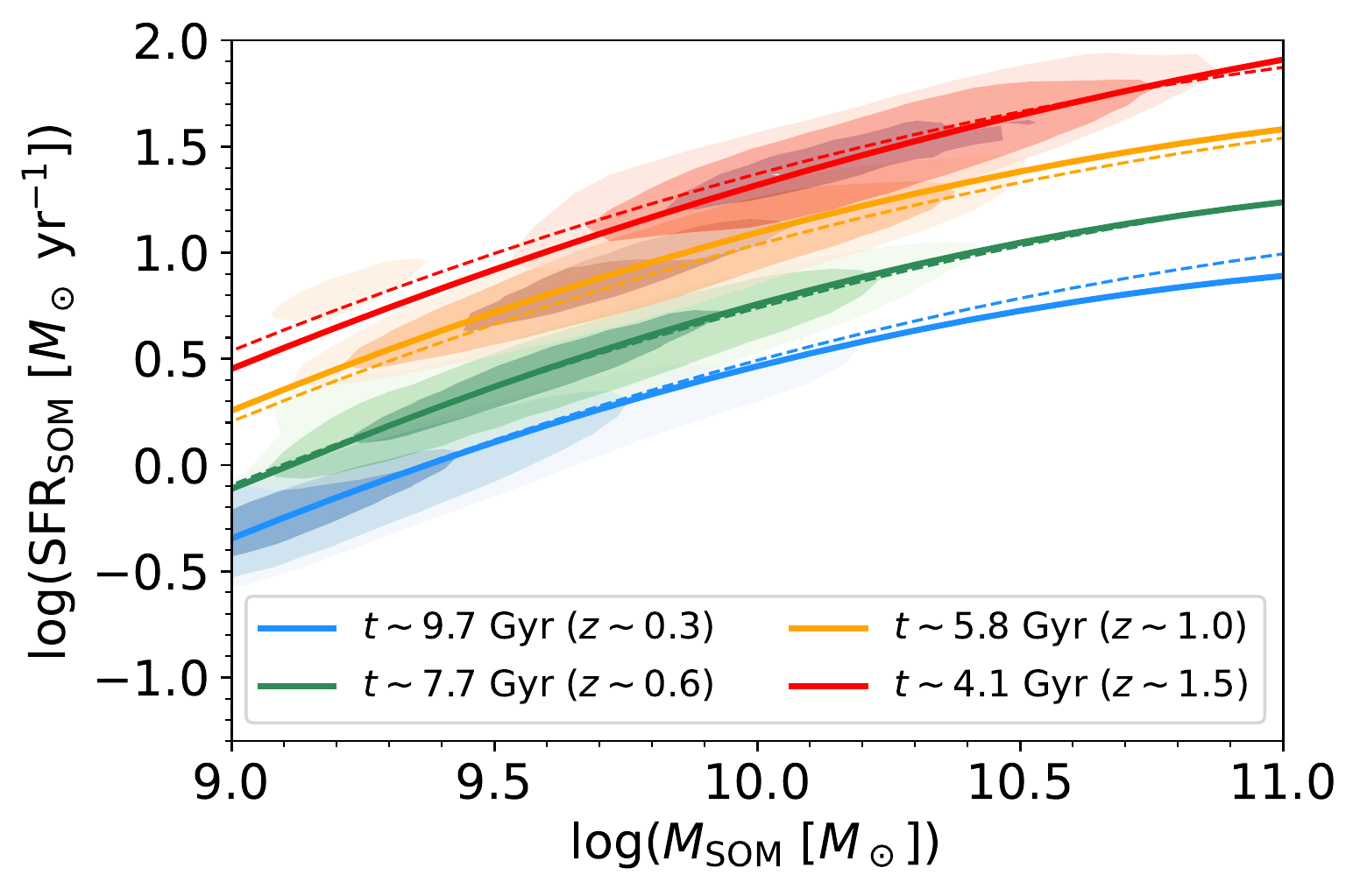}
    \caption{Evolution of the main sequence of star forming galaxies resulting from the SOM-based measurement of the SXDF2018 catalog. Redshift bins are chosen to correspond to four time steps separated by $\sim2$\,Gyr, whose color is indicated in the legend along with median redshift and cosmic age. In each bin, the main sequence is identified as the most prominent overdensity in the SFR$_\mathrm{SOM}$-M$_\mathrm{SOM}$ plane (shaded areas representing density contours).  Each overdensity peak is fit by Eq.~\ref{eq:schreiber} either independently (solid lines) or simultaneously at all $t$ (dashed lines).}
    \label{fig:ms_evo}
\end{figure}

With the SXDF2018 photometric redshifts, stellar masses, and SFRs obtained through the SOM we can trace the main sequence of star-forming galaxies (MS) and its evolution as a function of cosmic time. 
The SFR$_\mathrm{SOM}$ vs.\ $M_\mathrm{SOM}$ plane at different redshifts is shown in Fig.~\ref{fig:ms_evo}, where we choose the following bins in order to ensure homogeneous time intervals: $0.3<z\leq0.4$, $0.5<z\leq0.7$, $0.9<z\leq1.1$, $1.4<z\leq1.8$. The four bins are centered around median redshifts of $z=0.3$, $0.6$, $1.0$, and $1.5$ corresponding to $t=9.8$, 7.7, 5.7, and 4.2\,Gyr after the Big Bang. The time step between the last two bins is the only one shorter than 2\,Gyr because of the $z<1.8$ upper limit imposed by MIPS. Galaxies at $z\lesssim0.2$ are not included since COSMOS probes a small cosmic volume below that redshift, thus losing most of its statistical power. A more conventional binning is used later in this Section to compare to the literature. Similarly to the $\chi^2$ cut applied in many template-fitting studies \citep[see e.g.][]{bowler15} we remove the ``bad fit'' objects having $\Delta>5$. This threshold rules out nearly 1\% of the galaxies whose SED is not well represented by the weight of their own cell.

To locate the MS we apply a procedure inspired by the MS definition in \citet{renzini&peng15}. 
In that study\footnote{See in particular their figure 1.} the MS naturally emerges as a peak in the surface formed by \textit{all} the galaxies in the 3-D space SFR vs.\ mass vs.\ number of objects, without any a-priori separation between star forming and quiescent. In the \citeauthor{renzini&peng15} framework the MS core is the ``ridge'' of such a 3-D peak, a line that is stable whether or not the data set includes, e.g., post-starburst galaxies exiting the MS. A similar approach, not requiring a preselection of the star-forming sample, has been recently proposed in \citet{leja21}. The SOM method, however, is not entirely free from selection effects because galaxies falling far from the labeled cells do not obtain any SFR$_\mathrm{SOM}$ estimate by construction.  
Irrespective of that potential bias, which will be discussed in Sect.~\ref{subsec:discuss_ms}, the SXDF2018 sample includes a wide range of galaxies, from starbursts above the MS to galaxies with low specific star formation rates. 

To identify the ridge of the star forming sequence we apply the following procedure in every $z_\mathrm{SOM}$ bin. First, a Gaussian kernel density estimator\footnote{Namely, a \texttt{gaussian\_kde} method \citep{ScottMDE} from the \texttt{scipy} suite \citep{scipy}. } is used to build a probability distribution map of SXDF2018 objects in the $\log(\mathrm{SFR_{SOM}})$ vs.\ $\log(M_\mathrm{SOM})$ space. Each map is shown in Fig.~\ref{fig:ms_evo} with isodensity contours of different colors. Resampling them with a grid of points equally spaced by 0.05\,dex we find the ridge of the density peak (i.e., the ``backbone'' of the MS) and track its evolution over $\sim$6\,Gyr. The ridges in the four redshift bins are interpolated by the same function used in \citet{schreiber15}: 
\begin{equation}
    y = m - m_0 + a_0 r - a_1 [\mathrm{max}(0,m-m_1-a_2r)]^2\,,
    \label{eq:schreiber}
\end{equation}
where $r\equiv\log(1+z)$, $m$ is the logarithmic stellar mass in units of $10^9\,M_\odot$, and $y$ is the logarithmic SFR in $M_\odot$\,yr$^{-1}$. The other five parameters ($m_0,m_1,a_0,a_1,a_2$) are obtained via nonlinear least-square minimization using a trust region reflective algorithm\footnote{Implemented in \texttt{scipy} \citet[see also][]{STIR}.}. The best-fit values for the free parameters, listed in Table~\ref{tab:ms_fit}, result into the solid lines shown in Fig.~\ref{fig:ms_evo}. We also fit the four sequences together (dashed lines) obtaining the following parameters: $(m_0,m_1,a_0,a_1,a_2)=(0.75\pm0.04,0.0\pm0.2,3.2\pm0.2,0.17\pm0.04,0.0\pm0.5)$. The terms $m_1$ and $a_2$ being consistent with zero indicates that there is no need of second order corrections to describe the MS turnover at high masses. This is the major distinction with respect to the MS of \citet{schreiber15}, where the bending is more pronounced.  In fact, thanks to the stacking of \textit{Herschel} images,  \citeauthor{schreiber15} gain sensitivity to measure also galaxies with very low sSFR. The best-fit parameters they find are $(m_0,m_1,a_0,a_1,a_2)=(0.5\pm0.07,0.36\pm0.03,1.5\pm0.15,0.3\pm0.08,2.5\pm0.06)$. For sake of completeness, we also try a linear interpolation using the function $y=(\alpha_0 t+\alpha_1)\log(M/M_\odot)+(\beta_0 t +\beta_1)$ as done in \citet[][equation 17]{speagle14} but the goodness of fit is inferior to Eq.~\ref{eq:schreiber} and therefore it is not shown in the Figure. 

\begin{table}[]
     \caption{Best-fit values of the free parameters in Eq.~\ref{eq:schreiber} to describe the main sequence of star-forming galaxies at different redshifts. Parameters in the bottom row result from the simultaneous fit to all the four redshift bins.}
    \begin{tabular}{ccccccc}\hline\hline
        $z$ range & $\langle z \rangle$ &  $m_0$ & $m_1$ & $a_0$ & $a_1$ & $a_2$  \\ \hline
$0.30<z<0.40$ & $\langle 0.34\rangle$ & $0.44$ & $0.0$ & $0.7$ & $0.19$ & $0.0$ \\
$0.50<z<0.70$ & $\langle 0.60\rangle$ & $0.44$ & $0.1$ & $1.6$ & $0.20$ & $0.6$ \\
$0.90<z<1.10$ & $\langle 0.99\rangle$ & $0.38$ & $0.0$ & $2.1$ & $0.18$ & $0.0$ \\
$1.40<z<1.80$ & $\langle 1.53\rangle$ & $0.06$ & $0.0$ & $1.3$ & $0.14$ & $0.0$ \\
$0.30<z<1.80$ & -- & $0.75$ & $0.0$ & $3.2$ & $0.17$ & $0.0$ \\ \hline
    \end{tabular}  
    \label{tab:ms_fit}
\end{table}

\begin{figure*}
    \centering
    \includegraphics[width=0.99\textwidth]{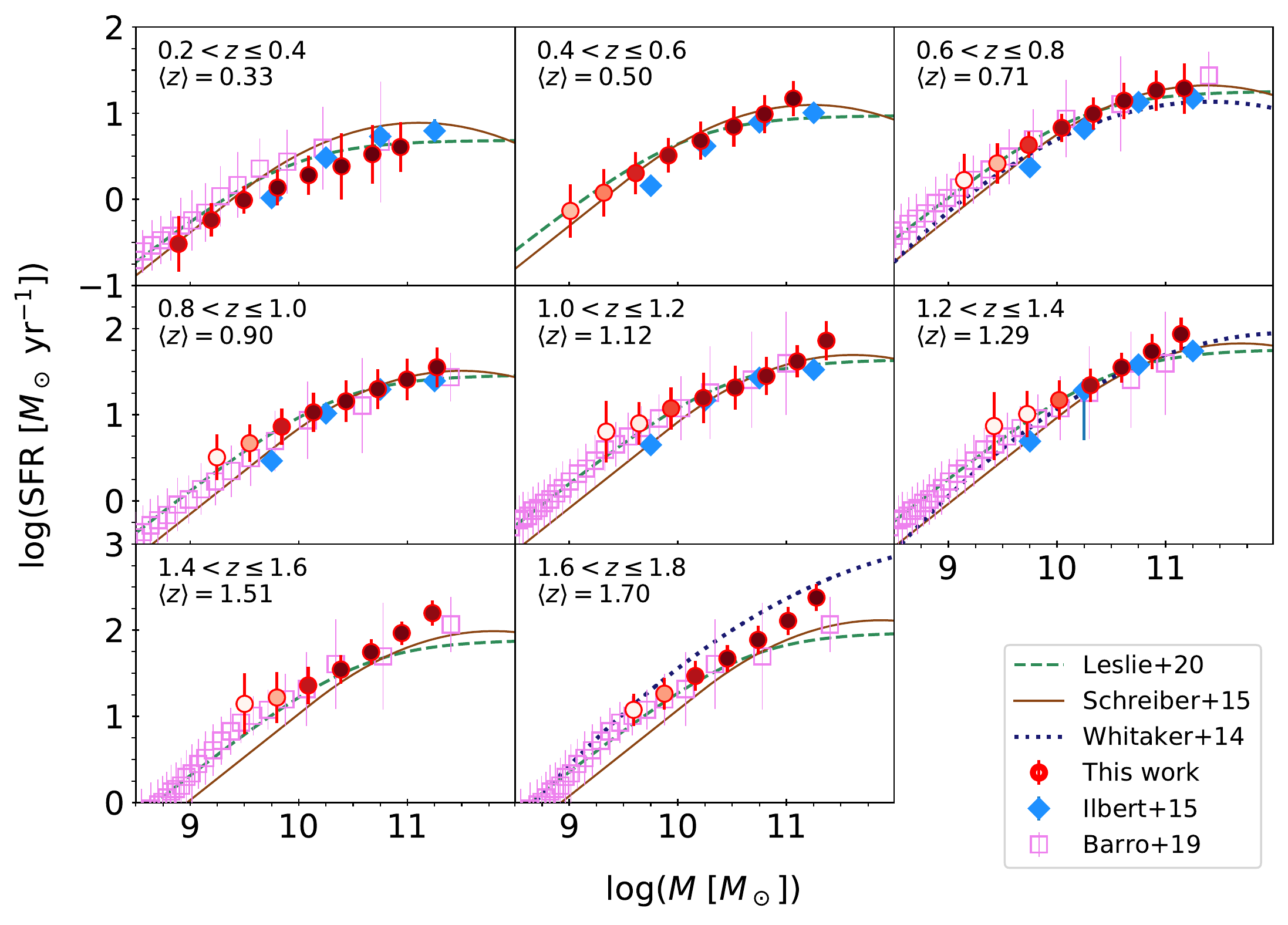}
    \caption{Redshift evolution of the MS of star forming galaxies (see Sect.~\ref{subsec:results_ms}) as determined in the present work (red circles), \citet[][blue diamonds]{ilbert15}, \citet[][pink squares]{barro19}, \citet[][blue dotted line]{whitaker14}, \citet[][brown solid line]{schreiber15}, and \citet[][green dashed line]{leslie20}. Our estimates stop at the $z$-dependent threshold for stellar mass completeness defined in \citet{weaver22}. Their error bars are the 16th-84th percentile range and each symbol is filled with a shade of red according to the completeness of the given mass bin; in particular, dark red symbols represent $>$90\% completeness, orange between 60 and 80\%, while white-filled circles have $\lesssim$50\% completeness (see Sect.~\ref{subsec:discuss_ms}). Results from the literature are shown only when the median redshift of the given study is sufficiently close to ours (which is indicated in the upper-left corner of each panel); the same measure may be repeated in more than one panel: e.g., the MS of \citet{barro19} at $0.5<z<1$ has a median redshift $\langle z \rangle = 0.8$ and it is compared to our data both at $0.6<z<0.8$ and $0.8<z<1.0$.  }
    \label{fig:ms_compar}
\end{figure*}

We also select from the literature some of the most relevant studies that have measured the MS between $z=0$ and 2, namely   \citet{whitaker14,schreiber15,ilbert15,barro19,leslie20}. To estimate galaxy SFRs those authors use different methods, all based on FIR data with the exception of \citet{leslie20} performing a radio analysis. Two of these studies deal with individual sources only \citep{whitaker14,barro19} while the other two include  undetected galaxies by stacking either 3\,GHz VLA \citep{leslie20} or \textit{Herschel} \citep{schreiber15} images. \citet{ilbert15} derive the MS from MIPS-detected galaxies but in an indirect way, through the SFR function.  
All the selected studies identify the MS locus by  binning their sample in  stellar mass and then computing the median SFR in each bin. Although such an approach is suboptimal from a mathematical point of view \citep{steinhardt&adam18} we decided nonetheless to proceed in the same way of the other authors to cohere with them.\footnote{
Before the actual comparison, however, we converted the results to the same framework as described in Appendix~\ref{appendix1}; see also \citet[][]{speagle14} for a thorough discussion about  how to homogenize miscellaneous data from the literature. 
}

The SOM results are in good agreements with  the other MS measurements (Fig.~\ref{fig:ms_compar}) showing that SFR$_\mathrm{SOM}$ and $M_\mathrm{SOM}$ estimates are accurate not only when evaluated  separately. Overall, the differences between our MS and previous results are comparable with the differences between them, and  smaller than the 1$\sigma$ error bars that we defined as the 16$^{th}$-84$^{th}$ percentile range in each mass bin. There are, however, two visible systematics, at both extremes of the probed mass range, which are worth noticing. At the  low-mass end, our median points slightly deviate from the linear extrapolation of the MS. This is related to  a sample incompleteness in those mass bins and a consequent skewness of the SFR distribution towards higher values. The incompleteness is shown in Fig.~\ref{fig:ms_compar} by white and orange filled circles, marking bins in which the fraction of galaxies with a defined SFR$_\mathrm{SOM}$ drops below $50$ and $70$\% respectively.\footnote{The fraction is calculated with respect to the total number of objects in the given stellar mass bin and classified as star forming by Eq.~\ref{eq:nuvrk}. } 
The other systematic effect concerns the most massive bin, which at $z>1$ is always located above the other studies (although within $\sim$1$\sigma$ from them, see Fig.~\ref{fig:ms_compar}). In general, our median SFRs do not show the high-mass turnover observed  in \citet{ilbert15}, \citet{lee15}, and subsequent studies (see discussion in Sect.~\ref{subsec:discuss_ms}). Besides that, the overall slope and redshift evolution of our MS agree with previous work confirming the reliability of the SOM method.

\section{Discussion}
\label{sec:discussion}

\subsection{SOM-based estimates vs.\ synthetic templates}
\label{subsec:discuss_sfr}

\begin{figure}
    \centering
    \includegraphics[width=0.99\columnwidth]{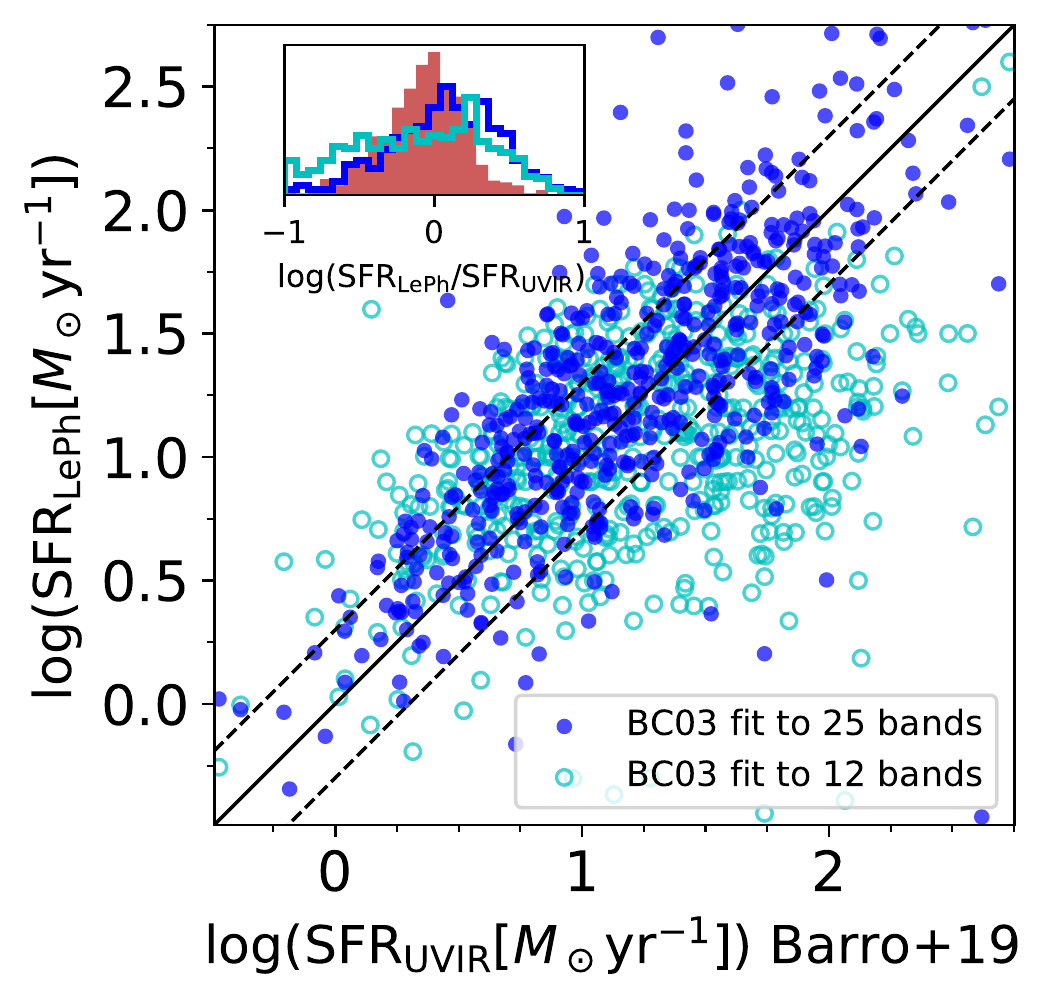}
    \caption{Comparison between \citet{barro19} and \texttt{LePhare} using the same sample of 608 galaxies also shown in the left panel of  Fig.~\ref{fig:sfr_barro_vs_som}. The same BC03 library is fit by \texttt{LePhare} to the full photometric baseline available in the SXDF2018 catalog (filled blue circles) and to the 12 filters only used in the SOM method (open cyan circles). Same colors are used in the inset for the histograms showing the respective $\log(\mathrm{SFR_{LePh}}/\mathrm{SFR_{UVIR}})$ distributions; the $\log(\mathrm{SFR_{SOM}}/\mathrm{SFR_{UVIR}})$ distribution (red histogram, same as the inset in the left panel of  Fig.~\ref{fig:sfr_barro_vs_som}) is also included as reference. }
    \label{fig:sfr_barro_vs_leph}
\end{figure}

To assess any improvement the SOM may provide over standard template fitting, the latter has to be compared to  reference estimates \citep{barro19} as it has been done with SFR$_\mathrm{SOM}$ (Sect.~\ref{subsec:results_sfr}). Figure \ref{fig:sfr_barro_vs_leph} shows such a test in the overlapping part of the SXDF field (see Appendix~\ref{appendix2} for COSMOS). When exactly the same 12-bands photometry is used (cf.~Table~\ref{tab:filters}) the scatter in $\log(\mathrm{SFR}_\mathrm{LePh}/\mathrm{SFR}_\mathrm{UVIR})$ is significantly larger than what found for the SOM-based method  (cyan circles in Fig.~\ref{fig:sfr_barro_vs_leph}).   
The  distribution becomes narrower, and less skewed, if additional colors (25 filters in total) are provided as input to \texttt{LePhare} (blue squares in Fig.~\ref{fig:sfr_barro_vs_leph}). Either way, the fit is performed without data points in the FIR regime. For the 12-bands fitting we run \texttt{LePhare} using the same version and set-up of COSMOS2020, while the fit to the extended photometry comes from \citet{mehta18}.  Therefore, the latter estimates not only  include ancillary data (medium- and narrow-band filters from Subaru and VISTA telescopes) but are also derived with a configuration of \texttt{LePhare} that is optimized for SXDF2018.  However, even in that case, the template-based estimates are less precise than SFR$_\mathrm{SOM}$.

The effective performance of our method is mainly due to the empirical collection of galaxy prototypes that the SOM creates during training \citep[see also the ``phenotypes'' in][]{sanchez&bernstein19}. Similarly to eigenvectors in Principal Component Analysis, the SOM weights are adapted to the observed sample so that \textit{by construction} their colors represent  realistic SEDs, to which physical properties are attached. These properties can  be derived from scaling relations or other empirical recipes, which despite their underlying assumptions, offer a  complementary approach to the use of stellar population synthesis models.   

In standard template fitting, the library is generated from theoretical models that might not be an accurate description of the observed targets -- or even of their parent population -- especially regarding star formation.  In fact, many template libraries  \citep[including the one in][]{weaver22} are built by using a simplistic SFH parametrization \citep[exponential-$\tau$ and delayed-$\tau$ models, see][]{ilbert13} with a limited number of time steps, which may be inadequate e.g.\ for starburst galaxies \citep{pacifici13}. Moreover, the physical parameter space is sampled by a coarse grid of stellar metallicity and $E(B-V)$ values, and there are limited options to add nebular emission line contamination \citep{pacifici15,yuan19}. Part of the parameter space covered by the models might have no correspondence in the observed universe (see the  discussion in \citealp{marchesini10,muzzin13} concerning dusty, passive galaxy templates). Moreover, the grid discretization may introduce severe biases, as investigated in \citet[][]{mitchell13}. Besides these approximations, the simplistic modelling of dust attenuation plays a major role  \citep[][]{chevallard13,chevallard16,laigle19}.

\begin{figure}
    \centering
    \includegraphics[width=0.99\columnwidth]{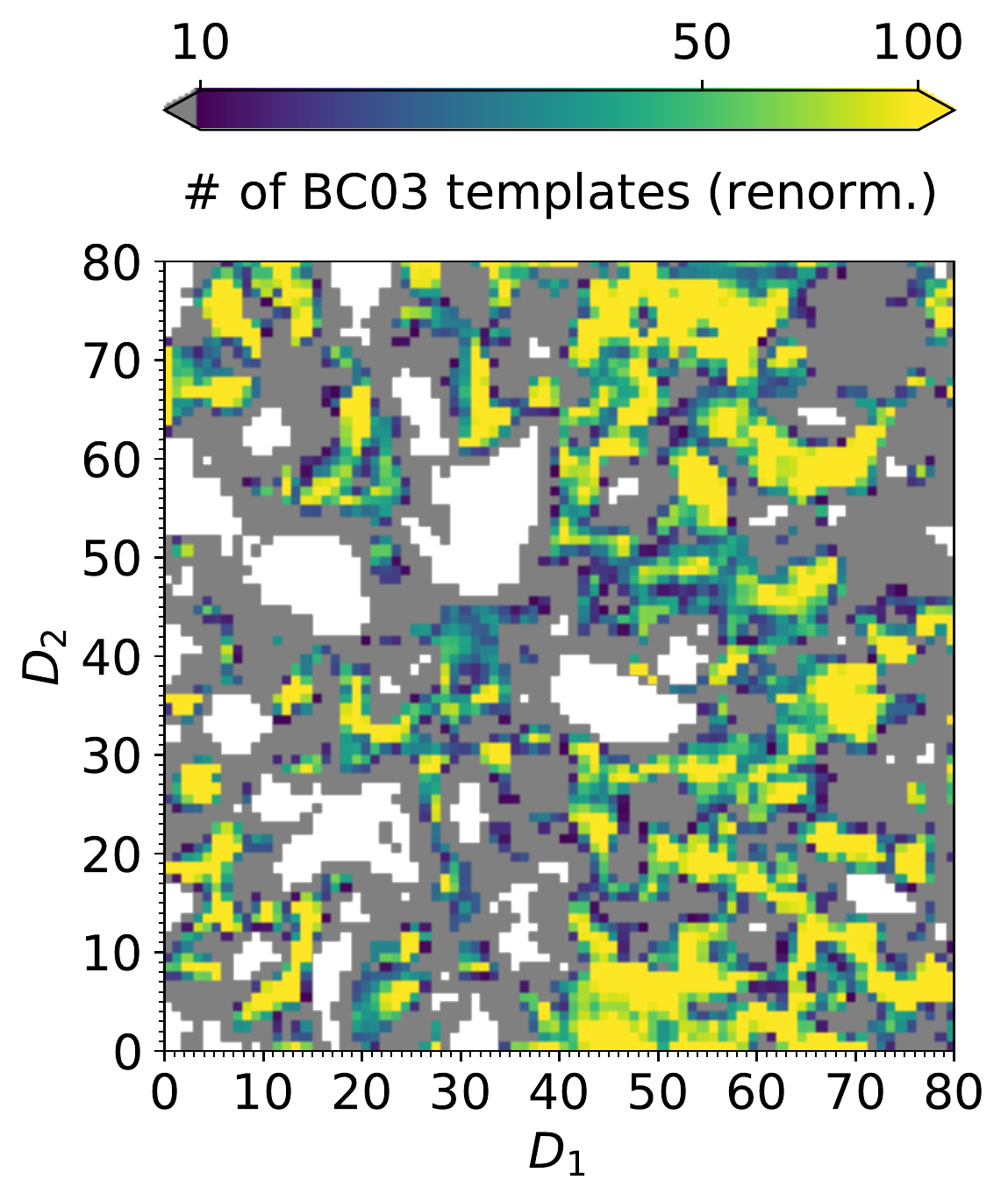}
    \caption{BC03 templates from the \texttt{LePhare} library projected into the SOM. The templates are shifted from $z=0.2$ to 1.8 with steps of $\Delta z=0.01$, and perturbed with Gaussian noise to mimic real photometry.  
    The SOM is color coded to show their cell occupation,  renormalized to the number density of COSMOS2020 galaxies to make this Figure  directly comparable with the left-hand panel of  Fig.~\ref{fig:som_train}.}
    \label{fig:som_bc03}
\end{figure}

To better visualize these concepts, we derive observed-frame colors for the BC03 templates in the  \texttt{LePhare} library, shifting them from $z=0.2$ to 1.8 with a linear increment of $\Delta z=0.01$.  Once projected on the COSMOS2020 SOM, these templates occupy only a fraction of the grid (Fig.~\ref{fig:som_bc03}) even after perturbing their photometry with Gaussian-shaped noise (0.05\,mag standard deviation for every band). Their distribution (number of templates per cell)  is very different compared to the one of COSMOS2020 galaxies (cf.\ Fig.~\ref{fig:som_bc03} and the left-hand panel of Fig.~\ref{fig:som_train}). Such a tension is not surprising because the sample of BC03 colors is built with a flat redshift distribution and no differentiation in the abundance of galaxy types.  This  has  an impact on the whole SED fitting process comparable to the bias from missing templates (empty cells in Fig.~\ref{fig:som_bc03}). Especially in codes like \texttt{LePhare}, which do not take into account prior probabilities, if a ``family'' of $(z,M,\mathrm{SFR})$ models is  over- or under-represented in the library then the  output quantities  will be biased too. 

The  severity of this last issue shall decrease in  future studies owing to the expected improvement of telescope surveys, since a higher photometric quality   shall result in more robust likelihood functions. In such a scenario, the likelihood would drive the SED solution  making the code less sensitive to the models' priors, but if a model were completely  missing from the library (not just over- or under-represented)  then the problem discussed above would persist in spite of the increased quality of input data. 

A treatment of probability distribution functions  including probability priors is presented in \citet{benitez00BPZ} and \citet{tanaka15} for redshift measurement only. The same Bayesian formalism is also the foundation of state-of-the-art SED fitting codes to estimate physical properties.  Notable examples are \texttt{BAGPIPES}\footnote{\citet{carnall18}.} and  \texttt{Prospector}\footnote{\citet{prospector_release17,leja17,johnson21}.}. Another distinctive feature of these  codes  is the capability of using SFHs in flexible bins  of time \citep[often designated as ``non-parametric'', see][]{leja19a}   or to include a large variety of SFHs from cosmological simulations \citep[as in \texttt{BEAGLE}, see][]{chevallard16}. Also, the implementation of parametric SFHs  has reached a level of complexity  higher than the previous generation of software \citep[see][]{carnall19}. This is however achieved  at the expenses of computational speed. In that regard, the advantage of SOM  is well illustrated in \citet{hemmati19}: the  time to process $10^7$ objects is less than 0.3 CPU hours, while a typical Bayesian fitting run \citep[e.g.,][]{mehta21} would take a similar amount of time  for a single object (V.\ Mehta, private communication). Such a trade-off may dramatically change in  the future, as cheaper machines (i.e., the possibility to rent or buy more CPU time) and more efficient Bayesian methods will become available. A promising example in this direction is the SED inference method proposed by \citet{hahn&melchior22} that employs simulation-trained neural networks to accelerate the estimate posterior probability distributions at the pace of 1 second per galaxy (which despite the dramatic improvement would still exceed the 0.3 CPU hours of the SOM to analyze our sample).

In recent work, the SFRs derived from the Bayesian codes mentioned above have been compared with other estimators. In \citet{carnall19}  galaxies in the GAMA survey  \citep[][]{baldry18} are analyzed with \texttt{BAGPIPES} and then compared to SFR measurements based on H$\alpha$ flux, showing a large scatter (see their figure  9) and a mass-dependent offset due to the fact the H$\alpha$ tracer is sensitive to more recent star formation episodes. The test with GAMA galaxies is performed at $z<0.1$, fitting bands up to IRAC channel 4.  These differences  prevent a straightforward comparison with our SFR$_{H\alpha}$ test (Fig.~\ref{fig:sfr_barro_vs_som}, right panel) which, however, shows more consistency even though it has been  performed over a larger $z$ range and without the use of channel 3 and 4.    

\begin{figure}
    \centering
    \includegraphics[width=0.99\columnwidth]{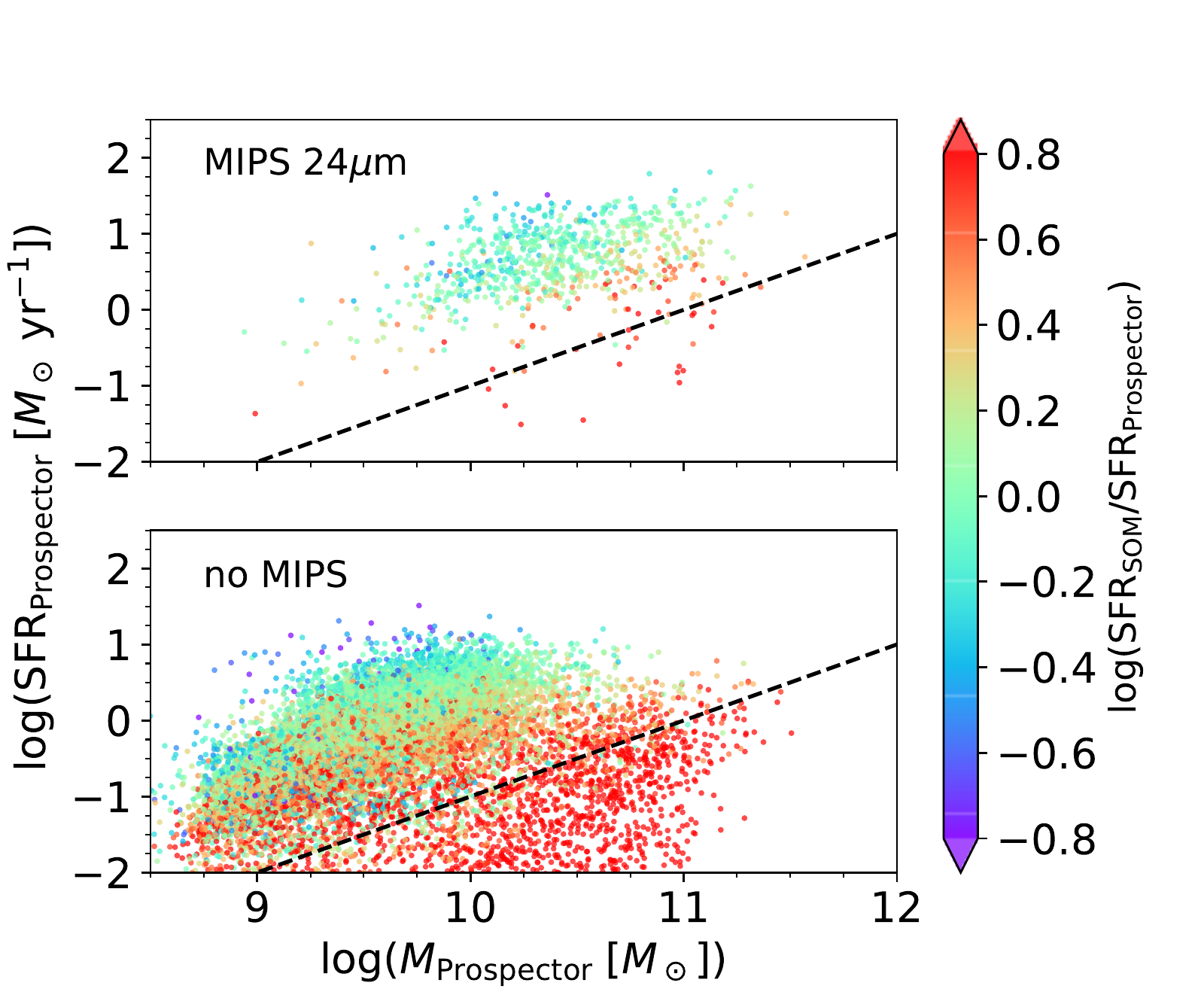}
    \caption{The SFR vs.\ $M$ plane of COSMOS2015 galaxies (colored circles) as it results from the SED fitting performed by the \texttt{Prospector} code \citep{leja21}. Only the objects cross-correlated to COSMOS2020 (0.6\arcsec searching radius) are shown, both those having a MIPS 24$\mu$m counterpart (in the \textit{upper panel}) and objects without FIR detection (\textit{lower panel}). Each data point is color-coded according to the difference in SFR between \texttt{Prospector} and the SOM estimator. A dashed line at constant  $\mathrm{sSFR}=10^{-11}\,\mathrm{yr}^{-1}$ is plotted in each panel to guide the eye. }
    \label{fig:prospector}
\end{figure}

Another recent study \citep{leja21}  thoroughly inspects physical quantities inferred by means of \texttt{Prospector} for 3DHST and COSMOS2015 galaxies up to $z\sim3$. Unlike the present analysis, their input photometry spans from UV to 24\,$\mu$m (when available) and  the emission in those bands is interpreted under the assumption of energy balance \citep[similarly to][]{dacunha08}. With this caveat in mind, we can  compare to \citet{leja21}  by examining the COSMOS galaxies \citeauthor{leja21} have in common with our study. Among the ones with a 24\,$\mu$m detection ($S/N>5$) we select 790  targets that have been  kept out-of-bag during the validation test in Sect.~\ref{subsec:meth_sxdf}. Another 16\,729 galaxies in our catalog  have no 24\,$\mu$m counterpart (or $<$5$\sigma$) but also match with a source in \citet{leja21}.  
For the former subsample, we find no significant systematics in $\log(\mathrm{SFR_{SOM}/SFR_{Prospector}})$ and a standard deviation of 0.3\,dex that is comparable with the typical SED fitting statistical uncertainties (Fig.~\ref{fig:prospector}, upper panel). With respect to the objects that are not detected in MIPS, we are able to confirm the findings in \citet{leja21}: when  $\mathrm{sSFR_{Prospector}}\gtrsim  10^{10}\,\mathrm{yr}^{-1}$ there is a good agreement between the two techniques, whereas below the MS the SFR$_\mathrm{SOM}$ values are systematically larger (Fig.~\ref{fig:prospector}, lower panel). The average discrepancy increases from a factor 2$\times$ to more than 10$\times$, as moving towards lower levels of specific star formation. Such a trend is coherent with \texttt{Prospector} non-parametric SFHs since their impact, compared to analytical descriptions of SFH,  is more pronounced for old galaxies that have exited the MS. 

Another reason behind the differences between the two methods is that galaxy models in  \texttt{Prospector} are the combination of various stellar populations. Such a flexibility corresponds to a variable scaling factor in the SFR-$L_\mathrm{IR}$ relationship, which can be adjusted on an object-by-object basis instead of the  constant $\mathcal{K}$ factor used to calibrate the SOM (Eq.~\ref{eq:SFR_UVIR}). In fact, the latter comes from studies based on simpler stellar population synthesis models. 
The discrepancy, however, could be removed by construction using \texttt{Prospector}  to label the SOM instead of the procedure summarized in Sect.~\ref{subsec:data_phys}. This option is further discussed in Sect.~\ref{subsec:discuss_othersoms}.

\subsection{Caveats in the SOM construction}
\label{subsec:discuss_ms}

The most entrenched limitations in the SOM method are the ones inherited from training and calibration samples.  In Sect.~\ref{sec:results} we observed that many galaxies  do not get a SFR$_\mathrm{SOM}$ assigned. The lack of estimates is due to the 24\,$\mu$m  flux limit, which  prevents the measurement of most of the low-SFR galaxies with $M<10^{10}\,M_\odot$, leaving several SOM cells without a SFR$_\mathrm{label}$.  As a consequence, the SFR  distribution in bins below $10^{10}\,M_\odot$ is skewed towards higher SFR values, and the observed MS departs from linearity (Fig.~\ref{fig:ms_compar}). This selection bias would affect any MS measurement derived from the COSMOS MIPS survey, irrespective of the method; with the SOM, the way data are visualized makes easier to identify this kind of issues. 
Incompleteness in the training sample is more difficult to quantify because one should characterize the parent population from which the sample is drawn. It is therefore  challenging to establish, before starting the SOM construction, whether a certain galaxy type has not been included in the training. Template fitting does not have the same problem, since that particular galaxy type can always be incorporated in the template library; the concern in that case is how accurately that galaxy is modeled. 

Another bias, more specific to the SOM method itself, is caused by the mixture of different galaxy types in the same cell. The main reason for such a contamination is the similarity of diverse SEDs after photometric errors are taken into account. This has been already shown in \citet[][see their figures 3 and 4]{speagle19}, although for a 5-band data set which is more susceptible to SED degeneracy.  In an ideal, noiseless universe, the SOM classification is extremely accurate, as shown in \citet{davidzon19} using a mock galaxy catalog from hydrodynamical simulations. Hence the necessity of using deep, state-of-the-art data to minimize observational errors and consequently the scatter introduced  in the SOM. The cells with a quiescent fraction $0.25<f_Q<0.75$  (5\% of the entire grid, see Fig.~\ref{fig:som_calib}) are another example of the SED  mixing, indicating that some galaxies with different $NUV-r$, $r-K_\mathrm{s}$ colors are located together inside the same cell. This bias may affect more seriously those calibrations relying on smaller samples, like SFR in the present study. In cells occupied only by a handful of MIPS galaxies, even a single interloper may have a strong impact on the resulting SFR$_\mathrm{label}$. One way to find them is to look at some property that is expected to be similar for all galaxies in the cell (e.g., their sSFR). If the inspected object is a (e.g., 2$\sigma$)  outlier of the distribution, then it can be flagged as a potential interloper. The suggested approach is only one of the possible solutions, and does not work e.g.\ in cells containing only one or two MIPS galaxies. Nonetheless, we test it by re-calibrating the SOM using sigma-clipped SFR$_\mathrm{UVIR}$ distributions. This effectively removes 2$\sigma$ outliers in the cells where sigma clipping is feasible. Since the general results of the analysis do not change, we conclude that contamination bias does not have a strong impact overall.   

Instead of relying on galaxy physical properties, like in the example above, one can find catastrophic errors in the SOM classification by inspecting galaxies with suspiciously  large  $\Delta$ or  $\chi^2$ distance (Eq.~\ref{eq:d_som} and \ref{eq:chi2_som}, respectively) as done in Sect.~\ref{subsec:results_ms}. Another way to mark  unreliable SFR$_\mathrm{SOM}$ estimates is according to the number of cells contributing to Eq.~(\ref{eq:sfr_som}): for some galaxies not all the $N$ neighbors may be labeled, potentially introducing a bias. Such a bias, if present, must be of second order because  removing  objects for which only one or two of the five neighbor cells have SFR$_\mathrm{label}$ defined, the MS locus does not change; the only detectable change is a  reduction in the error bars of Fig.~\ref{fig:ms_compar}.

\begin{figure}
    \centering
    \includegraphics[width=0.99\columnwidth]{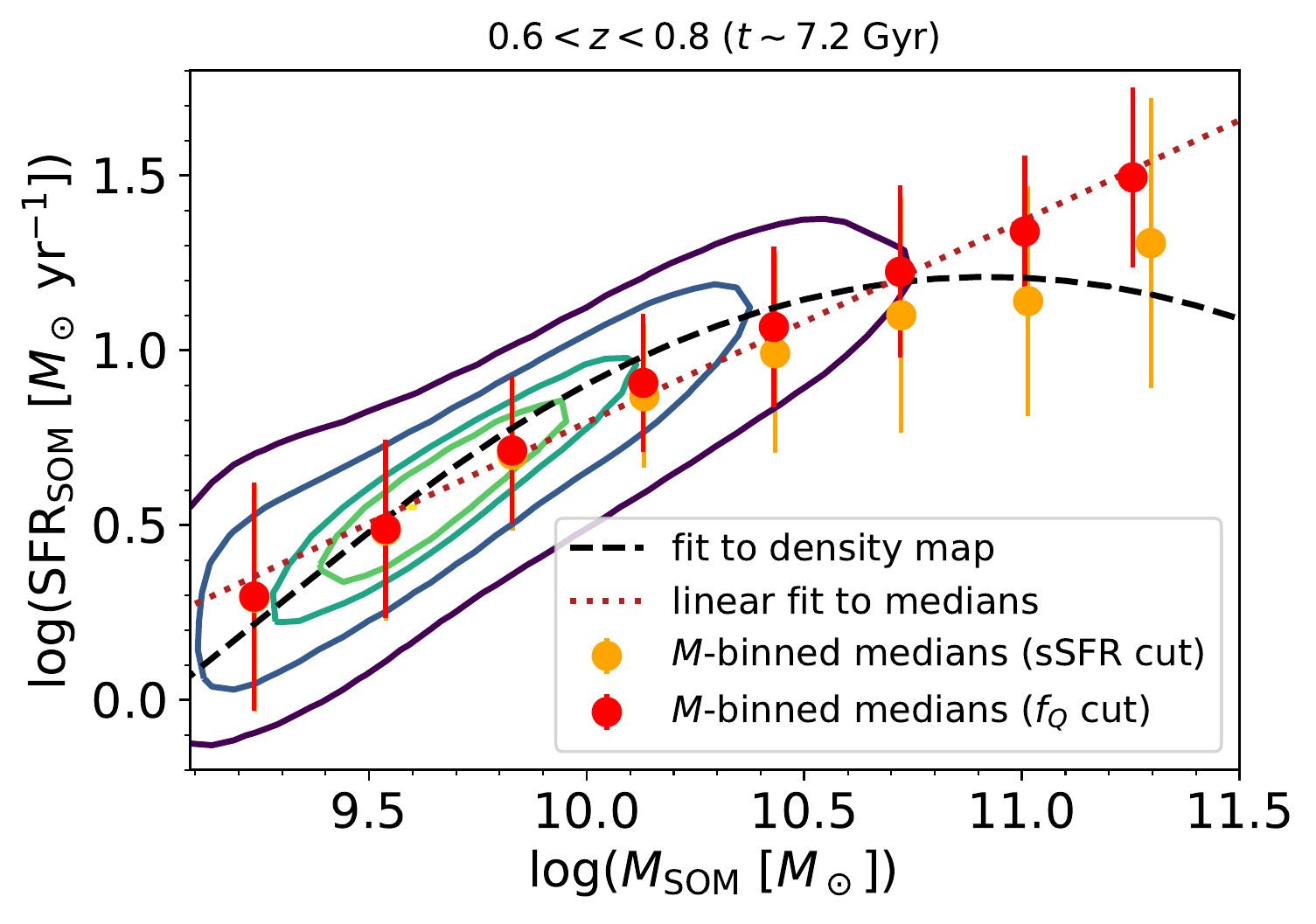}
    \caption{Solid lines show the contour density map for the distribution of SXDF2018 galaxies in the SFR-$M$ diagram; this is fit by Eq.~\ref{eq:schreiber} as described in Sect.~\ref{subsec:results_ms} (dashed line). Filled circles are SFR median in running bins of galaxies with SFR$_\mathrm{SOM}/M_\mathrm{SOM}>10^{-11}$\,yr$^{-1}$ (orange symbols) and $f_Q<0.5$ (red). In both cases, error bars are calculated with the 16th-84th percentile difference. The red circles are interpolated by a linear function (dotted line). }
    \label{fig:ms_z1}
\end{figure}

Certain sub-samples of galaxies are, however, affected by an ambiguous SOM classification  besides the obvious interloper cases. This is the case of the trend at the high-mass end of the MS, already highlighted in  Sect.~\ref{subsec:results_ms}.   Since there is no clear-cut separation between the  star-forming and quiescent population in the SOM, we impose a threshold ($f_Q<0.5$) that is somewhat arbitrary. Figure~\ref{fig:ms_z1} illustrates the issue at  $0.6<z<0.8$: massive galaxies with intermediate to low star formation, responsible for the MS turnover at $\sim5\times10^{10}\,M_\odot$, are excluded by the $f_Q<0.5$ criterion. A less conservative constraint ($\mathrm{sSFR}>10^{-11}\,\mathrm{yr}^{-1}$) does not remove those galaxies and would  
better identify the MS turnover, becoming  more in agreement with the  MS determined by the ``selection-free''  definition of  \citet[][see Sect.~\ref{subsec:results_ms}]{renzini&peng15}. 
The same is observed in the other redshift bins.     Figure~\ref{fig:ms_z1} also shows the general systematics induced by using  the median SFR in bins of mass: they locate the center of the MS slightly below the locus traced by the highest density in galaxy number density.

\subsection{Calibration with other tracers}
\label{subsec:discuss_othersoms}

The  estimates used as reference in Sect.~\ref{subsec:data_phys} are not the only option for calibrating our method. Relying on the same optical-to-FIR data, one can use alternate fitting codes (e.g., \texttt{Prospector}) instead of \texttt{LePhare}. Besides that, owing to the wealth of ancillary data in COSMOS, other proxies for stellar mass and star formation can provide  $M_\mathrm{label}$ and SFR$_\mathrm{label}$. This is particularly evident for the latter, which can be derived e.g.\ from H$\alpha$ nebular emission  or by the radio continuum. Other possibilities such as 21\,cm atomic hydrogen or sub-mm molecular gas transitions are not included in the present discussion for sake of simplicity, and we address the reader to  \citet[][]{kennicutt&evans12} for an exhaustive bibliography on the various star formation tracers.  Our goal in the present Section is not repeating the analysis with another sets of $M_\mathrm{SOM}$, SFR$_\mathrm{SOM}$ estimates but showing the capacity of different calibration samples to fill the SOM grid. 

A technique to derive SFR$_\mathrm{H\alpha}$ has been presented in Sect.~\ref{subsec:results_sfr}. Radio continuum emission also correlates  with galaxy star formation \citep{condon92} but we will not calculate SFR$_\mathrm{radio}$ here since we aim at discussing the availability of data rather than the final estimates. The latest, publicly available\footnote{\url{https://irsa.ipac.caltech.edu/data/COSMOS/tables/vla/}.} radio data for almost the entire COSMOS field come from interferometry in the 3\,GHz band with the ESO Very Large Array \citep{smolcic17b}. Instead  of extracting individual sources, as done with both MIPS and spectroscopic tracers, one could stack 3\,GHz map cutouts centered at the position of SOM galaxies belonging to the same cell. This is the procedure adopted by  \citet{leslie20}, which is similar to the  1.4\,GHz stacking described in \citet{karim11}. 

\begin{figure}
    \centering
    \includegraphics[width=0.99\columnwidth]{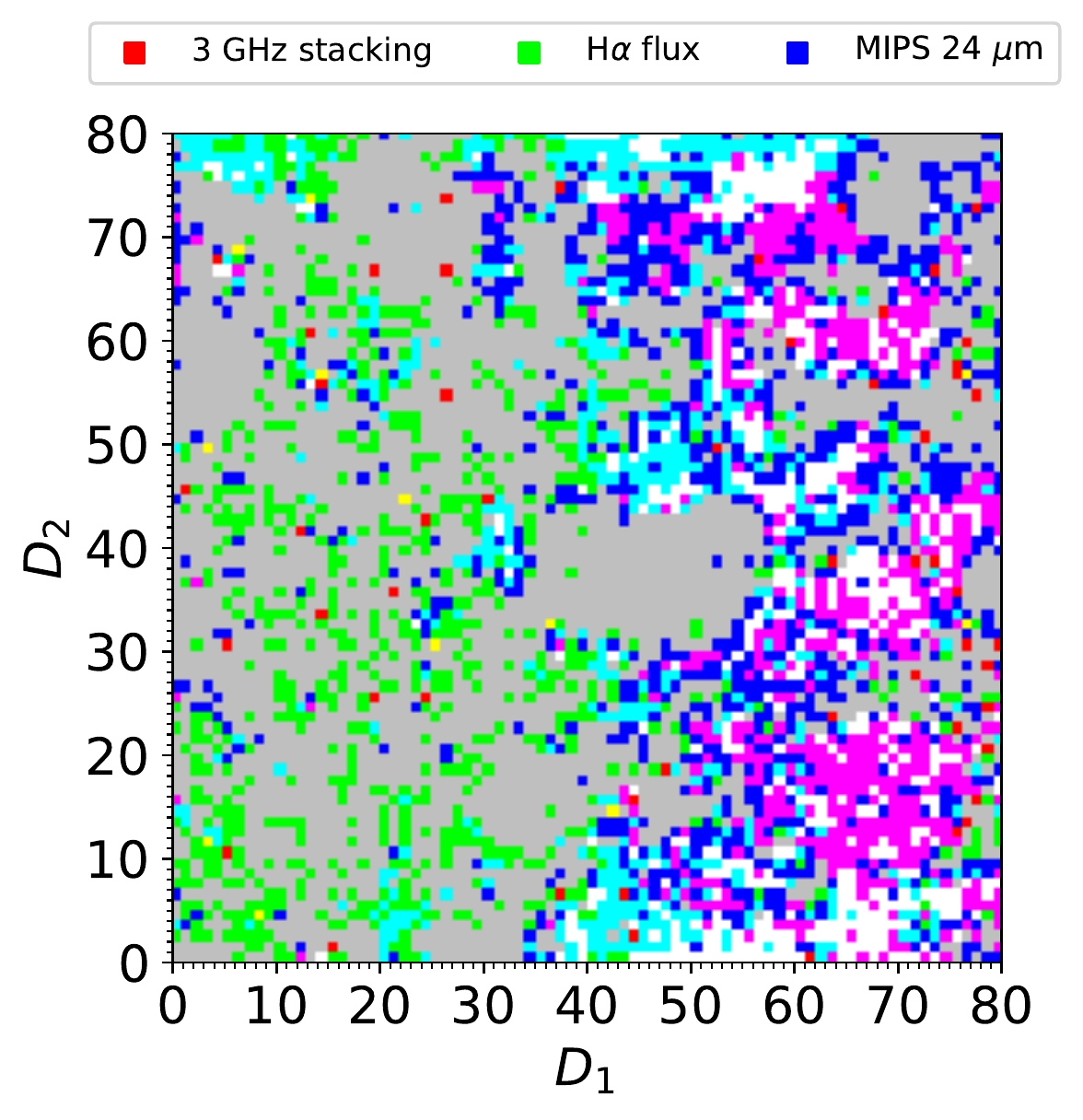}
    \caption{Different calibration samples that may be used to label SFR on the SOM. Cells containing galaxies from the  original MIPS sample (cf.\ Fig.~\ref{fig:som_calib}) are colored in blue; cells containing spectroscopic galaxies with an H$\alpha$ measurement and radio galaxies with a 3\,GHz stacking $S/N>5$ are colored in green and red  respectively. Colors combine by following the standard RGB rules, e.g.\ a cell that can be calibrated by both MIPS and H$\alpha$ (MIPS and radio) is colored in cyan (magenta). A handful of yellow pixels are cells that can be calibrated by only radio stacking and H$\alpha$, while white pixels are cells in which all the three star formation tracers are available. As the discussion in the main text is only about  coverage and  not the  actual SFR$_\mathrm{label}$ values, these are not shown in  the present figure. 
    }
    \label{fig:som_multicoverage}
\end{figure}

Figure~\ref{fig:som_multicoverage} shows the coverage of  potential calibrations with either H$\alpha$ or radio stacking, along with the actual FIR-based calibration already presented in Fig.~\ref{fig:som_calib} (bottom-left panel). We highlight only the cells that would benefit from a high-confidence  calibration, i.e.\ those containing at least one galaxy with H$\alpha$ flux $>2\times10^{-17}$\,erg\,s$^{-1}$\,cm$^{-2}$, or with $S/N>3$ in the 3\,GHz median stack. The figure of merit for radio stacking is done by following the recipe  (size of the cutouts, calculation of median flux, etc.) in \citet{leslie20}. 

The most evident feature of Fig.~\ref{fig:som_multicoverage} is the additional coverage provided by spectroscopic data, which can label 735 cells missed by the 24\,$\mu$m calibration. Most of them are in the area populated by low-mass galaxies because of the surveys' selection function targeting many  emitters  with $M_\mathrm{LePh}\simeq10^{9}\,M_\odot$. This shows how spectroscopic line detection can be complementary to FIR imaging.  The figure of merit is expected to improve after next-generation spectrographs will start operations. For example, H$\alpha$ is inside the wavelength range of the Multi-Object Optical and Near-infrared Spectrograph
\citep[MOONS,][]{cirasuolo14,taylor18} in low resolution mode from $z\sim0$ to 1.7,  with H$\beta$ (pivotal for dust correction) entering at $z\sim0.3$. The sensitivity and multiplex capability of the instrument\footnote{See specifications at  \url{https://vltmoons.org/}.} shall enable a more exhaustive labeling of the SOM grid \citep[see discussion in][]{davidzon19}. Also, the grism spectrograph on board of \textit{Euclid}\footnote{\url{https://www.euclid-ec.org/}} could be instrumental not only to fill the empty cells but also to dramatically increase the statistics: between 32\,000 and 48\,000 galaxies per square degree are expected to be observed at $0.40 < z < 1.8$   with an H$\alpha$ flux $>5\times10^{-17}$\,erg\,s$^{-1}$\,cm$^{-2}$  \citep[][]{pozzetti16}.  

The third tracer, based on 3\,GHz stacking,  provides SFR$_\mathrm{label}$  for 510 cells, corresponds to $\sim$20\% of the area actually  sampled by previous calibrations. Very few cells  (red pixels in Fig.~\ref{fig:som_multicoverage}) are newly labeled by the radio sample, partly due to the conservative $S/N>5$ cut we imposed: if instead we also allow stacked images with $3<S/N<5$ then the number of SOM cell probed by radio data nearly double. However, even in that case, the expansion with respect to the original MIPS calibration would be limited (only 65 additional SFR$_\mathrm{label}$) due to the FIR-radio correlation and its strong dependency on stellar mass \citep{delvecchio21}. Despite such a marginal expansion, it is still useful to have various tracers superimposed on the SOM for a better understanding of  the galaxy parameter space. For example, a difference between multiple  SFR$_\mathrm{label}$ values in the same cell may  help identify galaxy classes with rapidly varying SFH, leveraging the fact that those star formation tracers probe different time scales  \citep[which can be taken into account as in][]{sparre15}. 

In \citet{leslie20} the stacking technique is instrumental to probe the SFR-$M$ plane beyond the boundaries of the original \citet{smolcic17b}  sample, which is made by individual VLA sources  whose counterparts are mostly brighter than 22.5\,mag in IRAC channels 1 and 2. Moreover, the choice of adaptive binning allows the authors to always gather several hundreds of objects. In the finer grid of the SOM, galaxies that in \citet{leslie20} belonged to a single $(z,M)$ bin are now spread over several cells, ``diluting'' the stacking signal. A route to solve this problem goes through the use of all SOM cells, each one with an occupation probability weight \citep[as in][]{speagle19} instead of assigning the target galaxy to  its  best-fit (but low-$S/N$) cell only. 

As an aside, we notice that a stacking strategy would be more beneficial for the SFR$_\mathrm{UVIR}$ calibration. The MIPS sample used in Sect.~\ref{subsec:meth_calib} was pre-selected at $S/N>5$ in the 24\,$\mu$m band. A preliminary inspection of sources below that threshold suggests that the number of labeled cells may increase by $\sim30$\% (compared to Fig.~\ref{fig:som_calib}) by using median SFRs from 24\,$\mu$m stacking  \citep[calculated as in][]{magdis10}.

\subsection{Soundness checks using SOM}
\label{subsec:discuss_soundness}

\begin{figure}
     \centering
     \includegraphics[width=0.99\columnwidth]{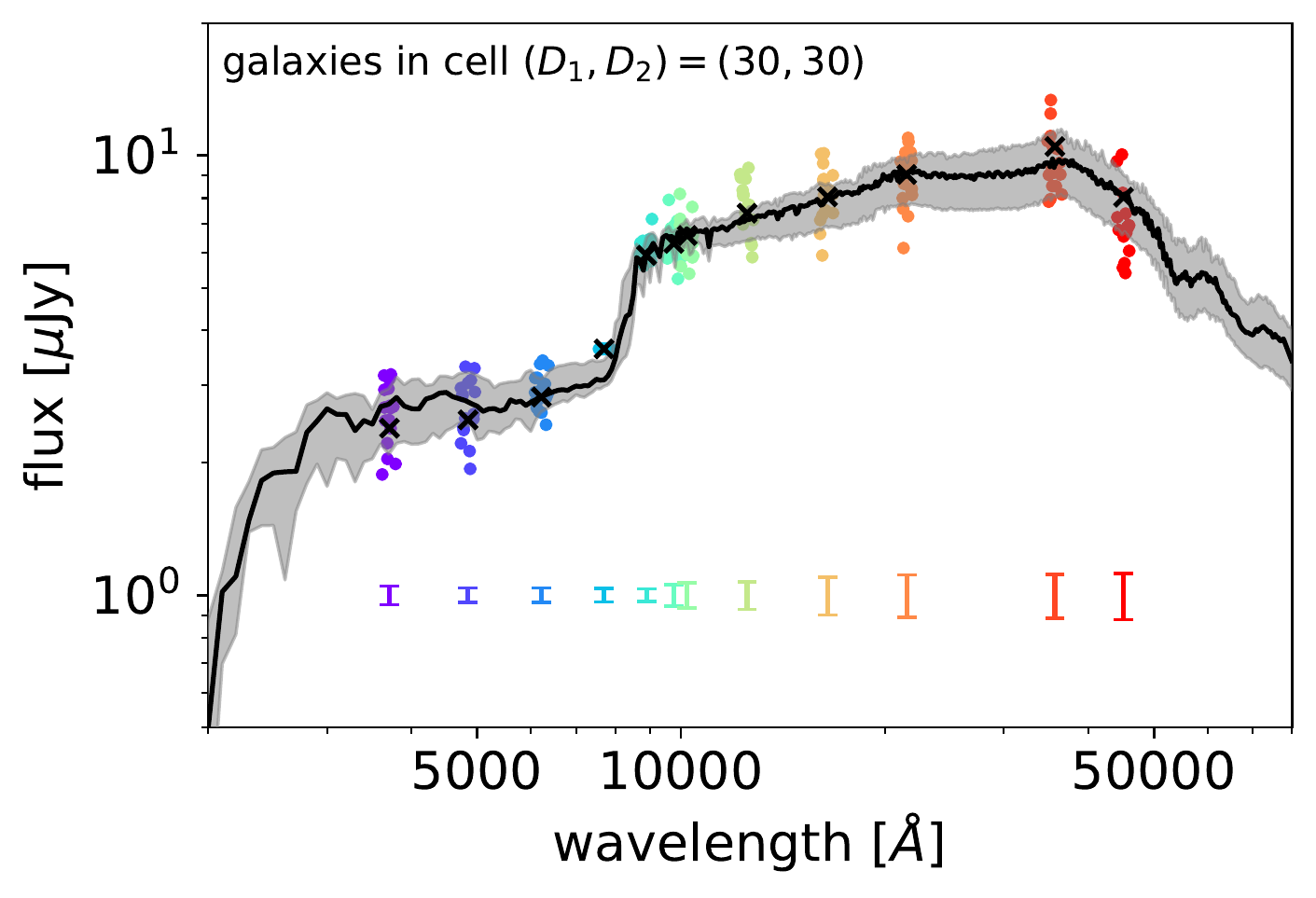}\\
     \includegraphics[width=0.99\columnwidth]{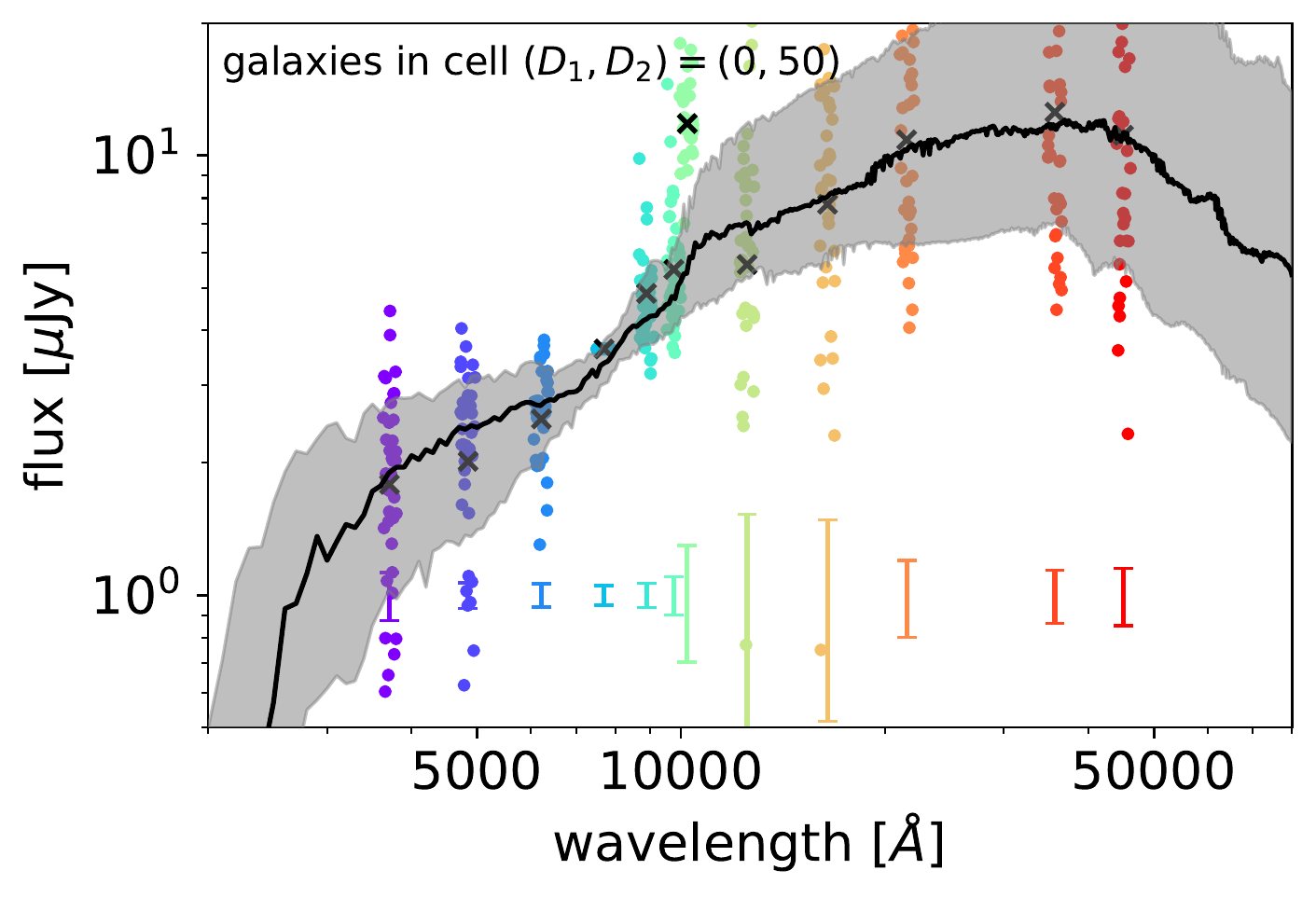}\\
     \includegraphics[width=0.99\columnwidth]{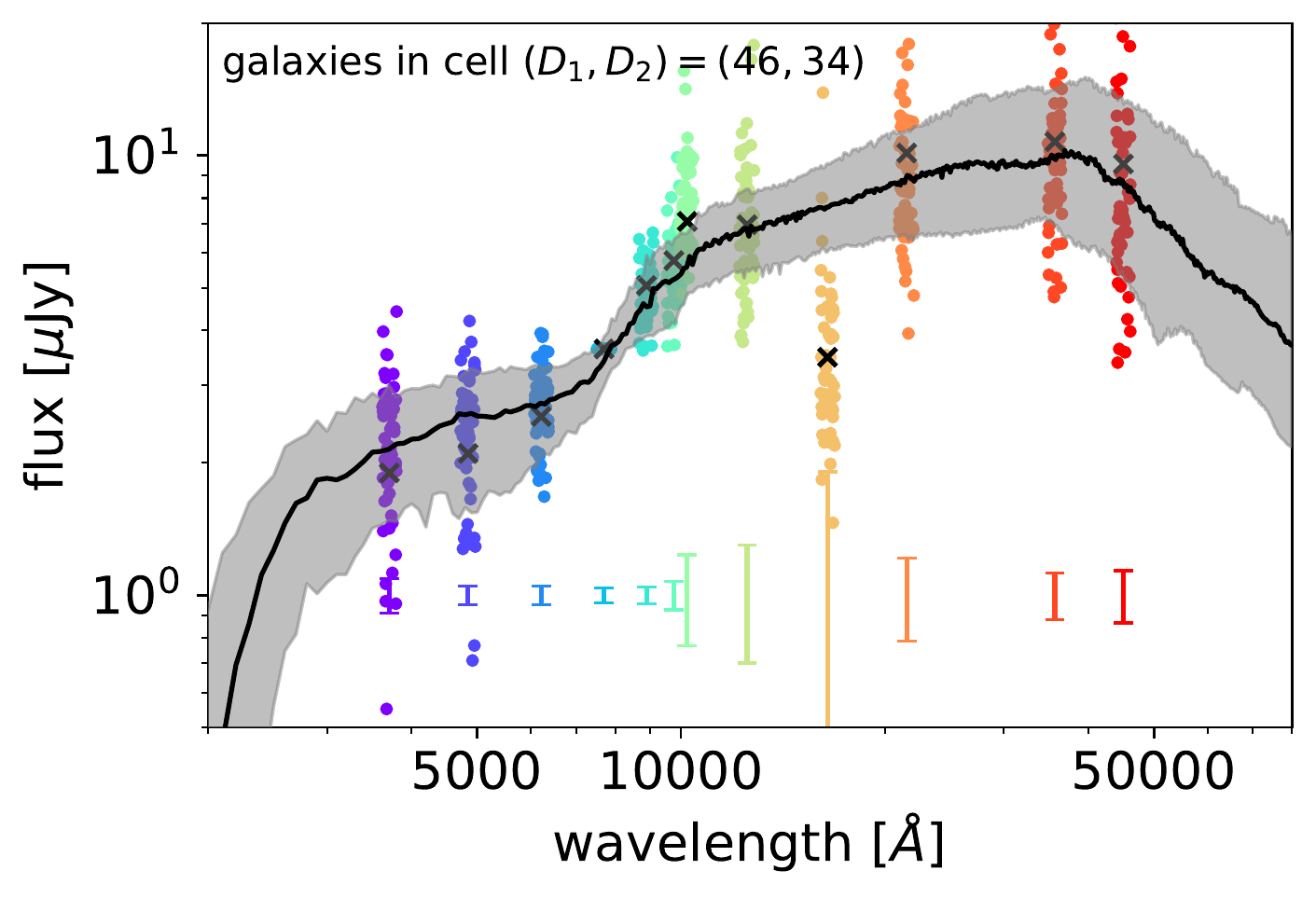}
     \caption{Inspection of three SOM cells with coordinates $(D_1,D_2)=[(30,30),(0,50),(46,34)]$ on the $80\times80$ gird. The first cell (\textit{upper panel}) has been randomly selected among those with a small galaxy-weight distance $\Delta$ (cf.\ Fig.~\ref{fig:som_train}) while the \textit{middle} and \textit{lower panels} display two cells located in  areas with large $\Delta$ values, to provide an insight on SOM shortcomings (see Sect.~\ref{subsec:discuss_soundness}). The number of galaxies in these cells is respectively 16, 39, 56.  
     In each panel, the solid line is the median of the BC03 spectral models  providing the best fit to each COSMOS2020 galaxy in the  \texttt{LePhare} analysis; before calculating the median, all the BC03 spectra (which include only the stellar component, not the nebular emission) have been rescaled to match 22.5\,mag in the $i$ band. Such a  rescaling also makes easier to compare spectral shapes from the different panels. The gray shaded area encloses the 16th and 84th percentiles of their distribution. The $y$-axis range in the upper panel is shorter to zoom in the tight 16th-84th dispersion, which would be barely visible otherwise.     
     Filled circles represent the observed photometry, color-coded for the different bands (see legend); for sake of clarity they have been horizontally shifted from the central wavelength of their band by a random offset within $\pm$3\%. In the bottom part of each panel we report the average error bars in the same color of their corresponding photometric band. Black crosses are the fluxes that would result from the cell's weight after training. 
     All symbols are renormalized to the  reference $i=22.5$\,mag. 
     }
     \label{fig:badcell}
 \end{figure}

None of the three tracers we considered are able to fill the central area of the SOM (around coordinates $D_1,D_2=50,40$) where the average intra-cell distance is particularly large  (Fig.~\ref{fig:som_train}, middle panel). This is another use of the SOM as a diagnostic for peculiar SEDs \citep[see also][where they apply dimensionality reduction to isolate catastrophic redshift errors]{hovis21}. 
The bottom panel of  Fig.~\ref{fig:badcell} illustrates this point through the  galaxy sample inside one of those problematic cells (the one with coordinates $D_1,D_2=46,34$). The sample includes 56 objects at $z\sim1.2$ with similar colors. When the SEDs are superimposed to their  \texttt{LePhare} best-fit templates, we observe that the UltraVISTA $H$ band is systematically below the flux predicted by \texttt{LePhare} (see lower panel of  Fig.~\ref{fig:badcell}). The corresponding photometric uncertainty is also significantly larger than the other UltraVISTA bands, always  resulting in a $S/N$ smaller than 2. Interestingly, Chartab et al.~(in prep.) also reach the same conclusion by analyzing $H$-band photometry with a different ML technique, where these objects escaped standard reliability checks in \citet{weaver22}. 

Another problematic region according to Fig.~\ref{fig:som_train} is the vertical stripe of cells with $D_1=0$ and $48<D_2<56$, from which we randomly  select  $(D_1,D_2)=(0,50)$ for inspection (middle panel of Fig.~\ref{fig:badcell}).  Galaxy SEDs inside that cell have a much larger dispersion than  non-pathological cells (as the one displayed in the upper panel of Fig.~\ref{fig:badcell} for comparison). Despite the scatter, these objects share a common characteristic, i.e.\ an extremely red  $y_\mathrm{HSC}-Y_\mathrm{VISTA}$ color\footnote{Note that central wavelength and shape of these two $Y$ bands are different  \citep[see figure 2 in][]{weaver22}.} and a blue $Y_\mathrm{VISTA}-J_\mathrm{VISTA}$. That color excess can be attributed to nebular emission line contamination in the  $Y_\mathrm{VISTA}$ band. Indeed, galaxies in $(D_1,D_2)=(0,50)$ are concentrated around two redshifts: $z\sim1.15$ and $z\sim1.7$, which approximately correspond to a $Y$-band flux enhancement due to H$\beta$ ($\lambda4863$) and O{\sc [II]} ($\lambda\lambda 3727,3729$) respectively. Hence, the SOM diagnostic power can serve at the same time to identify peculiar galaxy types like strong line emitters, but also to reveal technical issues such as photometric redshift degeneracy.

\section{Summary  and conclusions}
\label{sec:conclusions}

 Compared to the body of work focused on ML methods to measure galaxy redshifts, little progress has been made with respect to other physical parameters, which are equally important in galaxy evolution studies. 
 Here we continued the investigations presented in \citet{masters15}, \citet{hemmati19}, and \citet{davidzon19} to the point where we are now able to exploit an unsupervised ML algorithm such as the SOM to simultaneously derive galaxy redshift, stellar mass, and SFR. This particular use of the  SOM was already described in \citet{davidzon19}, but so far has been applied only on mock galaxy catalogs that, despite an advanced level of realism, cannot replace the complexity of a test with actual observations. 
  
 To this purpose, in the present work, we used the new COSMOS2020 galaxy catalog \citep{weaver22} up to $z=1.8$,  together with the rich archive of complementary data available for the COSMOS field. Not only the quality, but also the large size of COSMOS observations make them an ideal training sample for the SOM.
 In our method, after the SOM dimensionality reduction created a grid of 80$\times$80 cells, we labeled as many cells as possible with reference values of $z$, $M$, and SFR. These values are derived by means of ancillary data, in particular MIPS 24\,$\mu$m from the S-COSMOS survey to constrain galaxy star formation.   
 Once the SOM was calibrated, we projected other galaxies onto the grid, obtaining an estimate of their $z_\mathrm{SOM}$, $M_\mathrm{SOM}$, and SFR$_\mathrm{SOM}$ depending on the cell to which each target has been associated. We used SXDF galaxies, but any other galaxy sample can be measured as long as the targets have the needed features (in this case, broad band colors) to be projected into the COSMOS-calibrated  SOM. In our case, we also assumed photometric redshift are known in advance to preselect galaxies below $z<1.8$. This requirement can be disregarded, as the SOM proved to be efficient for redshift measurement too \citep{masters15}. However, the other techniques also fix the redshift of their targets before measuring the other physical parameters, therefore we did the same here for sake of comparison. 
 
 The SOM is made by a fixed number of cells, each one representing a specific combination of broad-band colors (i.e., a galaxy SED prototype). The galaxy properties in output are not strongly discretized because they result from averaging several cells  (Eq.~\ref{eq:m_som}-\ref{eq:sfr_som}). Nevertheless, the SOM  resolving power is lower than other ML methods \citep[in particular deep neural networks, see][]{simet21} that describe physical parameters with continuous functions. The SOM ``gridding effects'', however, are  counter-balanced by the advantage of building such a grid in an unsupervised fashion, and the flexibility of assigning different labels to it \textit{a posteriori}. As an aside, we note that other methods such as variational autoencoders \citep[see][for an application in the astronomical domain]{cheng20} may represent a synthesis between the two approaches by offering both an unsupervised (or semi-supervised) training and a regression analysis with continuous functions. 
 
We assessed the reliability of our procedure both by comparing to independent estimates and by means of a bootstrap technique (i.e., out-of-bag targets from our COSMOS sample). Most notably the galaxy SFR, generally difficult to recover from optical-NIR fitting, can be measured with good precision ($<$0.3\,dex scatter, see Fig.~\ref{fig:validation_oob} and \ref{fig:sfr_barro_vs_som}) and is consistent  with SFR indicators based on FIR \citep[only 0.05\,dex offset from][]{barro19} even though in our SOM method the input photometry stops at IRAC channel 2.  We also found a good agreement between the SOM main sequence of star formation and other MS studies from the literature, an indication that the $(z_\mathrm{SOM}, M_\mathrm{SOM}, \mathrm{SFR}_\mathrm{SOM})$ estimates provided by the SOM are \textit{simultaneously} accurate for most of the targeted objects.  

The computation of physical parameters for 208\,404  SXDF galaxies took about 0.1 CPU hours. In addition to the computational speed our method offers, other advantages shall be emphasized:

 \begin{itemize}
 \setlength\itemsep{0.5em}
 
     \item The quality of SFR$_\mathrm{SOM}$ estimates is better than the results obtained with \texttt{LePhare}, when  compared to SFR values  independently measured from UV-to-FIR photometry   \citep{barro19}. Concerning stellar mass estimates, we found that $M_\mathrm{SOM}$ and $M_\mathrm{LePh}$ are in overall good agreement;  small discrepancies between the two could not be solved  because of the lack of a third set  of  $M$ estimates to be used as   ``ground  truth''.  Without advocating that \texttt{LePhare} (or template  fitting in general) should be superseded by the novel method, our study demonstrates the benefits of a dual approach with ML and standard template fitting working in a complementary fashion. Indeed, as \texttt{LePhare} was instrumental in preliminary tests on the SOM \citep[here, but also e.g.\ in][]{hemmati19b} the latter can help to better understand caveats and limitations in template fitting (Sect.~\ref{subsec:discuss_sfr}). The key factor making our method competitive with respect to other state-of-the-art fitting codes is that it does not rely on the traditional library of synthetic templates. Such a library can be biased by the  theoretical modeling assumptions (stellar population synthesis, dust prescriptions, etc.) whereas this new fitting procedure is more data-driven, based on a set of galaxy prototypes coming from the SOM classification of observed SEDs (i.e., the cells and their weights).   
    
     \item The robust performance of SOM does not imply that its outcome is bias-free, e.g.\ systematics in the training sample may propagate to the target galaxy estimates, and also the SOM labels have underlying assumptions (see e.g.\ Eq.~\ref{eq:SFR_UVIR} for producing  SFR$_\mathrm{label}$). However the 2-D rendering of the SOM put us in a convenient position to solve the problem, as it allows an effective visual inspection of the parameter space. Sect.~\ref{subsec:discuss_soundness}  showed an example of such application, where we flag suspicious cells according to the average galaxy-weight distance inside them. We were able to identify not only issues due to corrupted photometry, but also galaxy types whose SED is intrinsically peculiar (e.g., because of nebular emission contamination). 
     
     \item Moreover, a deeper insight into the galaxy parameter space may result from the combination of multiple calibration samples. In fact, the labels used in the present analysis are not the only option in COSMOS. In Sect.~\ref{subsec:discuss_othersoms} we discussed two alternatives to SFR$_\mathrm{UVIR}$, since the star formation labels can also be derived from either direct  spectroscopic measurements (SFR$_\mathrm{H\alpha}$) or VLA 3\,GHz image stacking  (SFR$_\mathrm{radio}$). Having more than one tracer helps, at least to some extent, to calibrate a larger fraction of the SOM grid, but the most important advantage is the overlap of the different tracers: in a cell with multiple layers of physical information, 
     any similarity (or tension) between them is informative about the galaxy population inside that cell. 
     
 \end{itemize}

While discussing multiple calibration samples, we also mentioned the role that next-generation spectrographs like MOONS or the Prime Focus Spectrograph \citep[PFS,][]{pfs_pasj} will play in the near future. With those facilities, an ambitious programme such as the Complete Calibration of the Color-Redshift Relation \citep[C3R2,][]{masters17} may extend its scope to include the calibration of stellar mass and SFR over the entire area of the SOM grid. Nonetheless,  24\,$\mu$m calibration shall remain a compelling option as the Mid-IR Instrument (MIRI\footnote{MIRI specifications are from the official documentation at  \url{https://jwst-docs.stsci.edu/}.}) on board of the \textit{James Webb} Space Telescope will supersede MIPS images, also offering a choice among more broad bands (four of them between 15 and 25\,$\mu$m). The field of view of MIRI is only $74\arcsec \times 113\arcsec$, but in spite  of  that it should be possible to assemble a calibration sample of comparable size to the COSMOS one. In fact, owing  to the small cosmic variance of independent line of sights  \citep{moster11}, a hundred of well-separated pointings should be sufficient \citep[see][for a similar strategy with HST]{trenti11}. 

Irrespective of the preferred calibration strategy, the present work is intended as a blueprint  to develop ML-based software for upcoming large-area photometric surveys. Scanning a few thousands  square degrees (as planned by the \textit{Euclid} mission) or even the entire sky (like the \textit{Rubin} Observatory), those surveys will decrease dramatically the (already small) ratio  between well-characterized spectroscopic galaxies and objects only described by their broad-band colors. In the SOM, the former ones would only be needed for a training sample, proportionally much smaller than the number of targets.    Therefore, with  modest modifications, our method can be implemented e.g.\ in the \textit{Euclid}  pipeline, deriving precise physical properties for billions of  galaxies (Euclid Collaboration, in prep.). 
The fact that the method is computationally fast makes it particularly suited also for time-sensitive surveys, in  particular the Legacy Survey of Space and Time \citep[LSST,][]{lsst_overview}. 

SOM analysis of \texttt{LePhare} parameter space     (Fig.~\ref{fig:som_bc03}) also points towards another direction,  i.e.\ the development of ``hybrid  software'' that equips 
template fitting with manifold  learning tools in order to build and explore the template library more efficiently  
 \citep[see][]{speagle&eisenstein17,gilda21}. Moreover, galaxy models for measuring photometric redshift differ significantly from the ones used to derive  stellar mass and SFR. Attempts to realize a comprehensive set of templates, able to estimate simultaneously the aforementioned properties, have been recently pursued without exploiting ML algorithms   \citep[e.g.,][]{battisti19}. This issue is beyond the scope of the present study, as we assumed a known redshift for each galaxy -- not necessarily a high-precision estimate, but able to select $z<2$ targets with sufficient confidence  \citep[see also][]{hemmati19b}. Nonetheless, the SOM can also be used without $z_\mathrm{LePh}$ preselection by expanding the training to a larger redshift range \citep[up to  $z=4\!-\!6$,  as in][]{masters15,masters19,davidzon19} and including a stellar locus  \citep[e.g.][]{kim15,davidzon17}. In such a framework, to be developed in future studies, the number of high-$z$ interlopers would be negligible in most of the statistical applications.

\begin{acknowledgements}

The authors are grateful to the referee for the thorough and constructive comments they provided. 

ID would like to thank Eric Bell, Micol Bolzonella, Rebecca Bowler, Ivan Delvecchio, Joel Leja, Vihang Mehta  for useful discussions. The authors are grateful to Joel Leja also for providing \texttt{Prospector} data in a convenient digital format.

This work started during a meeting in Marseille, founded by  the French Agence Nationale de la Recherche for the project ``SAGACE''. This research is also partly supported by the Centre National d'Etudes Spatiales (CNES). 
ID, JRW, KJ, and OI  made use of the CANDIDE Cluster at the Institut d'Astrophysique de Paris and made possible by grants from the PNCG, CNES and the DIM-ACAV. The Cosmic Dawn Center is funded by the Danish National Research Foundation under grant No.\ 140. 

ID has received funding from the European Union's Horizon 2020 research and innovation program under the Marie Sk{\l}odowska-Curie grant agreement No.\ 896225. JRW acknowledges support from the European Research Council (ERC) Consolidator Grant funding scheme (project ConTExt, grant No.\ 648179). KM is grateful for support from the Polish National Science Centre via grant UMO-2018/30/E/ST9/00082. GEM acknowledges the Villum Fonden research grant 13160 “Gas to
stars, stars to dust: tracing star formation across cosmic time,” grant 37440 and “The Hidden Cosmos”.
DBS is grateful to Danmarks Nationalbank for their generous hospitality in the Nyhavn 18 residence during his extended visit to Copenhagen.

We warmly acknowledge the contributions of the entire COSMOS collaboration consisting of more than 100 scientists. The HST-COSMOS program was supported through NASA grant HST-GO-09822. More information on the COSMOS survey is available at \url{https://cosmos.astro.caltech.edu}.

A significant portion of this project  took  place  during  the COVID-19  global  pandemic; the authors would like to thank all those who made sacrifices in order for us to safely continue our research.   
       
\end{acknowledgements}

%
%
\bibliographystyle{aa} 
\bibliography{papers} 

\begin{appendix} 

\section{The SFR-$L_\mathrm{IR}$ relation in the literature }
\label{appendix1}

In the present study, and in those compared to it, infrared luminosity is used as a proxy of galaxy SFR. The SFR estimates considered in Sect.~\ref{sec:discussion} (but see also Fig.~\ref{fig:sfr_barro_vs_som} in Sect.~\ref{sec:results}) are based on one of the following scaling relations: 
\begin{itemize}
    \item $\mathcal{K}_\mathrm{K98}=1.72\times 10^{-10}$ from  \citet[][K98]{kennicutt98} assuming Salpeter IMF and using models of \citet{LeithererHeckman1995} for continuous bursts of age 10$-$100\,Myr.\footnote{Some articles report slightly different values, likely due to a different $L_\odot$, such as $\mathcal{K}_\mathrm{K98}=1.8\times 10^{-10}$ in \citet[][]{PerezGonzalez:2008p8736}.}
    \item $\mathcal{K}_\mathrm{B05}=0.98\times 10^{-10}$ from \citet[][B05]{bell05} with  \texttt{Pegase} models\footnote{\url{http://www2.iap.fr/users/fioc/PEGASE.html}} assuming Kroupa IMF and constant SFH. 
    \item $\mathcal{K}_\mathrm{M11}=1.48\times10^{-10}$ from \citet[][M11]{murphy11}, using models from \texttt{Starburst99\footnote{\url{https://www.stsci.edu/science/starburst99/docs/default.htm}}} with Kroupa IMF, $Z_\odot$, and continuous star formation up to 100 Myr age. 
\end{itemize}
%
%
Our goal is to set a common ground for the comparison:  results from the literature obtained through either the K98 or M11 calibration shall be converted to B05, which is the one used in the SOM method   (Eq.~\ref{eq:SFR_UVIR}). In our analysis   $\mathcal{K}_\mathrm{B05}=0.86\times10^{-10}$  because we convert the original value to Chabrier IMF following \citet{arnouts13}. 

Inconsistencies between studies may persist if  information about the scaling relation and/or IMF conversion is lacking in one of the articles. 
For example in \citet{whitaker12} the authors cite K98 as the reference for the value $0.98\times10^{-10}$ they use, which is actually $\mathcal{K}_\mathrm{B05}$.  
In \citet{barro19} the authors use $\mathcal{K}=1.09\times10^{-10}$ (Chabrier IMF) and quote K98 as the source. The same value is attributed  to BC05 elsewhere \citep[e.g.,][]{straatman16} creating some confusion. We deem \citeauthor{barro19} to be correct, as $1.09\times10^{-10}$ is indeed equal to  $\mathcal{K}_\mathrm{K98}\times 0.63$, with 0.63 being the conversion factor from Salpeter to Chabrier IMF provided in \citet[][]{madau&dickinson14}. 

The IMF conversion itself is another source of discrepancy. Using the prescription from \citet{madau&dickinson14},  $\mathcal{K}_\mathrm{B05}$ would become either $1.46\times10^{-10}$  or  
$0.92\times10^{-10}$ with Salpeter or Chabrier IMF respectively. The latter can be directly compared to the value provided in \citet[][]{arnouts13}, which is $\sim$10\% smaller.
However, there is not a unique prescription: in the literature one can find other IMF conversions resulting  e.g.\ into  $\mathcal{K}_\mathrm{B05}=1.33$ \citep{cowie&barger08} or $1.56\times10^{10}$  \citep{franx08}, both for a Kroupa to Salpeter IMF conversion. It  is important to underline that every  published value of  $\mathcal{K}_\mathrm{B05}$ (with Salpeter IMF) we found in the literature  remains  $\lesssim30\%$ smaller than $\mathcal{K}_\mathrm{K98}$, as also noted in B05. 
Also noteworthy is the fact that when $\mathcal{K}_\mathrm{M11}$ is retrieved from the review  of \citet[][table 1]{kennicutt&evans12}  the  $\mathcal{K}_\mathrm{K98}$ conversion quoted along with that scaling relation is potentially misleading.\footnote{ \citeauthor{kennicutt&evans12} explain in  section 3.8 that their conversion factor, namely $\dot M_\star / \dot M_\star$(K98), takes into account the difference between each reviewed study and K98 in both the IMF \textit{and} the star formation models. Unfortunately, this detail is not noticed in table 1 (which is located at the end of their article) and inattentive readers may misinterpret the values reported  in table 1. }  

The IMF conversion factor is usually derived from stellar population synthesis models, taking the ratio between a model assuming Salpeter and its analog with Chabrier (or Kroupa) IMF. Different conversion factors are originated from different assumptions to build these stellar populations. 
For example, they are often calculated after 100\,Myr of stellar evolution \textit{under the assumption of a constant SFH} \citep[e.g.,][]{madau&dickinson14}. This may not be appropriate for $L_\mathrm{IR}$, as the mean age of stars contributing to that emission is 5\,Myr \citep{kennicutt&evans12}. In that case, the conversion factor from Salpeter  to a bottom-lighter IMF) should be smaller.  

In conclusion, we acknowledge that these discrepancies at a few percent level play a secondary role in the analysis, compared to the impact of other systematic effects due e.g.\ to data reduction or to the photometric redshift uncertainty propagating into the $M$ and SFR error budget.  Nonetheless it is important to clarify any ambiguity in the literature when the opportunity can be taken, for future references but also for pedagogical purposes.

\section{Additional tests for SOM implementation}
\label{appendix2}

\begin{figure}
    \centering
    \includegraphics[width=0.99\columnwidth]{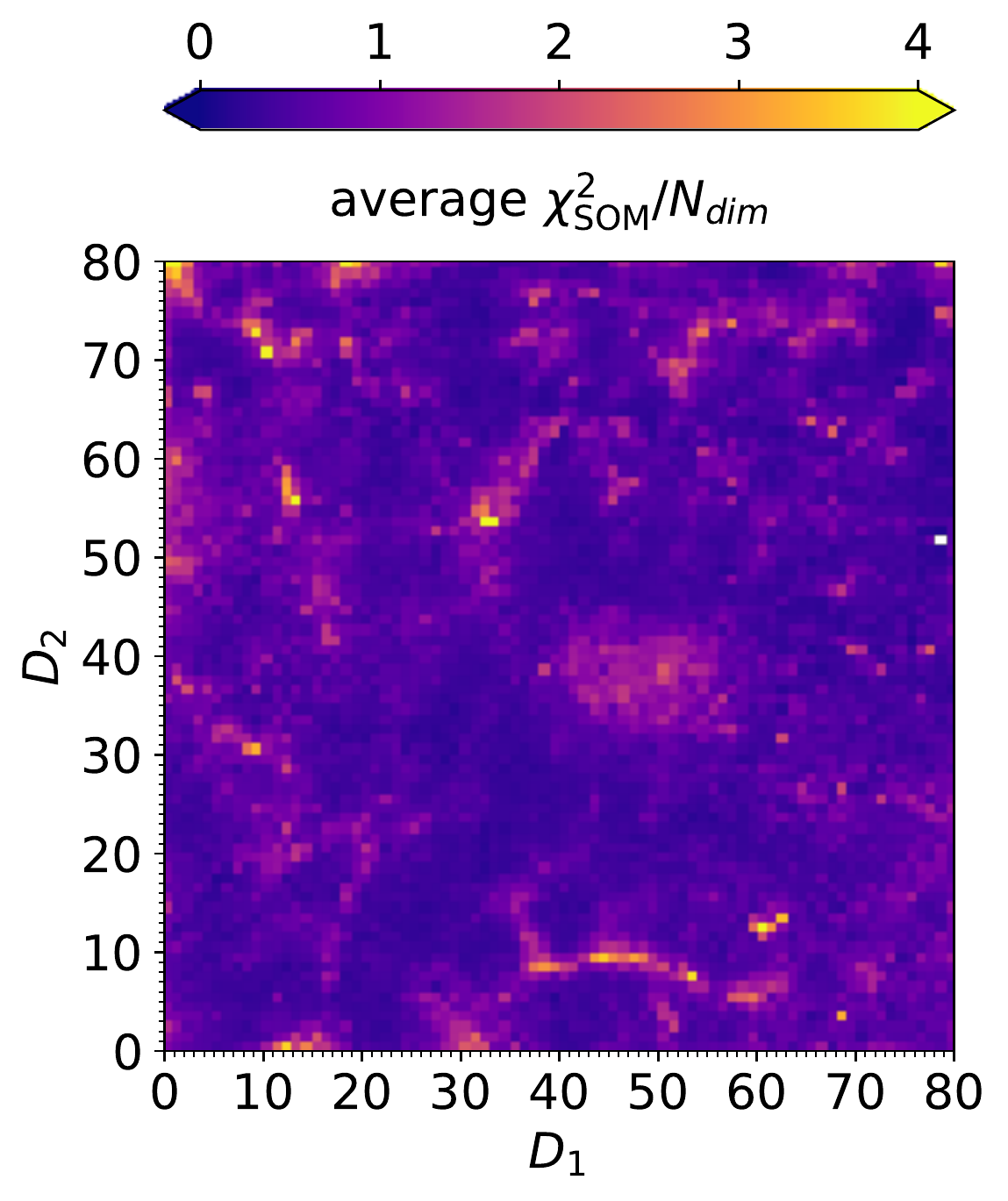}
    \caption{Similarly to middle panel of Fig.~\ref{fig:som_train}, this plot show the SOM grid color-coded according to the average $chi^2_\mathrm{SOM}$ (see Eq.~\ref{eq:chi2_som}) of galaxies in a given cell, normalized by the number of dimension of the manifold ($N_\mathrm{dim}=11$ colors).  }
    \label{fig:som_chi2}
\end{figure}

In Sect.~\ref{subsec:meth_basics} we computed $\Delta$ distances to quantify how well the SOM weights conform to the galaxy data manifold  (see Eq.~\ref{eq:d_som}). A variation of is  a $\chi^2$ distance 
\begin{equation}
    \chi^2_\mathrm{SOM} =  \sum_{N_\mathrm{dim}} \frac{(f_i - w_i)^2}{\epsilon_i^2},
\label{eq:chi2_som} 
\end{equation}
which takes into  account not only the difference between  observed color ($f_i$) and SOM weight ($w_i$) but also the photometric error of the former ($\epsilon_i$).  This alleviates some of the tension highlighted in Fig.~\ref{fig:som_train} (middle panel). In fact,  the average  $\chi^2_\mathrm{SOM}/N_\mathrm{dim}$ per cell (Fig.~\ref{fig:som_chi2}) has a  narrower spread and  regions that have been discussed in Sect.~\ref{subsec:discuss_soundness} as $\Delta$ outliers are not as prominently  highlighted here in Fig.~\ref{fig:som_chi2}  because the large $(f_i - w_i)^2$ values in those cells are compensated by the comparably large uncertainty in the  $f_i$ photometric measurement.

Thus, a combination  of $\Delta$ and $\chi^2_\mathrm{SOM}$ metrics is an efficient way to separate issues related to observational (statistical) errors vs.\ intrinsic failures of the  SOM when the weight is not  close enough to data despite the precision of the latter. Besides the average in each cell, we can also inspect the reduced $\chi^2$ (i.e.,  $\chi^2_\mathrm{SOM}/N_\mathrm{dim}$, assuming $N_\mathrm{dim}$ degrees of freedom) of individual galaxies to  find which ones are misrepresented by their weight. This happens e.g.\ at edges of the grid, where the SOM is forced to pack together sparse objects from the outskirt of the data distribution. A bad match between certain galaxies and the weight linked to them can also concern cells in the interior of the grid (see the cells marked in yellow in Fig.~\ref{fig:som_chi2}) since a non-linear manifold such as our $0<z<1.8$ galaxy sample is hard to describe with a relatively limited  network of weights. It is nonetheless reassuring that only 4.4\% of the COSMOS2020 galaxies overall have  $\chi^2_\mathrm{SOM}>24.7$; this should be the 99th  percentile  threshold of a $\chi^2$ distribution with $N_\mathrm{dim}=11$ degrees of freedom: the fact that more than 1\% of the sample exceeds the threshold can be ascribed partly to the SOM modeling, which is not expected to be a perfect description, but also to the fact photometric errors may not be entirely accurate in the COSMOS2020 catalog.\footnote{In fact, the native error bars from \texttt{SourceExtractor} were considered underestimated by \citet{weaver22}, and needed an adjustment with heuristic boosting factors. }

\begin{figure}
    \centering
    \includegraphics[width=0.98\columnwidth]{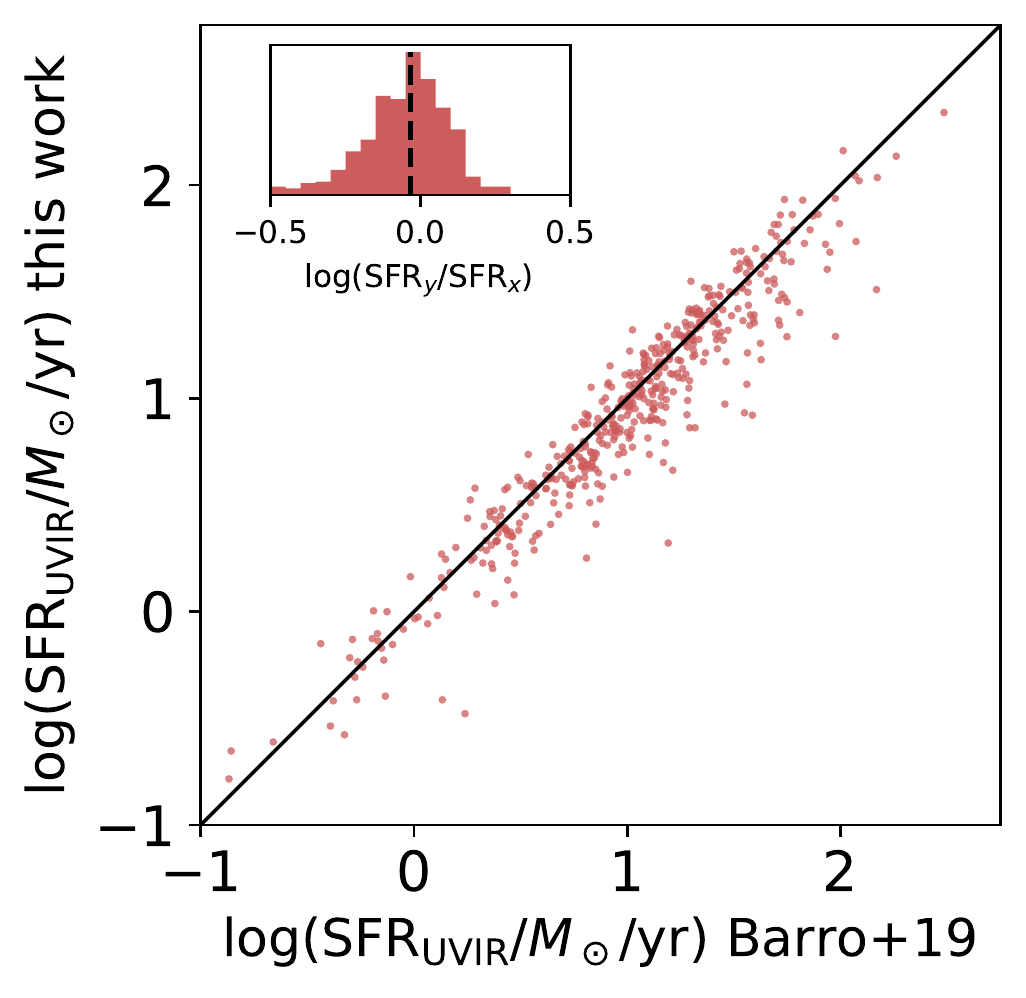}
    \caption{Comparison between two methods deriving SFR from rest-frame IR, both applied to 509 galaxies at $z<1.8$ in the COSMOS-CANDELS area (red dots). The method used in the present work follows \citet{ilbert15} and is summarized in Sect.~\ref{subsec:meth_calib} (see Eq.~\ref{eq:SFR_UVIR}). The SFR$_\mathrm{UVIR}$ estimates on the $y$ axis are described in \citet{barro19}.  A solid line shows the 1:1 relationship. The inset in the top-left corner shows a histogram of the difference between \citeauthor{barro19} (labeled SFR$_y$) and our estimates (SFR$_x$), and a vertical dashed line that marks the median offset at $-0.03$\,dex. }
    \label{fig:sfr_barro_vs_ilbert}
\end{figure}

Another test we implemented concerns  the \citet{barro19}  catalog, which  is used in this work as an independent source of information to verify the validity of SOM-based estimates. Since the SOM method is calibrated with SFR$_\mathrm{UVIR}$ measured as in \citet{ilbert15}, any difference between that and \citet{barro19} will propagate in the comparison between SFR$_\mathrm{SOM}$  and SFR$_\mathrm{UVIR}$ from \citet{barro19}.  Figure~\ref{fig:sfr_barro_vs_ilbert} shows our SFR$_\mathrm{UVIR}$ vs.\   \citeauthor{barro19} values (after homogenization, see Appendix~\ref{appendix1}). The latter are obtained by averaging four sets of FIR templates, including \citet{dale&helou02} which are the ones we used. Another important distinction between our approach and  \citeauthor{barro19} concerns objects with only the 24\,$\mu$m detection:  they use an  analytical prescription  \citep{wuyts11a} to convert the 24\,$\mu$m flux into $L_\mathit{IR}$, while we fit templates also in that case. It is also possible that the input SEDs differ, for example in \citet{barro19} the prior for the FIR photometric extraction and deblending  are the CANDELS sources. Nevertheless, the SFR estimates are in good agreement (Fig.~\ref{fig:sfr_barro_vs_ilbert}) with a contained scatter comparable to the typical (0.3\,dex) uncertainty of this kind of measurement. Our  SFR$_\mathrm{UVIR}$ values are slightly larger, by about 8\% (0.03\,dex in the logarithmic comparison in the Figure). 

\begin{figure}
    \centering
    \includegraphics[width=0.99\columnwidth]{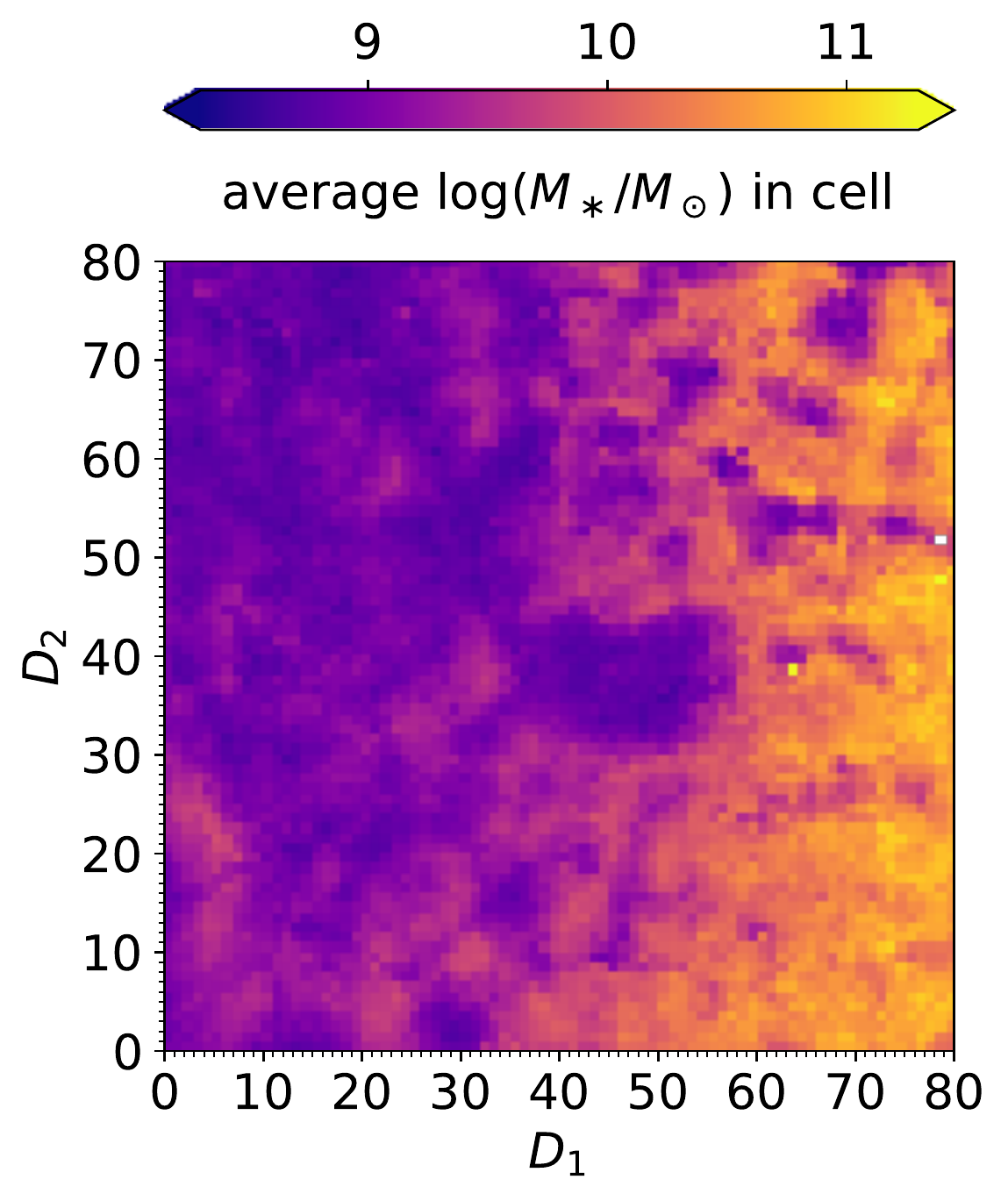}\\
  \includegraphics[width=0.99\columnwidth]{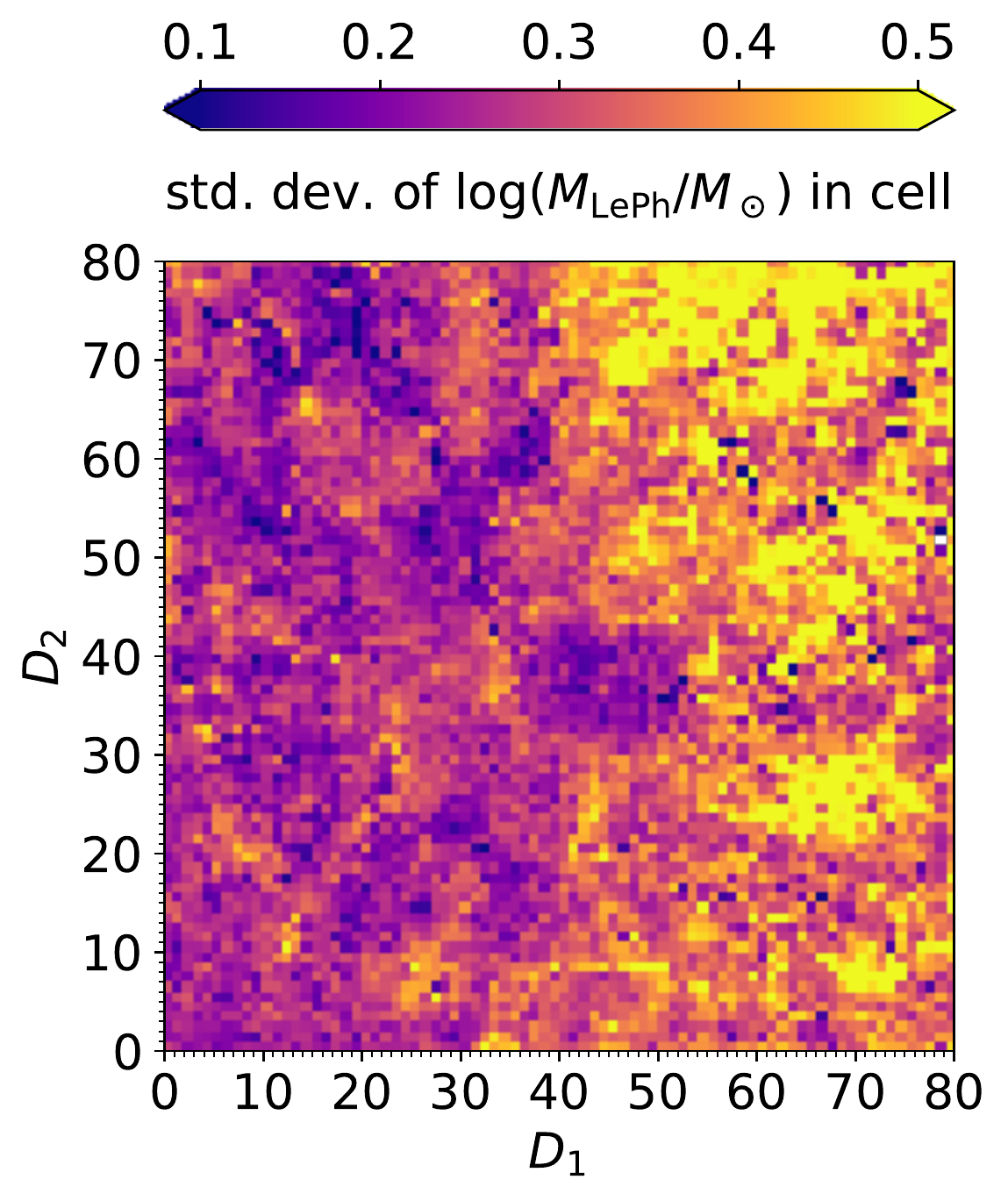}   
    \caption{\textit{Upper panel:}  average $\log(M_\mathrm{LePh}/M_\odot)$ of COSMOS2020 galaxies within the same cell,  coded according to the color bar above the SOM grid  (white cells are empty). \textit{Upper panel:}  standard  deviation  of  the  same  $\log(M_\mathrm{LePh}/M_\odot)$ distribution per individual cell,  coded according to the color bar above the SOM grid  (also in this case, empty cells are colored in white).  }
    \label{fig:som_mstar_map}
\end{figure}

Another test we made is summarized in Fig.~ \ref{fig:som_mstar_map}. The goal is to verify the ``clustering'' of low-mass galaxies in the SOM grid, which alleviate the scatter in the rescaling process of Eq.~(\ref{eq:mass_resc}). Even though we did not use individual fluxes as features, the clustering is a consequence of two combined  properties: the relationship between  colors and  $M$-to-light ratio \citep[e.g.,][]{bell&dejong00,bell&dejong01,zibetti09} and the distinctive $M$-to-light ratios that low-mass galaxies have compared to those at $M>10^{10}\,M_\odot$  \citep[e.g.][]{bundy06,moffett16,cui21}.

\end{appendix}

\end{document}